%% file: mnras_template.tex
\documentclass[fleqn,usenatbib]{mnras}

\usepackage{newtxtext,newtxmath}

\usepackage[T1]{fontenc}
\usepackage{ae,aecompl}

\usepackage{graphicx}	
\usepackage{amsmath}	
\usepackage{amssymb}	
\newcommand{\angstrom}{\textup{\AA}}
\usepackage{siunitx}

\title[Drivers of Type 1 AGN Diversity]{Exploring the Diversity of Type 1 Active Galactic Nuclei Identified in SDSS-IV/SPIDERS}

\author[Wolf et al.]{
Julien Wolf,$^{1}$\thanks{E-mail: jwolf@mpe.mpg.de}
Mara Salvato$^{1}$,
Damien Coffey$^{1}$,
Andrea Merloni$^{1}$,
Johannes Buchner$^{1}$,
\newauthor Riccardo Arcodia$^{1}$,
Dalya Baron$^{2}$,
Francisco J. Carrera$^{3}$,
Johan Comparat$^{1}$,
\newauthor Donald P. Schneider$^{4,5}$,
Kirpal Nandra$^{1}$
\\
$^{1}$Max-Planck-Institut f\"{u}r extraterrestrische Physik, Gie\ss enbachstra\ss e 1, 85748 Garching, Germany \\
$^{2}$School of Physics and Astronomy, Tel-Aviv University, Tel Aviv 69978, Israel \\
$^{3}$Instituto de F\'{i}sica de Cantabria (CSIC-Universidad de Cantabria), 39005 Santander, Spain \\
$^{4}$ Department of Astronomy and Astrophysics, The Pennsylvania State University, University Park, PA 16802, USA \\
$^{5}$ Institute for Gravitation and the Cosmos, The Pennsylvania State University, University Park, PA 16802, USA
}

\date{Accepted XXX. Received YYY; in original form ZZZ}

\pubyear{2019}

\begin{document}
\label{firstpage}
\pagerange{\pageref{firstpage}--\pageref{lastpage}}
\maketitle

\begin{abstract}
We present a statistical analysis of the optical properties of an X-ray selected Type 1 AGN sample, using high signal-to-noise ratio (S/N>20) spectra of the counterparts of the ROSAT/2RXS sources in the footprint of the SDSS-IV/SPIDERS (Spectroscopic IDentification of eROSITA Sources) programme.
The final sample contains 2100 sources. It significantly extends the redshift and luminosity ranges ($\rm z \sim 0.01- 0.80$ and $\rm L_{0.1-2.4 \, keV} \sim \rm 2.0 \times 10^{41} - 1.0 \times 10^{46} \,    \, erg \, s^{-1}$) used so far in this kind of analysis. By means of a Principal Component Analysis, we derive Eigenvector (EV) 1 and 2 in an eleven dimensional optical and X-ray parameter space, which are consistent with previous results. The validity of the correlations of the Eddington ratio $\rm L/L_{Edd}$  with EV1 and the black hole mass with EV2 are strongly confirmed 
These results imply that $\rm L/L_{Edd}$  and black hole mass are related to the diversity of the optical properties of Type 1 AGN. Investigating the relation of the width and asymmetry of $\rm H\beta$ and the relative strength of the iron emission $\rm r_{FeII}$, we show that our analysis supports the presence of a distinct kinematic region: the Very Broad Line Region. Furthermore, comparing sources with  a red-asymmetric broad $\rm H\beta$ emission line to sources for which it is blue-asymmetric, we find an intriguing difference in the correlation of the $\rm FeII$ and the continuum emission strengths. We show that this contrasting behaviour is consistent with a flattened, stratified model of the Broad Line Region, in which the $\rm FeII$ emitting region is shielded from the central source.

\end{abstract}




\input{01-Introduction}

\input{02-Data}
\input{03-AsymmetryDistribution.tex}
\input{04-PCA.tex}

\input{05-DerivedParameters.tex}

\input{06-Asymdriver.tex}
\input{07-BRSep.tex}
\input{08-Conclusions}

\input{09-Acknowledgements.tex}




\bibliographystyle{mnras.bst}
\bibliography{mybib} 





\appendix

\input{Appendix.tex}


\bsp	
\label{lastpage}
\end{document}

%% file: 01-Introduction.tex
 \section{Introduction}

The exploration of the multi-wavelength emission of Active Galactic Nuclei (AGN), including their spectral features, has motivated the development of unification schemes for this class of persistent and extremely luminous objects (e.g., \citealt{antonucci93,urry95, netzer15,Padovani17}, see also \citealt{Elitzur12} for some caveats about such unification schemes). Simplifying the complex and historically developed AGN zoo is an appealing idea, since most of the identified species share common spectral properties over large ranges of 
luminosity and redshift. In this context, two main sub-classes have been defined based on their optical emission line properties: Type 1, which have both broad and narrow optical/UV emission lines, and Type 2, which show only narrow lines. These two classes are unified via the hypothesis that the broad lines in Type 2 sources are obscured (e.g., \citealt{netzer08}; \citealt{hickox18}). There is nonetheless substantial variety in the properties within the Type 1 population. The broad lines encode information about the geometry and the kinematics of the inner regions of the system : the Broad Line Region (BLR, e.g., \citealt{rees89, peterson06, gaskell09}). Photoionised by the continuum emission of the central source, the BLR emits transition lines, which are expected to be broadened by the motion of the gas (e.g.,  \citealt{baldwin95, king16}). 
The virial broadening of the emission lines can, under certain assumptions, be used to estimate the mass of central super-massive black holes (SMBH). This mass estimation method is calibrated using estimates of the BLR radius from reverberation mapping. \citep{bahcall72,blandford82,peterson93,gebhardt00,kaspi05,bentz15,shen15}.  The response delay of the BLR emission to variations in the ionising continuum emission from the central source (i.e., the accretion disk) is used to determine the distance between the BLR and the nucleus, assuming a Keplerian velocity field.


 Broad, low-ionisation lines, such as $\rm{H\rm\beta}$ (rest-frame wavelength at $4862 \angstrom$) and the $\rm{{MgII\lambda2800\angstrom}}$ doublet\footnote{In an effort to simplify the notation, we will henceforth refer to this line as $\rm MgII$}, are generally thought to yield more reliable virial broadening estimators than, for instance, the high ionisation transition $\rm CIV\lambda1549\angstrom$ (e.g., \citealt{trakhtenbrot12, mejia16, mejia18}). This is in part due to the presence of prominent blueshifted components in the $\rm CIV\lambda1549\angstrom$ line profiles of strong accretors, which may not be related to the orbital motion of the gas and would affect the inferred velocities (e.g., \citealt{richards02,Zamanov02, marziani17, sulentic17}).

 However, $\rm H\beta$ lines possess profiles which are difficult to reconcile with simple Keplerian motion around the central black hole, and could significantly affect the black hole mass estimates \citep{negrete18}. Accurate black hole mass measurements are important for the development of galaxy evolution models. It is therefore critical to understand the mechanisms responsible for the observed broad line shapes. 
 
The debate on the actual BLR kinematics, physical processes and geometry has flourished through the past decades; the variety of the often asymmetric (sometimes double-peaked) broad lines and their variability has lead to an abundant diversity of competing models. It was not until recently that the kinematics of this region could be resolved for the first time, favouring a flattened BLR as an orbiting extension of the accretion disk \citep{gravity18}.

A flattened BLR geometry was initially supported by the relation between the width of $\rm H\beta$ and the radio core-to-lobe ratio.  \citep{wills86,jarvis06,brotherton15} and by the double shoulders and peaks observed in certain broad emission lines \citep{oke87,chen89a,chen89b,eracleous94,eracleous03,storch16}.  
The double peaked profile variability has been shown not to correlate with continuum changes. The peak shifts are therefore not a reverberation effect \citep{wanders96}.
The strong variability of the double-peaked emission profiles has also been discussed in the light of different disk models by \cite{lewis10}, which can account for some of the observed line profiles.

A potentially useful approach to constrain and explore the inner structure of AGN is to acquire an understanding of their spectral diversity. For instance, in their seminal paper \citet{Boroson92} performed a Principal Component Analysis (PCA) of a set of optical, radio and X-ray features of 87 Type 1 AGN.
Through the orientation of the first principal component (Eigenvector 1, EV1) in their parameter space, they established an anti-correlation of the equivalent width of the narrow $\rm [OIII]$ lines at $4959 \angstrom$ and $5007 \angstrom$ with the relative strength of the iron emission $\rm r_{FeII}=F(FeII\lambda 4570)/F(H\beta)$, and the full width at half maximum (FWHM) of the H$\beta$ line as central markers of diversity in their sample. 
A large range of subsequent studies have investigated the EV1 correlation planes (e.g., \citealt{sulentic00a, Sulentic00b, marziani01, shang03, Grupe04, kuraszkiewicz09, mao09, tang12}). In particular, \cite{sulentic00a,Sulentic00b} and \citet{marziani01} established the foundations of the four-dimensional EV1 (4DE1) formalism, 
including $\rm FWHM_{H\beta}$ and $\rm r_{FeII}$ as two of the main correlates of EV1 in \citet{Boroson92}.  
These two quantities are respectively related to the black hole mass (since $\rm FWHM_{H\beta}$ is used as virial broadening estimator) and the Eddington ratio (e.g., \citealt{Marziani03b, netzer07, sun15}). 

 The 4DE1 parameter space was extended with the soft X-ray photon index $\rm \Gamma_{soft}$ \citep{boller96} and the centroid shifts from rest-frame wavelength of $\rm CIV\lambda1549$ at $50\%$ fractional intensity \citep{sulentic07a}.   
 The careful exploration of this parameter-space led to a proposed two population paradigm in the low-redshift universe, with an empirical separation at $\rm FWHM_{H\beta}\approx 4000 \, km \, s^{-1}$ (\citealt{sulentic00a, Sulentic00b}, see section 5 of \citealt{marziani18} and references therein for a full review of the evidence supporting the population A/B classification.). \citet[henceforth Z10]{Zamfir10} studied the characteristics of the $\rm H\beta$ emission of $477$ optically selected AGN ($\rm z < 0.7$) in the optical plane of 4DE1 (i.e., $\rm FWHM_{H\beta}$ vs. $\rm r_{FeII}$). They reported that the line shapes distinctively changed along the sequence, the asymmetry and shifts scaling with both of the optical dimensions.
 
In this work, we explore the EV1 correlation space exploiting high signal-to-noise ratio optical spectra for the largest sample of  X-ray selected AGN used to date for this type of analysis.
 
The X-ray selection allows to construct a less biased sample of AGN, since the host X-ray related processes are believed to be particularly weak compared to high-energetic AGN emission. At the same time X-rays are able to penetrate the host galaxies and in general large column densities, thus making the sample more complete than with other AGN selection techniques \citep[e.g.,][]{Singh13,Padovani17}.
The AGN in our sample were detected by ROSAT \citep{Boller16, Voges:1999ju}, while their multi-wavelength counterparts were determined in \citet{Salvato17} and spectroscopically followed-up in  the SDSS-IV/SPIDERS program \citep{Dwelly17}. These sources are exclusively Type 1 AGN and were presented in \citet[][henceforth C19]{coffey19} where sources were selected according to the width of their broad lines.

The final sample contains 2100 sources, which is a factor of 4-18 larger than samples used in previous studies of this type. It significantly extends the probed redshift and luminosity ranges ($\rm z \sim 0.01- 0.80$ and $\rm L_{0.1-2.4 \, keV} \sim \rm 2.0 \times 10^{41} - 1.0 \times 10^{46} \,    \, erg \, s^{-1}$). We perform a PCA on the optical and X-ray features of these sources, which allows us to determine the markers of spectral diversity in our sample.
We specifically investigate the role played by the Eddington ratio and the black hole mass for the total variance in our data. The asymmetry of the broad $\rm H\beta$ contributes strongly to the orientation of EV1 through the chosen spectral parameter space. We thus follow in the footsteps of Z10 and connect our global statistical results to the shape of the broad $\rm H\beta$ emission line, which is discussed in the context of BLR stratification and low-ionisation outflows.

The paper is organised as follows: 

In Sec. \ref{Dataset} we briefly present the SPIDERS AGN value added catalogue of C19 and how we created our subs-ample. We also outline how the additional properties, which were not provided by the catalogue, were determined. In Sec. \ref{asymmetrydistribution} we describe the distribution of the emission line asymmetry index. 
The core results of the direct correlation analysis and PCA on our sample are presented in Sec. \ref{PCAandDirect}. We demonstrate how the derived physical parameters Eddington ratio and central black hole mass scale with the obtained principal components in Sec. \ref{EDDMBHsec}. 
The distribution of our sample in the EV1 plane and the possible presence of a distinct kinematic BLR region is investigated in Sec. \ref{asymdriversec}. We divided the sample in two distinct subsets according to the sign of their $\rm H\beta$ asymmetry indices. From the comparison of parameter correlations in the two sub-samples we detect an interesting contrast in the scaling relation between source luminosity and the equivalent width of the iron emission. 
This is discussed in Sec. \ref{BRSEPAB}. Conclusions are presented in Sec. \ref{sec:conclusions}

Throughout this work, we adopt a standard cosmology: $\Omega_{\rm M} = 0.3$, $\Omega_\Lambda = 0.7$ and $\rm H_0=70 \, km \,s^{-1} \, Mpc^{-1}$.

%% file: 02-Data.tex
\section{Data}
\label{Dataset}
The sample of Type 1 AGN analysed in this work is extracted from the SDSS-IV/SPIDERS AGN catalogue, presented in C19.
The original catalogue compiles spectral information through SDSS DR14\footnote{\href{https://www.sdss.org/dr14/data_access/value-added-catalogs/}{SDSS DR14 VACs}} \citep{abolfathi18}, for 7344 Type 1 AGN detected in the Second ROSAT All-Sky Survey ROSAT/2RXS, \citealt {Boller16}) and 1157 Type 1 AGN in the XMM-Newton Slew 1 catalogue (XMMSL1, \citealt{Saxton08}).

The multi-wavelength counterparts of these X-ray sources were determined via the Bayesian-based code NWAY by \citep{Salvato17}, using a prior \citep{Dwelly17} based on AllWISE data \citep{cutri13}.

Optical spectra of the counterparts were obtained from two different spectrographs, \href{https://classic.sdss.org/dr7/instruments/spectrographs/index.html}{SDSS} and \href{https://www.sdss.org/instruments/boss_spectrograph/}{BOSS}. These instruments cover different optical wavelength ranges: $\rm  3800 - 9200 \, \angstrom $ for SDSS (counting both channels, with a spectral resolution ranging from 1850 to 2200) and $\rm 3600 - 10400 \, \angstrom$ for BOSS (spectral resolution of 1560-2270 in the blue channel, 1850-2650 in the red channel). 
A technical summary of the SDSS-IV survey is provided by \citet{blanton17}. The instrument is presented in \citet{gunn06}. Finally, a detailed description of the SDSS and BOSS spectrographs is given by \citet{smee13}.

For the sources in the C19 catalogue, the authors fit the spectroscopy following a procedure detailed described in C19. Here we report only the key features.

In C19, the spectral regions centered around the $\rm H_\beta$ and $\rm MgII$ lines ($4420-5500\, \angstrom$   and $2450-3050\, \angstrom$, respectively) were fit with a multi-component continuum model and a series of Gaussian functions. The present work will specifically focus on the broad $\rm H\beta$ emission line and the narrow $\rm [OIII]\lambda 4959\angstrom$ and $\rm [OIII]\lambda 5007\angstrom$ forbidden transition lines. C19's fitting algorithm used up to four Gaussian functions to fit the $\rm H\beta$ line.  The best-fit model (and thus the required number of Gaussian components) was automatically selected on the basis of the Bayesian Information Criterion (BIC).  Of the four Gaussians, one accounts for the narrow core, while the remaining potential three components of $\rm H\beta$ are defined as broad if they fulfill the criterion: $\rm FWHM_{H\beta} > 800 \, km \, s^{-1}$. This relatively complex model allows one to trace not only typical broad bimodal profiles above $\rm \sim 1000 km \, s^{-1}$, but also distinct broader components (see C19, Section 8.1, and see Section \ref{asymdriversec} of this work). For instance \citet{marziani10} propose a broad line decomposition into broad, very broad and blue components, which the model defined in C19 may individually trace.

 Up to two Gaussians were used for each of the [OIII] lines: one fitting the narrow core and one tracing shifted wings.  
The continuum model was constructed from a power-law, a host galaxy component and an FeII template.
The iron emission and the galaxy contribution are obtained respectively from a normalized Narrow-Line Seyfert 1 I Zw 1 template \citep{Boroson92} and an early type galaxy template\footnote{\href{http://classic.sdss.org/dr5/algorithms/spectemplates/}{SDSS spectral cross-correlation templates, Template 24}}.  Morphological studies \citep[e.g.,][]{grogin05} have shown that AGN are mainly found in bulge dominated galaxies. Bulges are predominantly home to old star populations. C19 thus justify the use of an early-type galaxy template for the spectral host contribution by the fact that SDSS fibers sample only the central regions of targeted galaxies. The early-type approximation for the host contribution was also used in previous work (e.g. \citealt{calderone17}). The template has $\rm H\beta$ absorption features. Comparing a sub-sample with strong host contribution to a sub-sample with low host contribution at z<0.2, we found no significant impact of these features on $\rm H\beta$ line shape diagnostics.

The fitting parameters are listed in the catalogue of C19 along with monochromatic continuum and X-ray luminosities and derived parameters such  as bolometric luminosities, estimates for the black hole masses and Eddington ratios.

The single-epoch mass estimation method was used for the derivation of the black hole masses (\citealt{Vestergaard02,Mclure02,Vestergaard06,Assef11,Shen12} and \citealt{Shen13} for a review). The masses stem from the \citet{Assef11} calibration, which builds on the $\rm FWHM$ of $\rm H\beta$ and the BLR radius-luminosity relation from \citet{bentz09}. For this paper, we have reconstructed the best-fit models from the parameters listed in C19. Two examples of $\rm H\beta$ fit reconstructions are shown in Appendix \ref{sec:eqw} of the present work. Details of the broad $\rm H\beta$ line decomposition are also presented.

In the following section we describe the sample construction for our statistical analysis. 

\subsection{Sample construction and selection pipeline}
\label{sec:sample_construction}

We are primarily interested in Type 1 AGN spectral properties which can be associated with the physics of the BLR and the Narrow Line Region (NLR). We thus assembled a parameter subset inspired by the classical quasar PCA papers by \citet{Boroson92} and \citet{Grupe04}. They included emission properties of $\rm H\beta$, $\rm FeII$, $\rm [OIII]$ line at $\rm 5007 \angstrom$\footnote{ We choose to work with the [OIII] transition line at  5007 $\angstrom\,$  and not 4959 $\angstrom\,$, since this line will be less affected by interline fit contamination with $\rm H\beta$. We argue that any blue asymmetries and shifts in $\rm [OIII]\lambda5007\angstrom$  should be detectable in $\rm [OIII]\lambda4959$\angstrom. Throughout this paper, $\rm [OIII]\lambda5007\angstrom$  will be referred to as $\rm [OIII]$  } and $\rm HeII\lambda1640\angstrom$, as well as X-ray and optical monochromatic luminosities to trace the emission of the hot electron corona and the accretion disk.

The inclusion of X-ray luminosities forces a restriction  to just one of the two X-ray surveys compiled in the C19 catalogue, since 2RXS and XMMSL1 soft X-ray fluxes are measured in different bands, at different times, from two different satellites/instruments and different assumptions used for computing the flux (see \citealt{Saxton08,Boller16}). We confined the analysis to sources detected in 2RXS only (7344 sources), as they were six times more numerous than XMMSL1 sources in the SPIDERS AGN sample. More specifically we retained 2RXS sources, for which the X-ray flux in the $\rm 0.1-2.4 \, keV$ band is provided in C19 (7154 sources). The missing 190 sources have no X-ray flux measurement listed. They correspond to sources with low photon count-rates.

Since we use the monochromatic luminosity at $5100 \, \angstrom$ to trace the continuum emission, we limit the sample to z=0.8  so that the rest-frame $5100 \, \angstrom$ are covered by the spectral window.

The redshift  cut  ensures the presence of fits of the $\rm H\beta$ centered region. These constraints yield 5926 sources.

\begin{figure}
\includegraphics[width=8 cm]{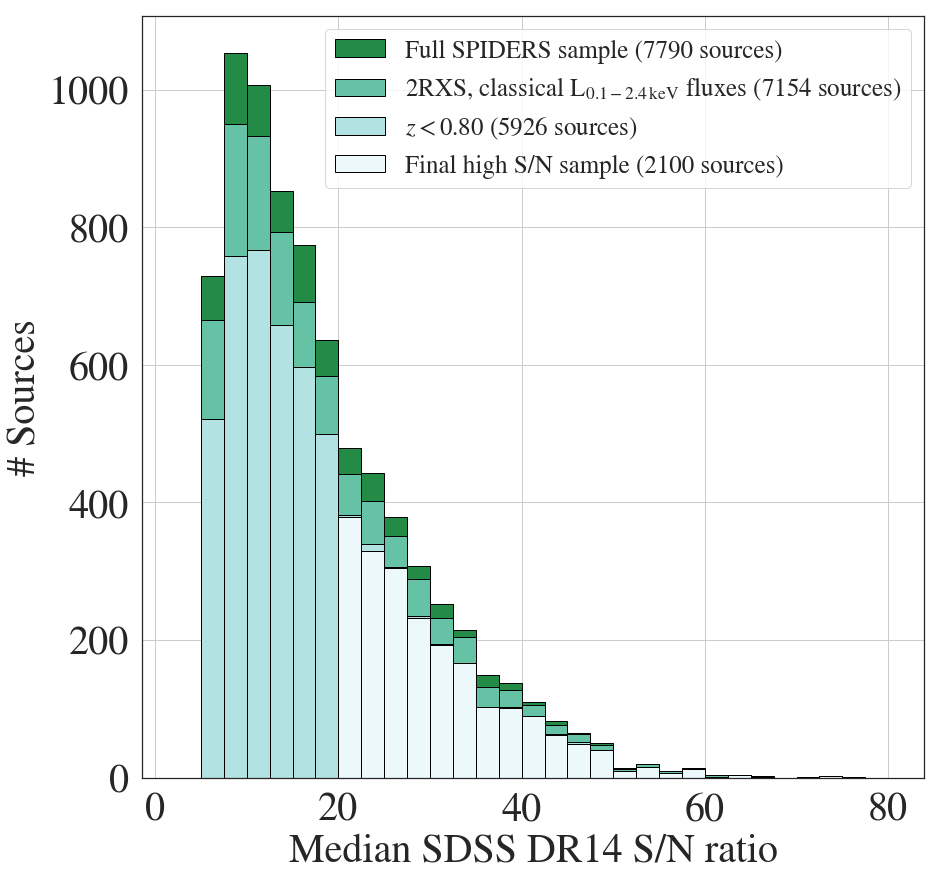}
\caption{Distribution of the median S/N ratios per resolution element of the  2100 SDSS DR14 sources in our final sample(white). Also presented are sub-samples of the full SPIDERS sample in C19 after two consecutive quality cuts: the presence of a $\rm 0.1-2.4 \, keV$ flux and the redshift restriction $\rm z < 0.8$. }
\label{fig:sn}
\centering
\end{figure}

We further limit our analysis to sources with good SDSS signal-to-noise ratio per resolution element in order to improve the significance of our results. Our analysis relies on the accurate measurement of line shapes, and low-quality spectra might strongly affect the fits.  For this reason we consider only sources with median S/N $\geq 20$. This critical restriction drops our total source count to 2124 objects.
The SDSS signal-to-noise ratio distribution of the original sample and its consecutive cuts are shown in Fig. \ref{fig:sn}.

Finally, 24 sources were excluded due to a lack of black hole mass estimates, extreme uncertainties on the fit parameters of $\rm H\beta$ or absent $\rm [OIII]$. A visual inspection of these sources' spectra revealed that these issues might be linked to strong continuum emission.

\begin{figure}
\includegraphics[width=8 cm]{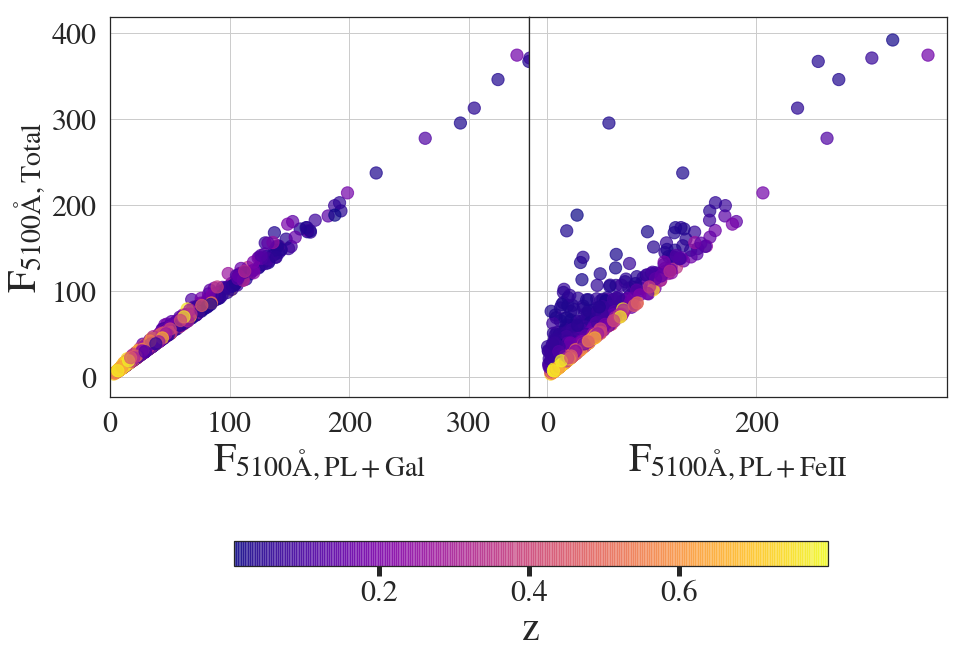}
\caption{The relation to the monochromatic fluxes at $\rm 5100 \angstrom$, as measured from the power-law model with FeII and host contributions ($\rm F_{5100\angstrom,Total}$), the power-law model with the FeII emission ($\rm F_{5100\angstrom,PL+FeII}$) and the power-law model with the host emission ($\rm F_{5100\angstrom,PL+Gal}$). The fluxes are in units of $\rm 10^{-17}\, erg\, cm^{-2}\,s^{-1}\, \angstrom^{-1}$.}
\label{fig:mon}
\centering
\end{figure}

 For the 2100 sources, the following subset of spectral properties was compiled (unless specified otherwise, all wavelengths, energies, equivalent widths are defined in the rest-frame):

\begin{itemize}
\item $\mathbf{FWHM_{H\beta}}$ : full width at half maximum of $\rm H\beta$ (broad component) $\rm [km\, s^{-1}]$
\item $\mathbf{F([OIII])/F(H\beta)}$ : flux ratio of $\mathbf{\rm [OIII]\lambda 5007 \angstrom}$ and $\rm H\beta$
\item $\mathbf{r_{FeII}= F(FeII)/F(H\beta)}$: flux ratio of $\rm H\beta$ and the total $\rm FeII$ (FeII emission in the $\rm 4434-4684\,\angstrom$ range)
\item $\mathbf{L_{5100{\angstrom}}}$ : optical continuum monochromatic luminosity at $5100$ \AA $\,$ from the power-law component (without host and FeII contributions)  $\rm [erg \, s^{-1}\angstrom^{-1}]$
\item $\mathbf{L_X}$ : soft observed X-ray luminosity in the 0.1-2.4 keV band $\rm [erg \, s^{-1}]$
\item $\mathbf{W(FeII)}$ : equivalent width of $\rm FeII$ ($ 4570$ \AA $\,$ blend)\footnote{cf. Appendix \ref{sec:eqw}\label{note1}} $\rm [\angstrom]$
\item $\mathbf{W([OIII])}$ : equivalent width of $\rm [OIII]\lambda 5007$\AA $\,$(full profile)\footnotemark[5]$\rm [\angstrom]$
\item $\mathbf{W(H\beta)}$ : equivalent width of the broad component of $\rm H_{\beta}$\footnotemark[5]$\rm [\angstrom]$
\item $\mathbf{W(HeII)}$ : equivalent width of $\rm HeII$ at $4686\,\angstrom$\footnotemark[5]$\rm [\angstrom]$
\item $\mathbf{\Delta \lambda _{[OIII]}}$ : asymmetry index for  $\rm [OIII]\lambda 5007$\AA \footnote{cf. Sec. \ref{sec:asymi}}
\item $\mathbf{\Delta \lambda_{ H\beta}}$ : asymmetry index for the broad component of $\rm H{\beta}$\footnotemark[6]
\end{itemize}

It is worth to highlight the following points:

$\bullet$ Sources for which the parameters relative to the $\rm FeII$ emission were not listed in C19, typically correspond to AGN with very weak FeII emission. We manually set their $\rm r_{FeII}$ parameters and iron equivalent widths to zero. We thus implicitly assume that their iron emission is too weak to be disentangled from the AGN continuum emission.

$\bullet$ All the spectral properties of our defined sub-set are available in C19 except for the equivalent widths and asymmetry indices which were derived as presented in Sec. \ref{sec:asymi} and Appendix \ref{sec:eqw}. 

$\bullet$ The $\rm L_{5100\angstrom}$ provided in C19 includes the host contribution. However, in the present analysis we aim at tracing the accretion disk emission with $\rm L_{5100\angstrom}$. We thus derived the monochromatic luminosity from the reconstructed power-law model at rest-frame, removing the FeII and host contribution.
The uncertainties are obtained from the errors on the normalization of the slope. 
The considerable degeneracy between the host, iron and power-law components at lower redshift ($z<0.2$) due to the stronger host contribution makes it challenging to correctly perform an AGN-host decomposition. It is the principal limitation of deriving monochromatic luminosities directly from the fitted power-law model.
The relative contributions of the FeII complex and the host galaxy to the monochromatic flux a $\rm 5100 \angstrom$ are illustrated in Fig. \ref{fig:mon}. While the FeII contribution at $5100 \angstrom$ to the total monochromatic flux is marginal over the complete redshift range, the host galaxy strongly contributes to the monochromatic flux at lower redshifts. 
 
$\bullet$ The X-ray fluxes were obtained by C19, following \citet{Dwelly17}, from an absorbed power-law ($\rm \Gamma =2.4$). In addition to these classical fluxes, C19 provide Eddington bias corrected 2RXS fluxes using a Bayesian prior (\citealt{kraft91,Laird08,Georgakakis11}). The Bayesian flux  estimates reach unrealistic low values  for sources with low count rates. C19 allow up to a factor ten difference with the classical fluxes. Sources for which the Bayesian fluxes do not meet this requirement only have classical flux measurements. We chose to use the uncorrected X-ray flux measurements in this work, since using Bayesian fluxes would reduce the size of our final sample by a factor of two. We note that a test run of the statistical pipeline presented in this work on a smaller sample with Bayesian soft X-ray luminosities yielded similar results to those presented here for the entire sample using the nominal fluxes.

\begin{figure}
\includegraphics[width=8 cm]{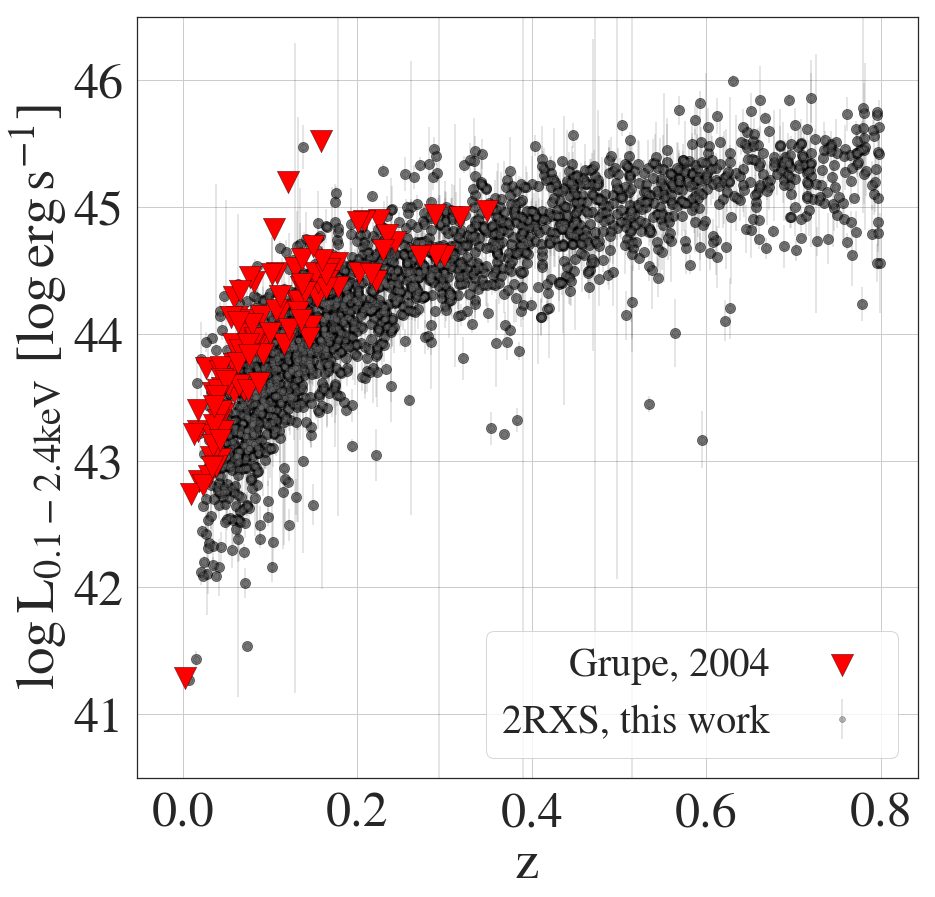}
\caption{ The observed soft X-ray luminosity-redshift for our 2100 2RXS sources. The red triangles are the 110 AGN used in \citet{Grupe04}, which all lie at the brighter end of our sample for equivalent redshifts. Our sample significantly extends both the redshift and the luminosity ranges. }
\centering
\label{fig:lumi_z}
\end{figure}

Fig. \ref{fig:lumi_z} presents the soft X-ray luminosity-redshift distribution of our sources, and compares our sample to that of \citet{Grupe04}.  Our sample spans a range of soft X-ray luminosities from $\rm 1.9 \times 10^{41}   \, erg \, s^{-1}$ to $\mathbf{\rm 9.9 \times 10^{45} \, erg \, s^{-1}}$, with redshifts up to $0.8$. Compared to \citet{Grupe04}, we sample to lower luminosities and higher redshifts. The larger sample will in particular allow placement of more stringent constrains on the relation of EV1 and EV2 to physical parameters (see Fig. \ref{fig:ledd_1}).

\begin{table}
\begin{tabular}{cc}
\hline
\begin{tabular}[c]{@{}c@{}}Sample cuts\\ (Initial SPIDERS AGN sample: 7790 sources)\end{tabular} &  remaining sources \\ \hline
2RXS flux                                                                                        & 7154                        \\
$\rm z < 0.8$                                                                                    & 5926                        \\
$\rm S/N \geq 20$                                                                                & 2124                        \\
Robust asymmetry index/black hole masses                                                         & 2100                       
\end{tabular}
\caption{Summary of the sample construction pipeline}
\label{Tab:tablecuts}
\end{table}

\subsection{Measuring asymmetry in emission lines}
\label{sec:asymi}

This section presents the derivation of a key emission line diagnostic:  the asymmetry index.

  Emission lines in AGN show asymmetries which  encode precious information on the geometry and kinematics of the emitting region. In the case of broad emission lines, asymmetry arises from the relative displacement of the profile's peak and base component's centroid wavelengths. 
 The peak wavelength of the profile is in practice measured at a high fraction of the broad component's intensity (without the narrow core) and is related to the classical broad component \citep{brotherton94,popovic02,Adhikari16}. The emission line's profile base is, as its name indicates, measured at lower fractional intensities and is expected to trace a distinct, red- or blue shifted, sometimes broader component. 
 In \citet{marziani10} and subsequent works of this group, the authors distinguish blueshifted and broad redshifted components of the full profile ('BLUE' and Very Broad Component 'VBC'; for a review see Section 6.2 in \citealt{marziani18}). The shifted VBC is believed to originate from the presence of a distinct emitting region in the BLR: the Very Broad Line Region (VBLR). The VBLR is expected to be composed of highly ionised gas situated even closer to the central black hole. The concept of BLR stratification arose from reverberation studies. \citet{Peterson86} reported variability in the profiles of $\rm HeII 4686\angstrom$ and $\rm H\beta$ in NGC 5548, most notably the appearance of broader line components while the continuum luminosity increased.    Measuring the asymmetry of broad emission lines and including the indices in a statistical analysis therefore offers an approach to investigate how the relative kinematics of assumed layers in the BLR impacts the diversity of observed Type 1 AGN.
 
 One can also use the asymmetry index to trace the relative displacement of narrow emission lines and their shifted wings. This is particularly useful to trace the presence of outflows in the NLR (e.g., \citealt{Heckman81,Zamanov02,zakamska16,Wang17,Rakshit18} ). In this case the displacement of peak and base components is measured for the full profile (i.e. narrow core and shifted wings).
 
 The asymmetry index was not included in C19 and was determined here starting from the reconstruction of the lines, using the Gaussian models listed in the catalogue. In this work we derived the following asymmetry indices:
 
\begin{itemize}
\item $\rm \Delta\lambda _{H\beta}$ for the broad component of $\rm H\beta$. Following C19, we retain the Gaussians with $\rm FWHM > 800 \, km \, s^{-1}$ as broad components. From the four Gaussian components used to fit the full $\rm H\beta$ profile, one was fixed by C19 at $\rm FWHM < 800 \, km\, s^{-1}$ to force the fit of the narrow emission line. Up to three Gaussians were thus used to fit the broad profile.

\item $\rm \Delta\lambda _{[OIII]}$ for the full profile of $\rm [OIII]\lambda5007{\angstrom}$. Here both narrow and broad components are used for the measure.
 \end{itemize}

We use the same method to measure these two indices. The description of $\Delta \lambda$ for any emission line presented here was first introduced by \citet{Heckman81}, and is further detailed in \citet{winkler14}.
 
\begin{figure}
\centering
\includegraphics[width=7.0 cm]{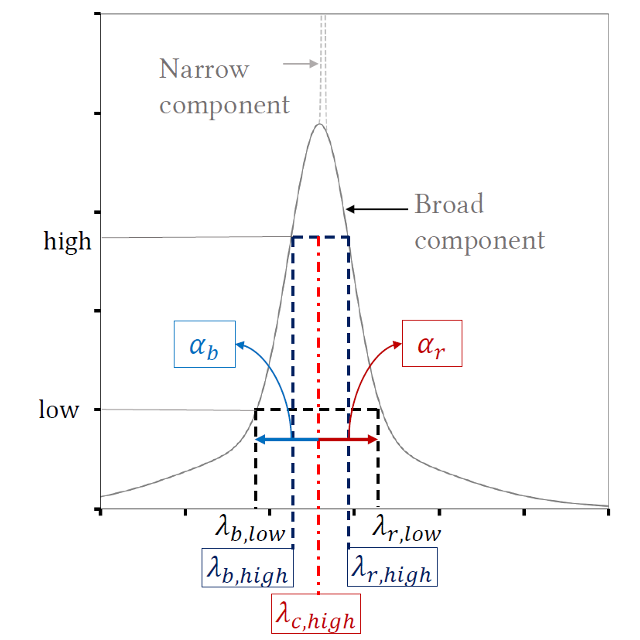}
\caption{ Diagram for the determination of the asymmetry parameter using the \citet{Heckman81} definition. Adapted from \citet{winkler14}. For $\rm H\beta$ the broad component is used without the narrow line. For $\rm [OIII]$ the asymmetry of the full profile (narrow and wing) is measured. }
\label{fig:asymexample}
\centering
\end{figure}

 This method requires the identification of the wavelengths at which the line reaches $\mathbf{15 \%}$ and $80 \%$ of its maximum intensity, which required the reconstruction of the multi-Gaussian line fits as described in Appendix \ref{sec:eqw}. We determined the maximum of the full line profile (or alternatively the broad components for $\rm H\beta$) and measure $\mathbf{15 \%}$
and $80\%$ of this maximum. A schematic view of this measurement is presented in Fig. \ref{fig:asymexample}. The asymmetry index ($\rm \Delta \lambda$) is defined by:

\begin{align}
\label{eq:asym}
\Delta \lambda &= \frac{\alpha_{\rm{b}} - \alpha_{\rm{r}}}{ \alpha_{\rm{b}} + \alpha_{\rm{r}}} = \frac{(\lambda_{\rm{c,high}}-\lambda_{\rm{b,low}})-(\lambda_{\rm{r,low}}-\lambda_{\rm{c,high}})}{(\lambda_{\rm{c,high}}-\lambda_{\rm{b,low}})+(\lambda_{\rm{r,low}}-\lambda_{\rm{c,high}})} \\
 &= \frac{\lambda_{\rm{b,high}}+\lambda_{\rm{r,high}} - \lambda_{\rm{b,low}}- \lambda_{\rm{r,low}}}{\lambda_{\rm{r,low}}-\lambda_{\rm{b,low}}}
\end{align}

Here $\lambda_{\rm{b,high}}$ and $\lambda_{\rm{r,high}}$ are measured at $80\%$ fractional intensity and $\lambda_{\rm{b,low}}$ and $\lambda_{\rm{r,low}}$ at $\mathbf{15 \%}$ fractional intensity. $\lambda_{\rm c,high}$ is the wavelength of the center of the line at $80 \%$ fractional intensity, i.e., the midpoint between $\lambda_{\rm{b,high}}$ and $\lambda_{\rm{r,high}}$. The parameters $\alpha_{\rm b}$ and $\alpha_{\rm r}$ are the distances from  $\lambda_{\rm c,high}$ to the blue and red edge of the line at $\mathbf{15 \%}$ fractional intensity.

A negative (positive) $\Delta \lambda$ corresponds to a blue-ward (red-ward) asymmetry, and its value ranges from -1 to 1. This measure is not sensitive to the continuum emission (cf. correlation matrix shown in Section \ref{thematrix}). We propagated the uncertainties of the $\rm FWHM$ of the fits into the measure of asymmetry and obtain typical errors of $\rm \sigma_{\Delta\lambda} \sim 0.05$. An important caveat is that using the uncertainties of the $\rm FWHM$ from the full profile, does not account for the lower, distinct and possibly more contaminated line components. 

Examples of  fit reconstructions of red- and blue-asymmetric $\rm H\beta$ from our sample are displayed in Appendix \ref{sec:red_blue_vbc}, alongside their Gaussian decomposition.

%% file: 03-AsymmetryDistribution.tex
\section{Impact of fit contamination on the \texorpdfstring{$\rm H\beta$}{Hbeta} and \texorpdfstring{ $\rm [OIII]$}{[OIII]} asymmetry index distribution }
\label{asymmetrydistribution}

\begin{figure}
\includegraphics[width=8 cm]{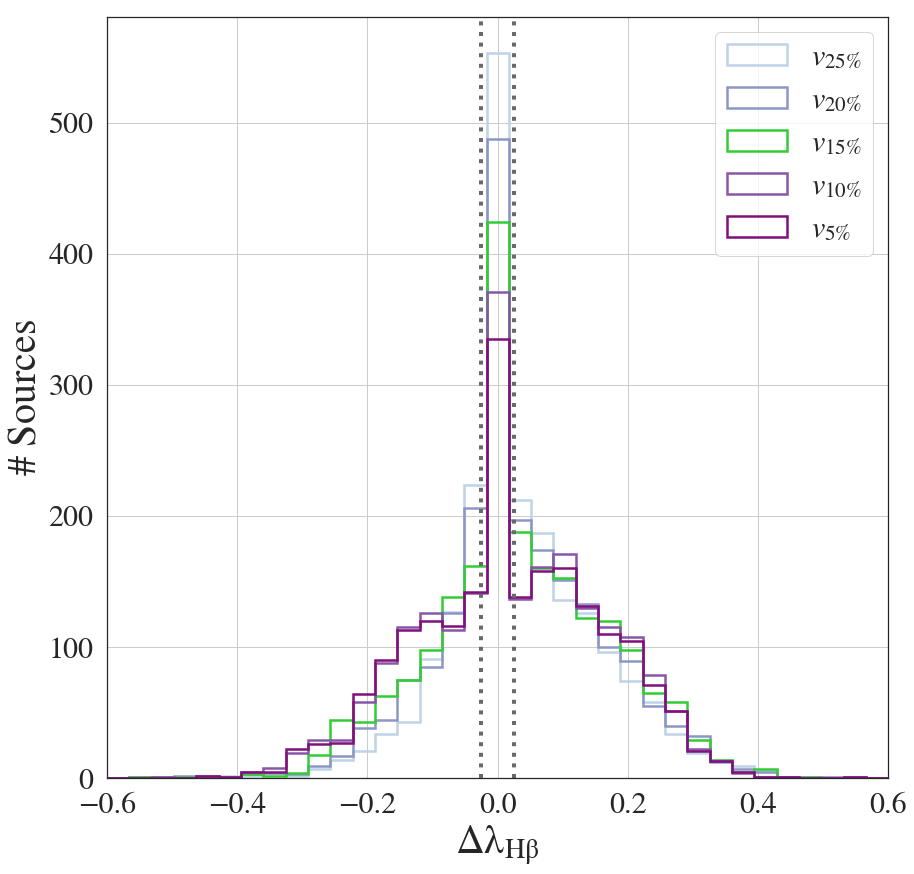}
\caption{The $\rm H\beta$ asymmetry distribution in the final source sub-sample for different levels of base fractional intensity: $5\%$, $10\%$, $15\%$, $20\%$ and $25\%$. A clear inflection redward of the symmetric $\rm H\beta$ profiles arises for bases measured at lower intensities. Typical $1\sigma$ errors are shown centered at zero by two dotted vertical bars. $\rm \Delta\lambda_{H\beta}$ values derived for a fractional base intensity of $\mathbf{15 \%}$ are used in this paper (green histogram).  }
\label{fig:asym_seq}
\centering
\end{figure}

\begin{figure}
\includegraphics[width=8 cm]{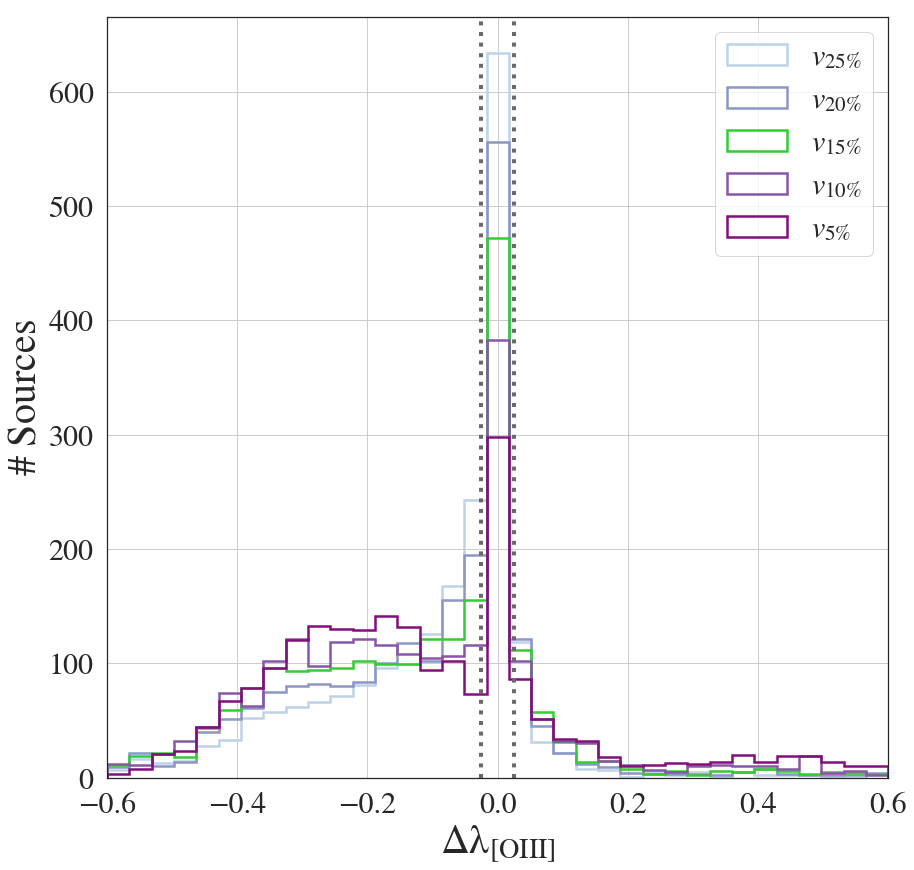}
\caption{The $\rm [OIII]$ asymmetry distribution for different levels of base fractional intensity: $5\%$, $10\%$, $15\%$, $20\%$ and $25\%$. A clear inflection can be detected here, bluewards of the symmetric $\rm [OIII]$ profiles for bases measured at lower intensities. Typical $1\sigma$ errors are shown centered at zero by two dotted vertical bars.The green histogram was obtained from $\rm \Delta\lambda_{[OIII]}$ measurements with base fractional intensity at $\mathbf{15 \%}$ as used in this work. }
\label{fig:asym_seq_2}
\centering
\end{figure}

The distributions for $\rm H\beta$ and $\rm [OIII]$ asymmetry indices are shown respectively in Fig. \ref{fig:asym_seq}-\ref{fig:asym_seq_2}.  The indices clearly follow very different distributions for the two emission lines. The following statistics were derived for the distributions of asymmetries with base centroids measured at $25 \%$ fractional intensity (i.e., the blue histograms in Fig. \ref{fig:asym_seq}-\ref{fig:asym_seq_2}).  The asymmetry indices of the $\rm [OIII]$ are predominantly blue-ward (skewness of distribution: $\gamma = -0.54$), which is consistent with the presence of a blue-shifted wing detected in a two-Gaussian model. The mean value of $\rm \Delta \lambda_{[OIII]}$  is $\rm \sim - 0.10$. 1546 sources with $\rm [OIII]$ asymmetry below zero and 547 above zero are counted, confirming the excess of blue asymmetric [OIII] lines. The distribution of the $\rm H\beta$ asymmetry appears positively asymmetric (skewness of distribution $\gamma = 0.36$). The mean value of $\rm \Delta\lambda_{H\beta}$ is $+0.03$. An excess of red-asymmetric $\rm H\beta$ emitters is clearly measured (1244 sources out of 2100 with $\rm \Delta\lambda_{H\beta}> 0$).

 The fractional intensity at which the lower velocity shift is measured for the asymmetry index has important repercussions on the overall distribution of asymmetries as Fig. \ref{fig:asym_seq}-\ref{fig:asym_seq_2} demonstrate. The distribution of $\rm \Delta \lambda_{H\beta}$ and $\rm \Delta \lambda_{[OIII]}$ are displayed for different fractional intensities at which $\lambda_{\rm{b,low}}$ and $\lambda_{\rm{r,low}}$ are set : $5\%$, $10\%$, $15\%$, $20\%$, $25\%$. While we always detect an excess in red-ward asymmetry for the $\rm H\beta$ profiles, the lower the percentage at which $\lambda_{\rm{b,low}}$ and $\lambda_{\rm{r,low}}$ are measured, the more the asymmetry index distribution is bimodal, with a clear separation between the symmetric and the red asymmetric $\rm H\beta$ modes. 
 It is unclear if this red-ward excess in $\rm H\beta$ asymmetries is influenced by  $\rm [OIII]$ and continuum contamination in the fit.
 The bimodality of the $\rm \Delta\lambda_{H\beta}$ distribution measured at lower base intensities, however, clearly indicates a preferred displacement between peak and base component which could arise from the near systematic contamination by $\rm [OIII]$ or its blue wing.
 
 A similar effect is observed for the distribution of the [OIII] asymmetries: it becomes double peaked, with a distinct second mode appearing blue-ward of the symmetric sources for lower base intensities.
  For the measurement of the asymmetry index, we set the base of the lines at $\mathbf{15 \%}$ fractional intensity, in order to avoid the detection of overlaying line models. This choice is motivated by two aspects of the distributions seen in Fig. \ref{fig:asym_seq}-\ref{fig:asym_seq_2} : $\mathbf{15 \%}$ fractional intensity, the peak of the symmetric sources is more sensitive to low intensity, shifted profile components are detected by the measurement than at $25 \%$ and $20\%$. Furthermore, for this measurement configuration (peak at $80 \%$ and base at $\mathbf{15 \%}$ fractional intensity), the bimodality which we identify as a signature of model degeneracy is not yet detected.      
 
\begin{figure}
\includegraphics[width=8 cm]{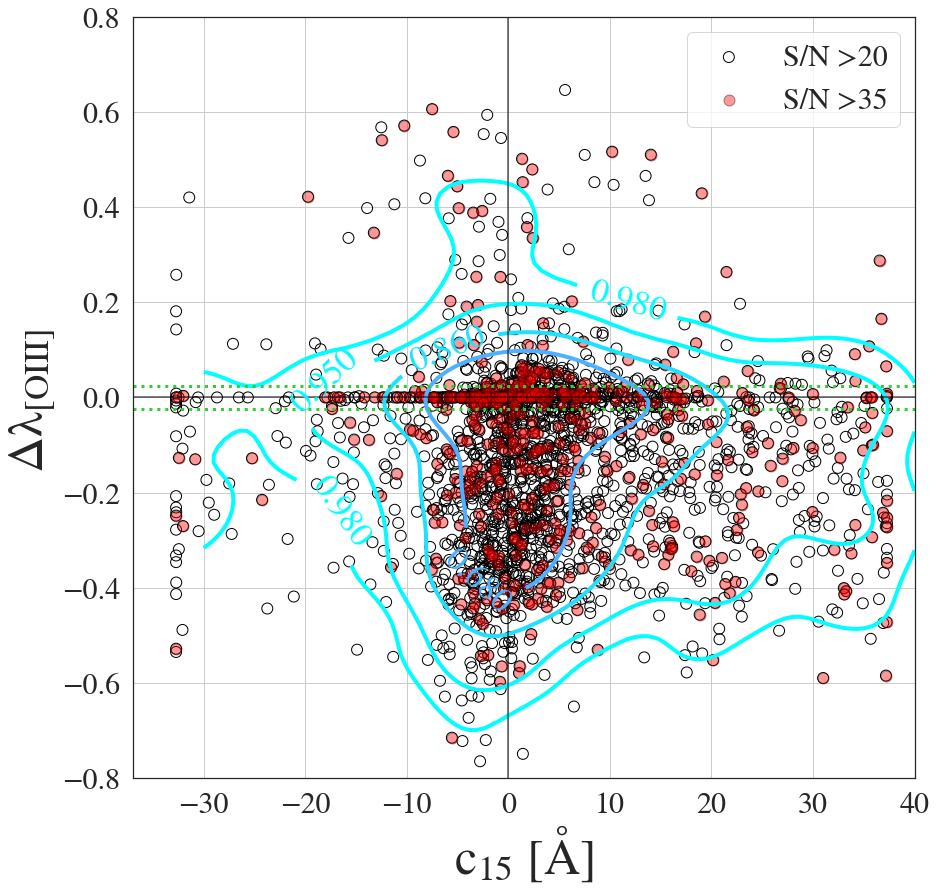}
\caption{The asymmetry index of $\rm [OIII]$ as a function of the $\rm H\beta$ centroid shift at $\mathbf{15 \%}$ fractional intensity. The excess of sources in the lower right quadrant suggests an inter-line contamination for redshifted $\rm H\beta$ components and blueshifted $\rm [OIII]$ components. The filled circles mark the sources with S/N above 35. The green dotted horizontal lines indicate typical $1\sigma$ uncertainties in the asymmetry parameter. The density contours containing $68\%$, $86\%$, $95\%$ and $98\%$ of all the sources in our sample were obtained from a Gaussian kernel density estimate.}
\label{fig:contam}
\centering
\end{figure}

 To further investigate the risk of model degeneracy, the dependency of the $\rm [OIII]$ asymmetry index on the shifts of the $\rm H\beta$ base component $\rm \mathbf{c_{15}}$ (the centroid shift at $\mathbf{15 \%}$ fractional intensity with respect to the rest-frame wavelength) is presented in Fig. \ref{fig:contam}. We clearly observe an excess of sources in the quadrant of blue-asymmetric $\rm [OIII]$ lines and redshifted $\rm H\beta$ base components. The coloured markers correspond to high signal-to-noise ratio sources, which also appear to preferentially populate the lower right quadrant in the figure.
 1196 sources out of 2100 ($57\%$) have $ \rm \Delta\lambda_{[OIII]}<0$  and $\rm \mathbf{c_{15}>0}$. In the high signal-to-noise regime, i.e. S/N > 35, 278 out of 498 sources ($56 \%$) have $\rm \Delta\lambda_{[OIII]}<0$  and $\rm \mathbf{c_{15}>0}$. A binomial test showed that the clustering in the lower right quadrant is indeed significant. For  both, $\rm \mathbf{c_{15}}$ and $\rm \Delta\lambda_{[OIII]}$, the null hypothesis that the sources distribute equally on both sides of zero was rejected with p-values $\rm p < 10^{-6}$\footnote{Throughout this work, $\rho$ denotes correlation coefficients, while $\rm p$ stands for p-value}.
 
 This result underlines the possibility of model degeneracy in the $\rm H\beta$ region. In their study of Extremely Reddened Quasars (ERQs), \citet{perrotta19}, recently presented rare sources for which $\rm H\beta$ and $\rm [OIII]$ appear to blend (see their Fig. 1).
 Disentangling the presence of a shifted Very Broad Component from fit contamination would offer a cleaner window on the kinematics of the BLR, however, the fitting procedure in C19 did not allow for such a decomposition.

%% file: 04-PCA.tex
\section{Statistical Analysis}
\label{PCAandDirect}

 This section presents the statistical analysis performed on the selected optical properties of our sample of 2100 SPIDERS AGN, listed in Section \ref{Dataset}.

\subsection{Direct correlation}
\label{thematrix}

\begin{figure}
\includegraphics[width= 9 cm]{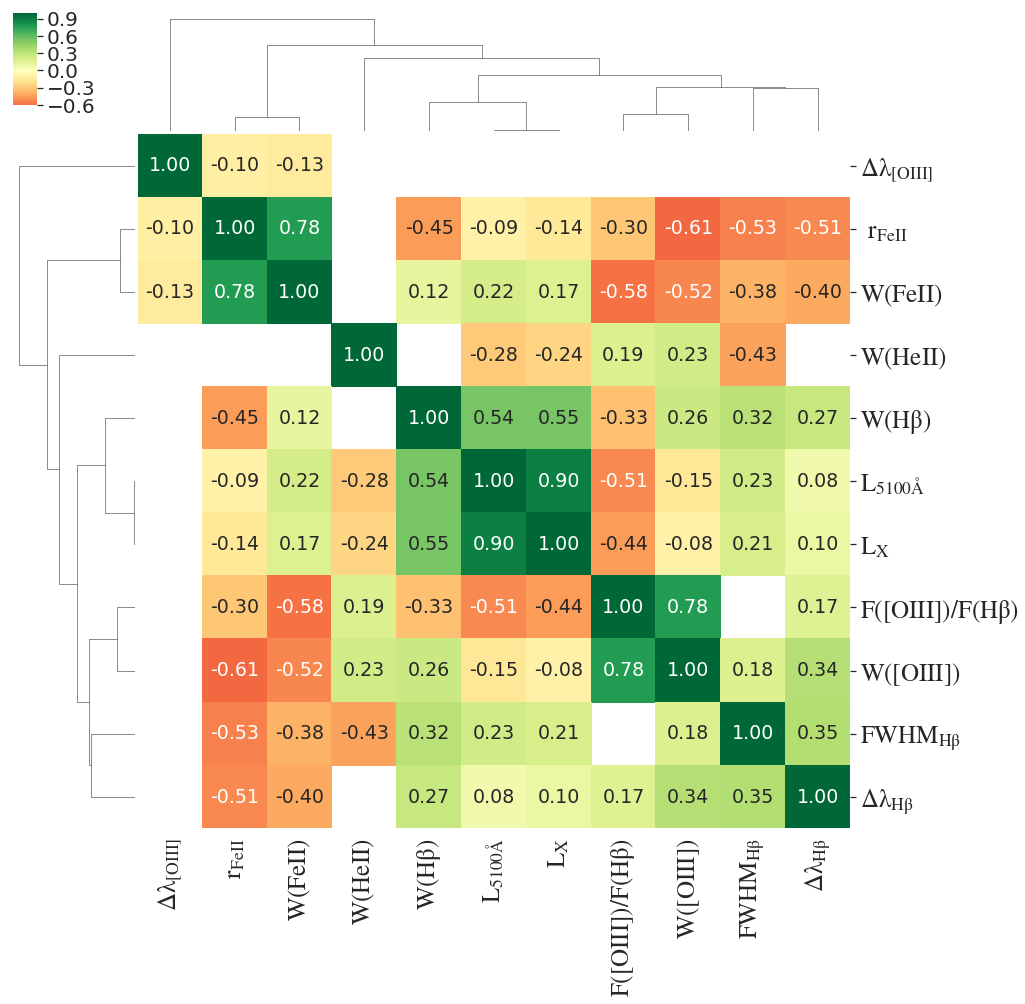}
\caption{Spearman rank correlation matrix for our parameters.
A positive (negative) coefficient indicates a positive (negative) monotonic correlation. Statistically insignificant correlations (with p-values $>0.05/55$) have been masked. The matrix was hierarchically clustered using correlation as a distance measure. A dendrogram displays the average linkage of the parameters. Previously reported correlations are confirmed.}
\label{fig:final_corr}
\centering
\end{figure}

Our initial step is to generate a Spearman rank correlation matrix for our sample; the resulting matrix is shown in Fig. \ref{fig:final_corr}. Statistically insignificant correlations, defined by imposing a threshold of $\rm p_{max} = 0.05/55$ on the p-values, were masked. The threshold $\rm p_{max}$ applies the Bonferroni correction for multiple statistical hypothesis testing (e.g., \citealt{Haynes2013}).

Positive (negative) coefficients indicate a positive (negative) monotonic relation between two parameters. We have included a dendrogram (tree diagram) which clusters our data hierarchically using correlation as distance metric (for a review see \citealt{baron19}). The clustered correlation matrix reveals structure in our parameter-subset.

\begin{figure}
\includegraphics[width= 9 cm]{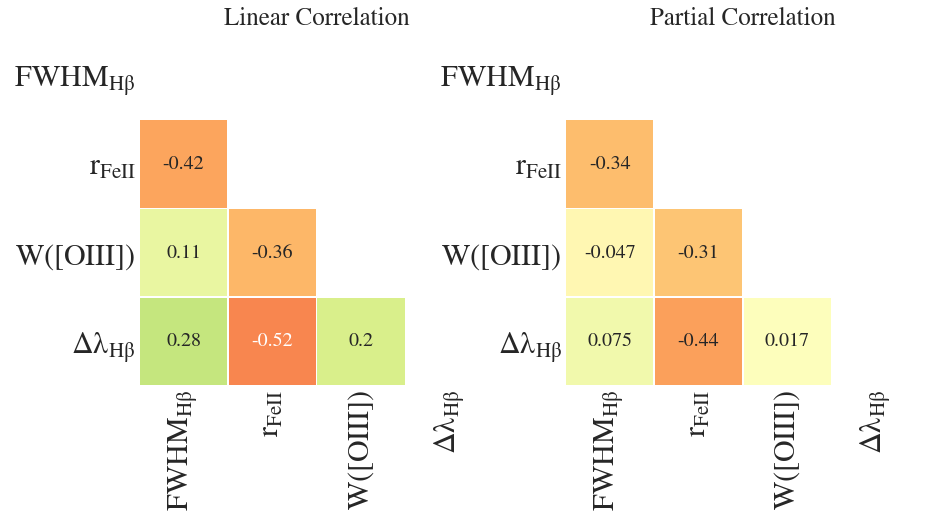}
\caption{Left panel: Pearson Correlation coefficients for the parameter subset : $\rm FWHM_{H\beta}$, $\rm W([OIII])$, $\rm r_{FeII}$, $\rm \Delta \lambda_{H\beta}$.
Right panel: partial correlation coefficients (correlation coefficients, when the effects of confounding variables are removed).}
\label{fig:partialcorr}
\centering
\end{figure}

We confirm individual correlations also previously reported in \citet{Boroson92}, \cite{Grupe04} and \citet{Shen14}. The correlations all have significance levels $\rm p < 10^{-6}$.
\begin{itemize}
    \item A clear anti-correlation ($\rho = -0.53$) arises between $\rm FWHM_{H\beta}$ and $\rm r_{FeII}$. This behaviour is the well-known anti-correlation between the first two dimensions of the 4DE1 (e.g., \citealt{sulentic00a,Sulentic00b,Marziani03a,Marziani03b, sulentic07b}, for a review see \citealt{marziani18} and \citealt{sulentic15}). In addition, the equivalent width of iron $\rm W(FeII)$ and $\rm FWHM_{H\beta}$ are anti-correlated ($\rho =-0.38$), as expected.
    \item Similarly, an even stronger anti-correlation ($\rho = -0.61$) is measured between the equivalent width of $\rm [OIII]$  and $\rm r_{FeII}$, in agreement with the main EV1 anti-correlation.
    \item The X-ray and continuum luminosities, $\rm L_X$ and $\rm L_{5100{\angstrom}}$, are related to the emission line properties in a similar manner. There is a strong correlation ($\rho = 0.54$ for $\rm L_{5100{\angstrom}}$ and $\rho = 0.55$ for $\rm L_X$) with the equivalent width of $\rm H\beta$. The correlation between $\rm L_{5100{\angstrom}}$ and $\rm W(H\beta)$ is  consistent with photoionisation models in which increasing emission of the central engine results in a more luminous BLR. 
    $\rm L_X$ and $\rm L_{5100\angstrom}$ decrease strongly  ($\rho = -0.44$  and $\rho = -0.51$ respectively) with increasing flux ratio $\rm F(OIII)/F(H\beta)$.
    \item  The asymmetry index $\rm \Delta\lambda_{H\beta}$ is positively correlated to the equivalent width of $\rm [OIII]$ ($\rho = 0.34$)   and the $\rm FWHM_{H_\beta}$ ($\rho = 0.35$). We also report its relatively strong anti-correlation with $\rm r_{FeII}$ ($\rho = -0.51$) and $\rm W(FeII)$ ($\rho = -0.40$). Despite using a different convention for the asymmetry index, \citet{Boroson92} describe a similar correlation behaviour of $\rm \Delta\lambda_{H\beta}$.
    \newline
    
    In order to study the correlations with the $\rm \Delta\lambda_{H\beta}$ parameter more closely we perform a partial correlation analysis of $\rm FWHM_{H\beta}$, $\rm W([OIII])$, $\rm r_{FeII}$ and $\rm \Delta \lambda_{H\beta}$ (e.g., \citealt{baba04}). For this exercise, we first generate the Pearson correlation coefficients for this subset of parameters, then measure the strength of each of these correlations while controlling for the other confounding variables. In the simplest case, when only one confounding variable is accounted for, this is achieved by performing a linear regression for each of the two test variables with the confounding variable. The partial correlation of the two test variables is then obtained by measuring the correlation of the residuals resulting from the linear regressions.  The results are shown in Fig. \ref{fig:partialcorr}. When we control for $\rm FWHM_{H\beta}$ and $\rm r_{FeII}$, the correlation between $\rm W([OIII])$ and $\rm \Delta\lambda_{H\beta}$ significantly decreases (we measure a drop $\rho_{P} = 0.20$, $\rm p < 10^{-7}  $   to $\rho_{P,partial} = 0.017$, $\rm p = 0.45$). The p-value of the latter partial correlation is above the corrected significance threshold: $\rm p=0.05/6$. Similarly, when we control for $\rm W([OIII])$ and $\rm r_{FeII}$, the strength of the $\rm FWHM_{H\beta}$ vs. $\rm \Delta \lambda_{H\beta}$ correlation decreases ($\rho_{P} = 0.28$, $\rm p < 10^{-6}$ to $\rho_{P,partial} = 0.075$, $\rm p \sim 10^{-4}$). The anti-correlation between $\rm \Delta \lambda_{H\beta}$ and $\rm r_{FeII}$ is, however, less affected when we control for $\rm FWHM_{H\beta}$ and $\rm W([OIII])$ ( $\rho_{P,partial} = -0.44$,  $\rm p < 10^{-6}$ instead of $\rho_{P} = -0.52$, $\rm p < 10^{-6}$ ). These results demonstrate that $\rm \Delta \lambda_{H\beta}$ parameter is principally related to the relative strength of the iron emission and is marginally linearly correlated to $\rm FWHM_{H\beta}$, indicating that the broader profiles tend to be more red-skewed, as also suggested by \citet{Marziani13b}.      
\end{itemize}

The key result so far is not the rediscovery of correlations mapped out by e.g., \citet{Boroson92, Grupe04, Shen14}, but the fact that these relations hold for type 1 AGN in general, up to a redshift limit of at least $\rm z = 0.80$ and for luminosities up to $\rm L_X \sim 10^{46} erg.s^{-1}$.

 A new insight provided by the correlation matrix is the  correlation behaviour of $\rm \Delta\lambda_{[OIII]}$ with parameters related to the BLR emission :

\begin{itemize}
    \item The asymmetry index of [OIII] appears to be marginally related to the rest of the parameters, suggesting  the absence of kinematic linkage between the inner region of the AGN and the NLR. This result is in contrast to \citet{Zamanov02}, who reported evidence for correlation between the shifts of the high ionisation lines $\rm CIV\lambda 1549 \angstrom$ and the shifts of the [OIII] lines, inferring a possible linkage between the NLR and BLR. 
\end{itemize}

\subsection{Principal components}

The second step consisted of running a PCA on the standardized data set to determine which parameters are contributing most to the total variance of our sources. It is particularly important to scale the parameters to unit variance since PCA is sensitive to the variance of the parameter distributions.  PCA is an orthogonal linear transformation which is often used as a dimensionality-reduction algorithm (for a review see \citealt{jolliffe16}). It yields the eigenvectors in parameter-space which point in the direction of total maximal variance in the dataset. For an $\rm N\times M$ dataset, the first component is found by minimising the distance between points in the M-dimensional parameter space and their orthogonal projections onto an M-dimensional vector, which simultaneously increases the variance of the projected points.
The second component is chosen to be orthogonal to the first and is determined in the same manner given this condition. The associated eigenvalues measure the amount of explained variance in each principal component (eigenvector). The variables (our parameters) are linked to the principal components by linear coefficients. These coefficients yield the amount of a variable's variance explained by the component.\footnote{In practice, we made use of a SCIKIT-LEARN implementation of PCA \citep{scikit11}.} While PCA is a very common tool in astronomy it comes with several disadvantages, such as the linearity of the dimensionality reduction, which might not capture the full complexity of our data. One must in general be careful with the choice of the particular dimensionality reduction one uses and its interpretation \citep[e.g., see][]{baron19}.

\begin{figure}
\includegraphics[width=8 cm]{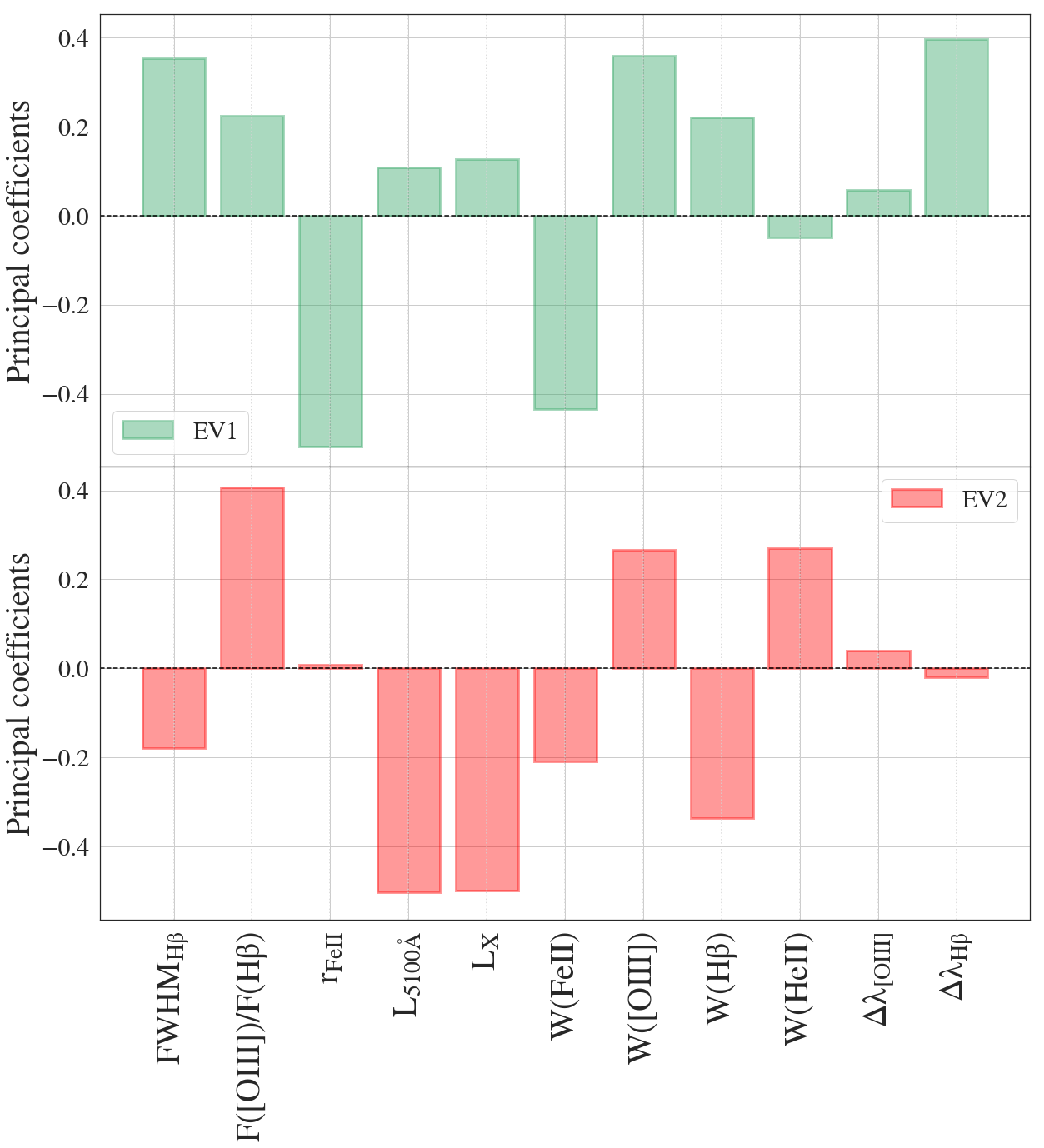}
\caption{The component coefficients (factor loadings) defining the first and second principal components : EV1 (green bars) and EV2 (red bars). EV1 appears to be heavily dominated by the anti-correlation of $\rm FWHM_{H\beta}$ and the strength of the $\rm [OIII]$ emission with $\rm r_{FeII}$. A relatively large linear coefficient links the $\rm H\beta$ asymmetry to EV1. EV2 is heavily dominated in equal measure by the X-ray and optical luminosities. }
\label{fig:ev1f}
\centering
\end{figure}

The first two principal components are presented in Fig. \ref{fig:ev1f}.  The bar diagrams show the values of the correlation coefficients which link the parameters (our variables) to the given component. The first and second principal component respectively explain $25.2\%$ and $20.3\%$ of the total variance in the selected parameter subset.

The first principal component (Eigenvector 1, EV1), is anti-correlated to the strength of the iron emission and correlated to $\rm FWHM_{H\beta}$ and the (relative) strength of the $\rm [OIII]$ emission, i.e., the diversity in the selected optical features of our SPIDERS AGN sub-sample is dominated by the anti-correlation of $\rm W([OIII])$ and $\rm W(FeII)$. The Balmer profile asymmetries $\rm \Delta\lambda_{H\beta}$ are noticeably correlated to EV1, which corroborates the findings of Z10, who stress that the asymmetry parameter could in essence be used as surrogate 4DE1 parameter. The authors note, however, that $\rm \Delta\lambda_{H\beta}$ is also substantially orienting Eigenvector 2 (EV2). This behaviour is not observed in our PCA results. Our EV1 is consistent with early results in  \citet{Grupe04}.

The second principal component, EV2, is strongly dominated by the X-ray and optical continuum luminosities. The flux ratio $\rm F([OIII])/F(H\beta)$ is correlated to EV2. The equivalent width of $\rm H\beta$ once again significantly contributes to the orientation of the principal component. EV2 is also consistent with \citet{Grupe04}. 


%% file: 05-DerivedParameters.tex
\section{Black Hole Mass and Eddington ratio}
\label{EDDMBHsec}
We extend the analysis to derived AGN properties: the black hole mass and the Eddington ratio. 
In \citet{Boroson92} EV1 was related to the Eddington ratio $\rm L/L_{Edd}$, tracing the accretion power of the observed SMBH. Their EV2 was largely dominated by luminosity and related to the Baldwin effect (\citealt{baldwin77,baldwin78,dietrich02}). The Eddington ratio has indeed long been considered a prime candidate to explain the observed diversity in optical AGN features (e.g., \citealt{ Sulentic00b, yuanwill03,Grupe04, kuraszkiewicz09, Grupe11, Shen14, bon18}).

The FWHM of broad emission lines serves as proxy for the line of sight velocity of the BLR gas, which in combination with the distance to the black hole yields an estimate for its mass :
\begin{equation}
     M_{\rm{BH}} = \frac{r_{\rm{BLR}}f_{\rm{BLR}}  \rm {FWHM_{BLR}}^2}{G}
 \end{equation}
 $f_{\rm{BLR}} $ denotes the geometric form factor of the BLR. The distance $r_{\rm{BLR}}$ from the SMBH to the line-emitting region is obtained from the delay in response of the line fluxes to continuum changes. The $\rm FWHM_{BLR}$ of a line from the emitting region corresponds to the radial velocity component of the BLR gas. The Keplerian velocity $\rm v_{Kepl}$ of the gas is obtained by correcting for projection effects encoded in the geometric form factor : $\rm {v_{Kepl}}^2=f_{\rm{BLR}}  \rm {FWHM_{BLR}}^2$. One can construct an expression of $\rm f_{\rm{BLR}}$, which takes in account the source orientation $\theta$ and the shape of the velocity field described by the motion of the gas in the BLR (e.g., equation 12 in \citealt{negrete18}), by making assumptions about the structure of the disk.

The black hole masses and the Eddington ratio for our sample are derived from $\rm H\beta$ using the calibration developed in \citet{Assef11}, who based their estimates on the BLR radius-luminosity relation of \citet{bentz09}.

\begin{equation}
    \log \left ( \frac{M_{\rm BH}}{M\odot}\right) = A + B \log \left( \rm \frac{\lambda L_\lambda}{10^{44}\rm erg\,s^{-1}} \right) + C \log \left(\rm \frac{FWHM_{H\beta}}{1000\, \rm km\,s^{-1}} \right)
    \label{eq:bhm_equations}
\end{equation}

$\rm L_\lambda$ corresponds to the monochromatic luminosity at $5100 \, \angstrom$ and the parameters $A$, $B$ and $C$ are derived from reverberation mapping studies: $A=0.895$, $B=0.52$ and $C=2$. 
 From these black hole masses, C19 provided estimates of the Eddington ratio following:
 
 \begin{equation}
    \rm L_{Bol}/ L_{Edd} = \frac{L_{\rm Bol}}{4\pi cGM_{\rm BH}m_p / \sigma_T} 
 \end{equation}

where G is the gravitational constant, c the speed of light, $\rm m_p$ the proton mass and $\rm \sigma_T$ the Thomson scattering cross-section. 
The bolometric luminosities $\rm L_{Bol}$ used here are obtained from the bolometric corrections presented in \citet{richards06} : $\rm L_{Bol}= 9.26 L_{5100\,\angstrom}$. This factor is in principle dependent on $\rm L_{5100\,\angstrom}$, but for the purpose of this work, this assumption is not crucial.

\begin{figure*}
    \includegraphics[width=8 cm]{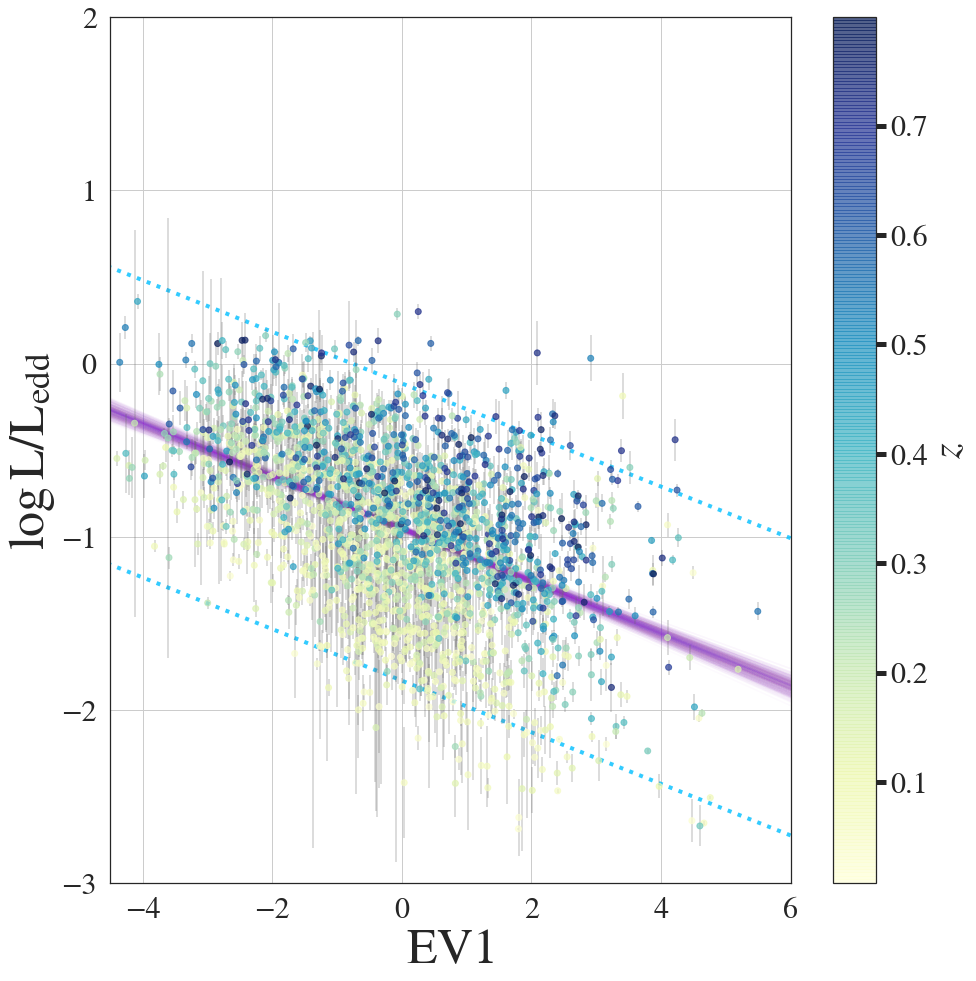}
    \includegraphics[width=8cm]{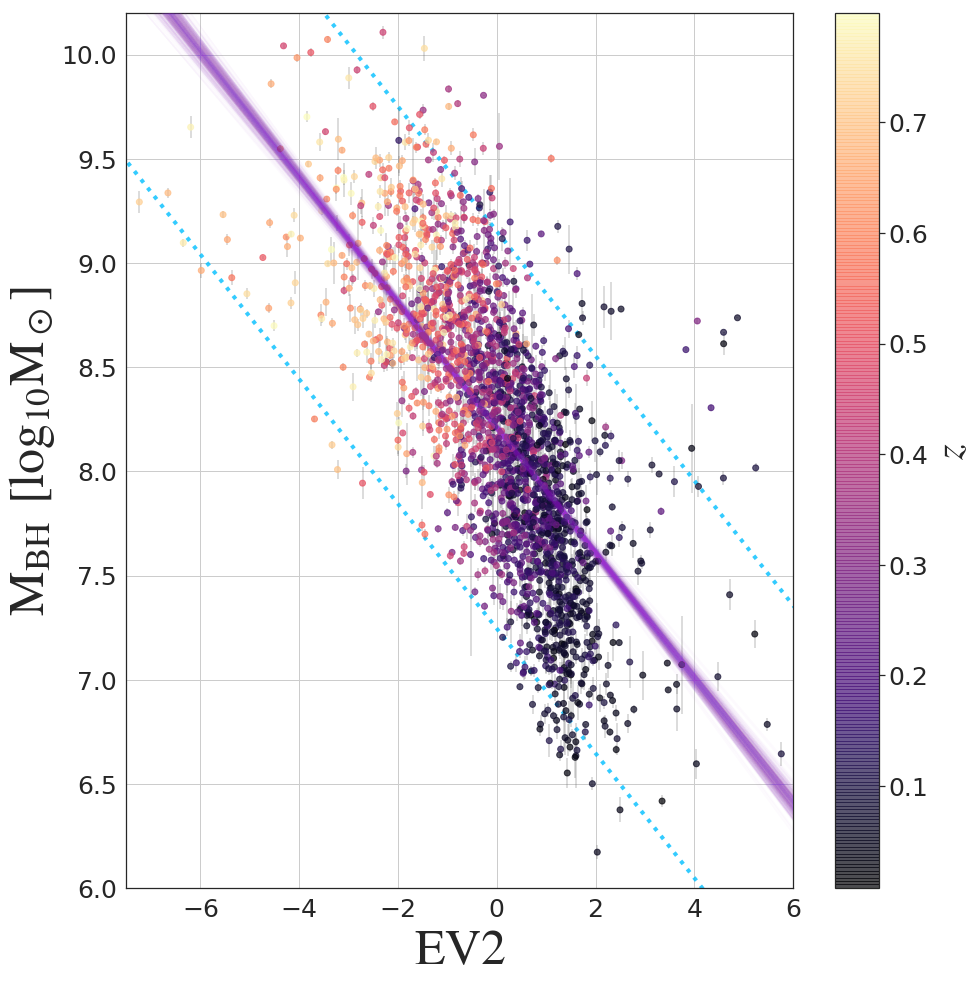}
    \caption{\textit{Left panel}: The Eddington ratio has an exponential correlation to EV1. The $95\%$ confidence contours are represented by the dashed blue lines. \textit{Right panel}: The black hole mass is correlated to EV2. Again the $95\%$ confidence contours are shown in blue. The linear regressions were performed with \textit{Linmix} \citep{kelly07}. }
    \label{fig:ledd_1}
    \centering
\end{figure*}

Fig. \ref{fig:ledd_1} displays how the Eddington ratio and black hole mass scale with the projection of the sources onto our newly determined EV1 and EV2. 

\begin{figure}
\includegraphics[width=8 cm]{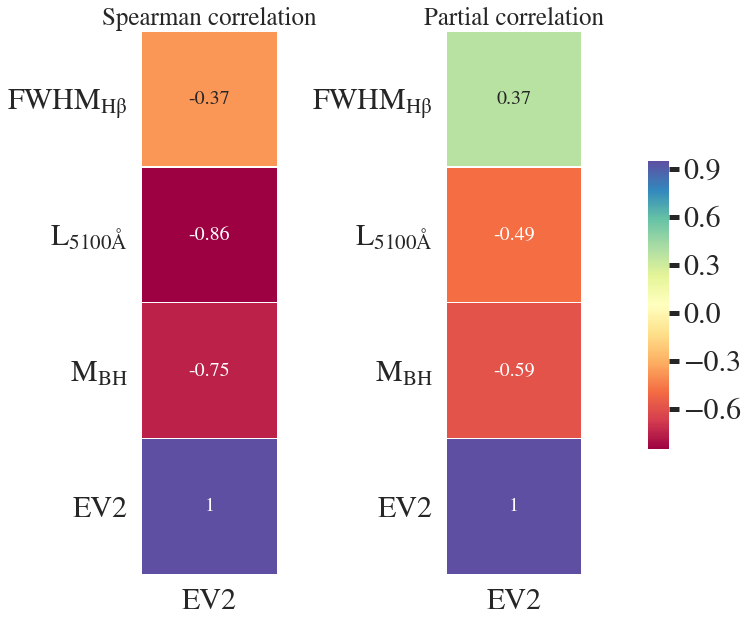}
\caption{The monotonic correlation coefficients of $\rm EV2$ with $\rm M_{BH}$,  $\rm L_{5100 \, \angstrom}$ and $\rm FWHM_{H\beta}$ for our sample are shown: simple Spearman correlation in the left column and partial Spearman correlation in the right column. For the computation of the partial correlation of a parameter pair, we control for the other confounding variables. The $\rm M_{BH}$ vs. $\rm EV2$ anti-correlation slightly drops when we control for $\rm L_{5100 \, \angstrom}$ and $\rm FWHM_{H\beta}$, while partial $\rm M_{BH}$ vs. $\rm EV2$ anti-correlation remains significantly strong $\rm p = -0.59$.} 
\label{fig:testpartialev1}
\centering
\end{figure}

We can strongly corroborate that EV1 correlates with the Eddington ratio.
An even stronger anti-correlation is that of EV2 with black hole mass. However, the $\rm \log \, M_{BH}$-EV2 relation breaks down at lower redshifts ($z \sim 0.2$). In order to determine to which extend this correlation is a side-product of the $\rm L_{5100\angstrom}$ and $\rm FWHM_{H\beta}$ principal coefficients in EV2, we once again perform a partial correlation test. We start by measuring the Spearman rank correlation of EV2 with black hole mass, while controlling for $\rm L_{5100 \, \angstrom}$ and $\rm FWHM_{H\beta}$. In Fig. \ref{fig:testpartialev1} the Spearman correlation coefficients measured for $\rm \log \, M_{BH}$, $\rm EV2$, $\rm L_{5100 \, \angstrom}$ and $\rm FWHM_{H\beta}$ are displayed in the left panel. In the right panel  of Fig. \ref{fig:testpartialev1}, the partial correlation coefficients results are presented. As already discussed in Section \ref{PCAandDirect}, this exercise consisted in performing a (multi) linear regression of EV2 and $\rm \log \, M_{BH}$ with the confounding variables. In this version we compute the Spearman correlation coefficients of the residuals of these two multi-linear regressions. EV2 is still strongly anti-correlated to the black hole mass ($\rho = -0.56$). This is not surprising since in Eq. \ref{eq:bhm_equations}, $\rm \log \, M_{BH}$ is a non-linear function of $\rm FWHM_{H\beta}$ and $\rm L_{5100 \, \angstrom}$, while the Eigenvectors obtained from PCA are obtained through linear orthogonal transformations of the initial parameter. The monotonic anti-correlation measured in the residuals might thus simply arise from the non-linearity of Eq. \ref{eq:bhm_equations}. More generally, this result implies that one can construct a linear combination of the parameters which dominate the orientation of EV2 (see Fig. \ref{fig:ev1f}) in order to estimate the black hole mass at redshifts $z>0.2$.


We propose that EV2, through its anti-correlation to black hole mass, might be related to the evolution of the broad line AGN population. Over the observed redshift range, the median black hole mass increases with decreasing redshift. This trend  a combination of two effects: the 2RXS flux limit resulting in the Malmquist bias and the decreasing number density of high black hole masses at lower redshifts. This downsizing of black hole masses across cosmic time remains a matter of debate (cf. downsizing models, e.g., \citealt{fanidakis12}).

We conclude that the Eddington ratio and the black hole mass are related to the principal components which explain $\sim 45\%$ of the total variance in our data.

%% file: 06-Asymdriver.tex
\section{Asymmetry of the broad \texorpdfstring{$\rm H\beta$}{Hbeta} emission line; a marker of Type 1 AGN diversity}
\label{asymdriversec}

EV1's direction in parameter space has been used in the past to establish an analog to the stellar Hertzsprung-Russel diagram for the Type 1 AGN population. In the plane spanned by $\rm FWHM_{H\beta}$ and $\rm r_{FeII}$ (i.e., the first two dimensions of the 4DE1)  the domain occupied by Type 1 AGN has been presented as a quasar main sequence (\citealt{marziani01, sulentic11}). We investigate the role played by $\rm H\beta$ line asymmetries in this framework. Since they appear to be one of the central sources of variance in our sample, they can be source of optical diversity of Type 1 AGN.

\begin{figure}
\includegraphics[width=8.0cm]{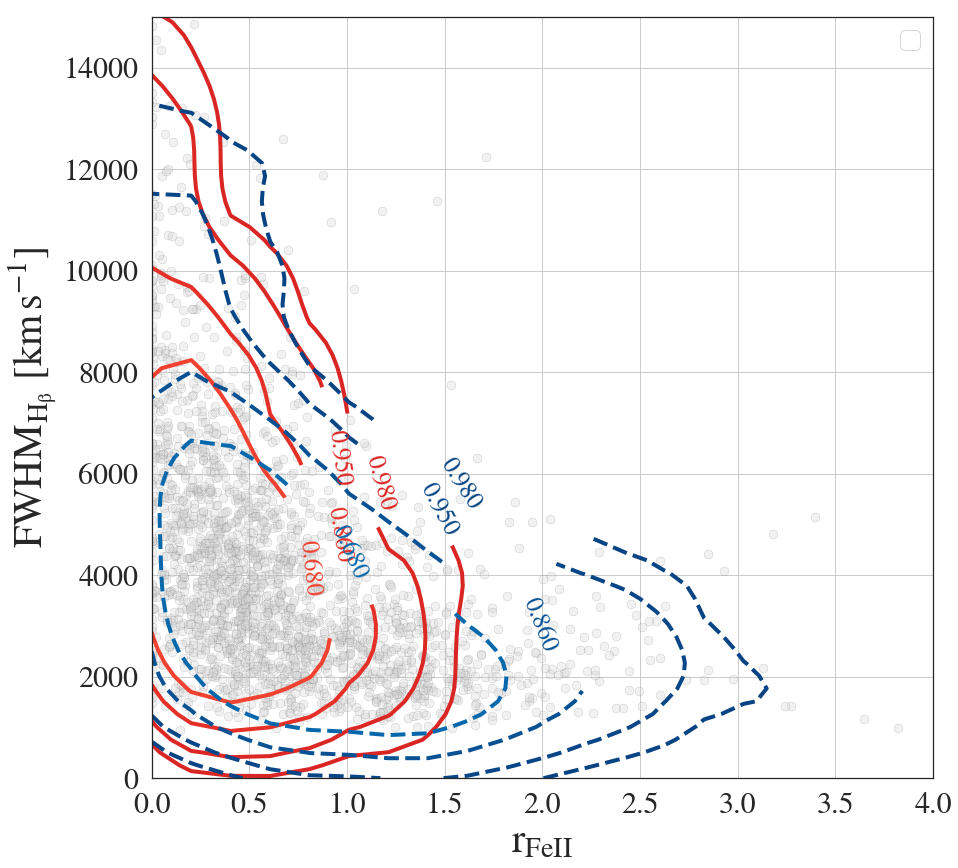}
\caption{The horizontal trend of the $\rm H_\beta$ asymmetry in the EV1 plane is made clear in the figure above. The contours derived from the KDE, delimit the areas in which $68\%$, $86\%$, $95\%$ and $98\%$ of the data points are confined. These contours reveal two very different occupation domains for red- and blue asymmetric Balmer emitters : sources with red-asymmetric $\rm H\beta$ are confined at low $\rm r_{FeII}$ values, while sources with blue asymmetric $\rm H\beta$ seem to extend over the full EV1 sequence. }
\label{fig:ev1plane}
\centering
\end{figure}

Fig. \ref{fig:ev1plane} presents the sample in the  $\rm FWHM_{H\beta}$ vs. $\rm r_{FeII}$ plane, also called EV1 plane. The sources have been separated in two subsets: blue and red asymmetric $\rm H\beta$ emission. More specifically these two sub-samples were constructed according to : $\rm \Delta \lambda _{H\beta} > 0.07$ and $\rm \Delta \lambda_{H\beta} < -0.07$. These criteria exclude symmetric sources and account for the typical uncertainties of the asymmetry index ($\rm \sigma_{\Delta \lambda_{H\beta}} \sim  0.05$). Contours of the bivariate Gaussian kernel density estimates (KDE, e.g., \citealt{Silverman86}) for each of the subsets are also indicated. The contours delimit the areas containing $68\%$, $86\%$, $95\%$ and $98\%$ of the data points.\footnote{The density contours shown throughout the rest of this work are defined in the same way.}

The sample occupies an L-shaped form in the plane. Using simulations, C19 demonstrate that the absence of high $\rm FWHM_{H\beta}$ and high $\rm r_{FeII}$ sources is to an extent due to limitations of the spectral fitting method.
Most of the sources with blue asymmetric $\rm H\beta$ emission profiles and those with red-asymmetric profiles appear to occupy different domains of this sequence. 
While the AGN with blue-ward asymmetric $\rm H\beta$ spread over the full $\rm r_{FeII}$ range and a large portion of the $\rm FWHM_{H\beta} $ range, the red-ward asymmetric sources are concentrated at lower values of $\rm r_{FeII}$, while dominating the higher segment of $\rm FWHM_{H\beta} $. This result is essentially suggesting that high accretors, with relatively large $\rm r_{FeII}$ values (e.g., \citealt{panda19}), show mainly blue-ward asymmetric $\rm H\beta$ profiles. We also observe a considerable overlap of red and blue asymmetries for sources with moderate $\rm FWHM_{H\beta}$ and $\rm r_{FeII}$. 

\begin{figure}
    \centering\includegraphics[width=8 cm]{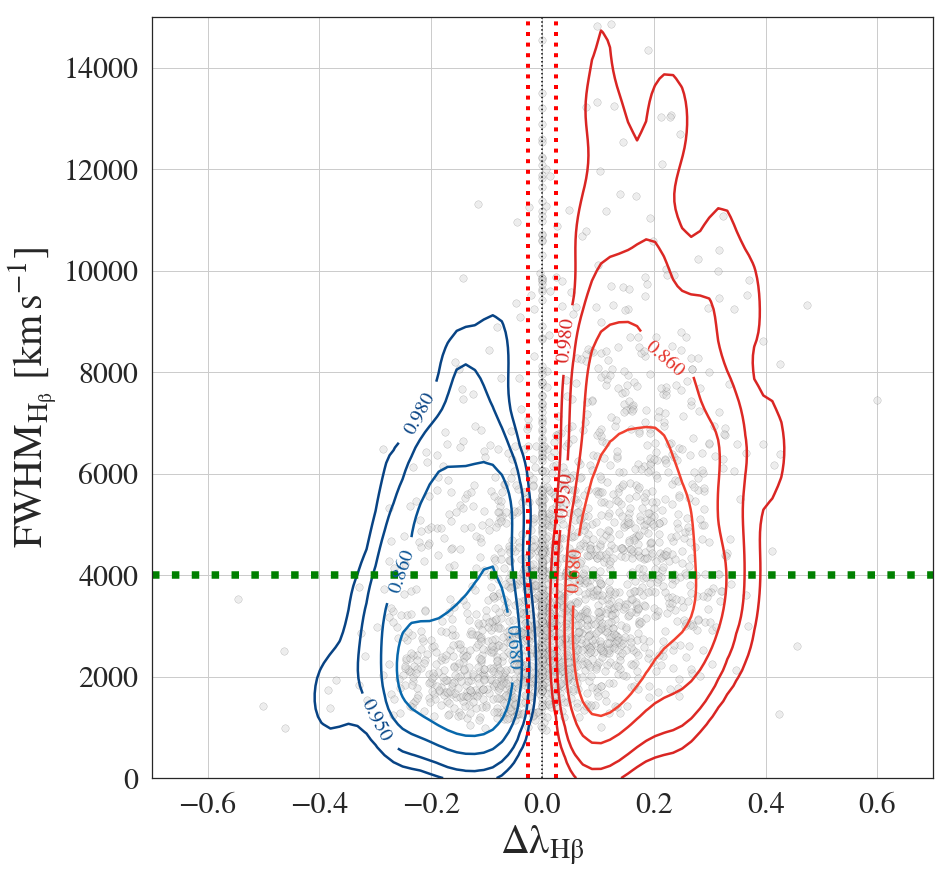}
    \caption{FWHM of $\rm H\beta$ as a function of the Balmer asymmetry index. The kernel density contours separate blue- from redward asymmetric $\rm H\beta$ emitting sources. The green dotted line indicates the separation in Pop.A/B at $\rm FWHM\_{H\beta} \sim 4000 \, km \, s^{-1}$. Typical $1\sigma$ errors are shown centered at zero by two dotted vertical bars.  }
    \label{fig:fwhm_asym}

    \centering\includegraphics[width=8cm]{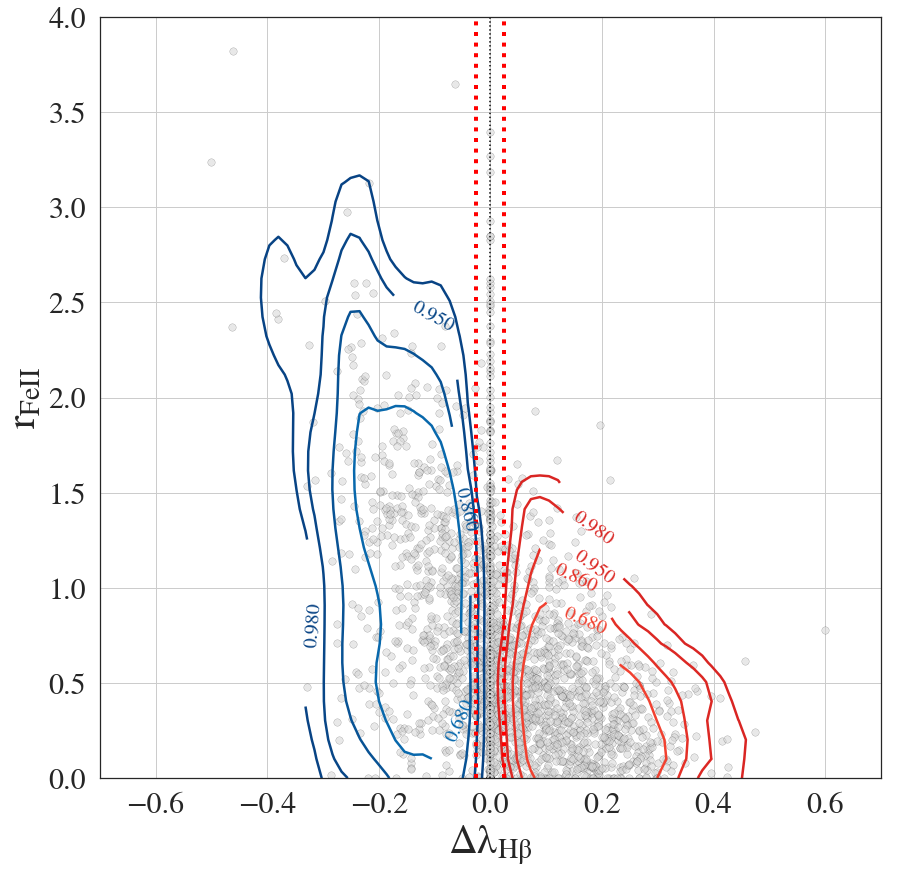}
    \caption{$\rm r_{FeII}$ as a function of the Balmer asymmetry index. The kernel density contours separate blue- from redward asymmetric $\rm H\beta$ emitting sources. The typical $1\sigma$ error in $\rm \Delta \lambda $ are presented by two dotted vertical bars. }
    \label{fig:rfeii_asym}
\end{figure}

 In their in-depth study of $\rm H\beta$ line profiles of $ \sim 470$ low-z SDSS (DR5) quasars, Z10 show that low accretion rates sources (low $\rm r_{FeII}$) possess a typically red-asymmetric $\rm H\beta$ profile, while high accretion rate sources tend to prefer blue-asymmetric $ \rm H\beta$ profiles. Figs. \ref{fig:fwhm_asym}-\ref{fig:rfeii_asym} can directly be compared to Fig. 6a and 9a presented in their work: 
 
 \begin{itemize}
     \item The size of our sample allows us to extend their results by affirming that blue asymmetric $\rm H\beta$ profiles can be found in sources with relatively high $\rm FWHM_{H\beta}$ ($\sim 8000 \, km\,s^{-1}$) and along the full EV1 sequence. There is, however, a clear decrease in blue asymmetric Balmer profiles at $\rm FWHM_{H\beta} \sim 4000 \, km \, s^{-1}$, consistent with the low-redshift separation in Population A/B introduced by \citet{sulentic00a}. An excess of red-asymmetric $\rm H\beta$ profiles is indeed observed for larger widths. Sources with lower $\rm FWHM_{H\beta}$ do not, however, display more symmetric profiles. We thus argue that there is no evidence for a systematic relationship between $\rm FWHM_{H\beta}$ and $\rm \Delta \lambda$. The excess of red asymmetric $\rm H\beta$ emission for larger widths might be the signature of an additional redshifted broad emission component. 
     \item Fig. \ref{fig:rfeii_asym} strongly confirms the decreasing trend presented in figure 5 of \citet{Boroson92} and figure 9a of Z10. Combined with the Spearman correlation coefficient of $\rho \sim -0.5$ reported in the previous section, we can state that, up to redshift of $z \sim 0.8 $, low accretors tend to show more red asymmetric profiles, with a significant decrease of red-asymmetric $\rm H\beta$ emitters beyond $\rm r_{FeII} \sim 1.5$. While blue-asymmetric profiles are reported over the full $\rm r_{FeII}$ range, the shifts become more pronounced at higher values of the iron-$\rm H\beta$ flux ratio.
 \end{itemize}

Z10 discuss the VBLR-emission as potential origin of typically red asymmetries in $\rm H\beta$ (\ref{sec:asymi}). In a layered model of the BLR, the asymmetry index traces the displacement between the the $\rm H\beta$ broad component and its very broad component. 
 $\rm H\beta$ lines often show a broad, redshifted, low-intensity feature in their profiles (\citealt{Peterson86,corbin95, brotherton96}). The presence of a distinct emission region, the VBLR (\citealt{popovic04,marziani10}), has been proposed to explain the characteristic red wing of broad Balmer emission. The VBLR is expected to be located in the inner region of the AGN: The typical widths of the very broad component in $\rm H\beta$ indicate much higher Keplerian velocities than the classical broad line region. The presence of a VBC in the profile shapes has to be corrected for when broad low ionization lines are used as virial broadening estimator \citep{Marziani13b}. In Appendix \ref{fig:vbc_example}, three sources of our sample, which have one broad Gaussian component with $\rm FWHM > 10.000 km \, s^{-1}$, are presented.

\begin{figure}
\includegraphics[width=8.0cm]{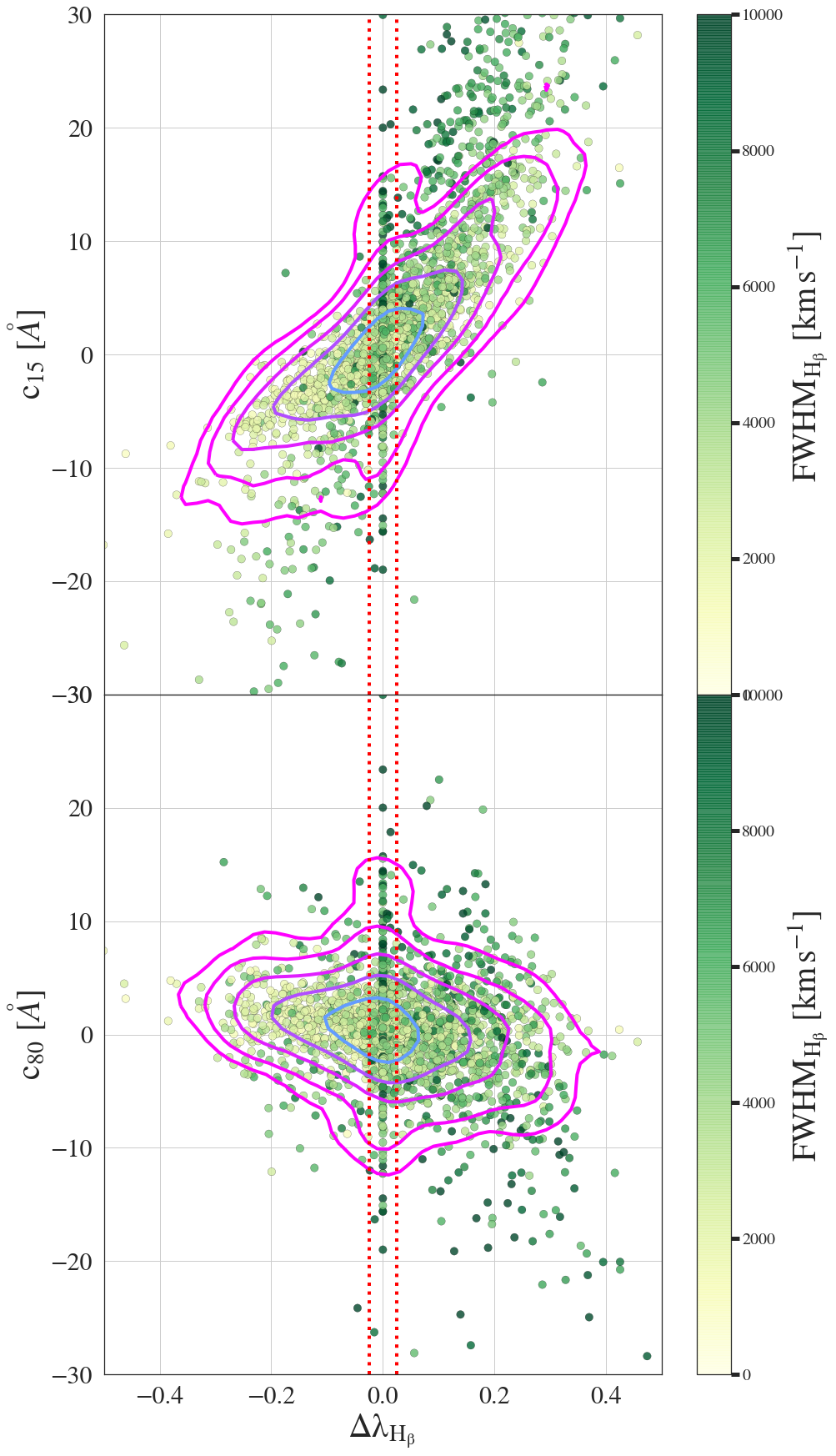}
\caption{The $\rm H\beta$ centroid shifts at $15 \%$ and $80\%$ fractional intensity as a function of $\rm H\beta$ asymmetry. $\rm \mathbf{c_{15}}$ appears more tightly related to $\rm \Delta\lambda_{H\beta}$ than $\rm c_{80}$. The kernel density contours show the sub-sample of sources, for which $\rm H\beta$ has not been fit with broad Gaussians ($\rm FWHM < 10.000 \, km \, s^{-1}$). Typical $1\sigma$ uncertainties in the measurement of $\rm \Delta \lambda_{H\beta}$ are indicated as vertical red dotted lines.}  
\label{fig:centroid}
\centering
\end{figure}

In order to further characterise the kinematics of two potentially distinct emitting regions, we compute the centroid shifts of the $\rm H\beta$ profiles at $\mathbf{15\%}$ ($\rm \mathbf{c_{15}}$) and $80\%$ ($\rm c_{80}$) fractional intensity. Fig. \ref{fig:centroid} displays the correlations between these shifts and the asymmetry index of $\rm H\beta$, colour-coded according to the $\rm FWHM_{H\beta}$. The shifts with respect to the rest-frame at $80\%$ fractional intensity are negatively related to the asymmetry index, i.e., the more blue shifted the top of the profile, the more red-ward asymmetric the complete profile and vice-versa. Symmetric profiles range over the full shift range.     

For the centroid shifts at $\rm \mathbf{c_{15}}$ fractional intensity, an opposite and stronger trend is observed. $\rm \mathbf{c_{15}}$ correlates positively with $\rm \Delta\lambda_{H\beta}$: a red-ward asymmetric $\rm H\beta$ profile can be associated to a redshift of the broad component. As expected, sources with symmetric $\rm H\beta$ span over the full $\rm \mathbf{c_{15}}$ range.
The highest values of $\rm FWHM_{H\beta}$ occur for red asymmetric $\rm H\beta$ consistent with the results in Fig. \ref{fig:fwhm_asym}.

\begin{figure*}

    \includegraphics[height = 7.0 cm]{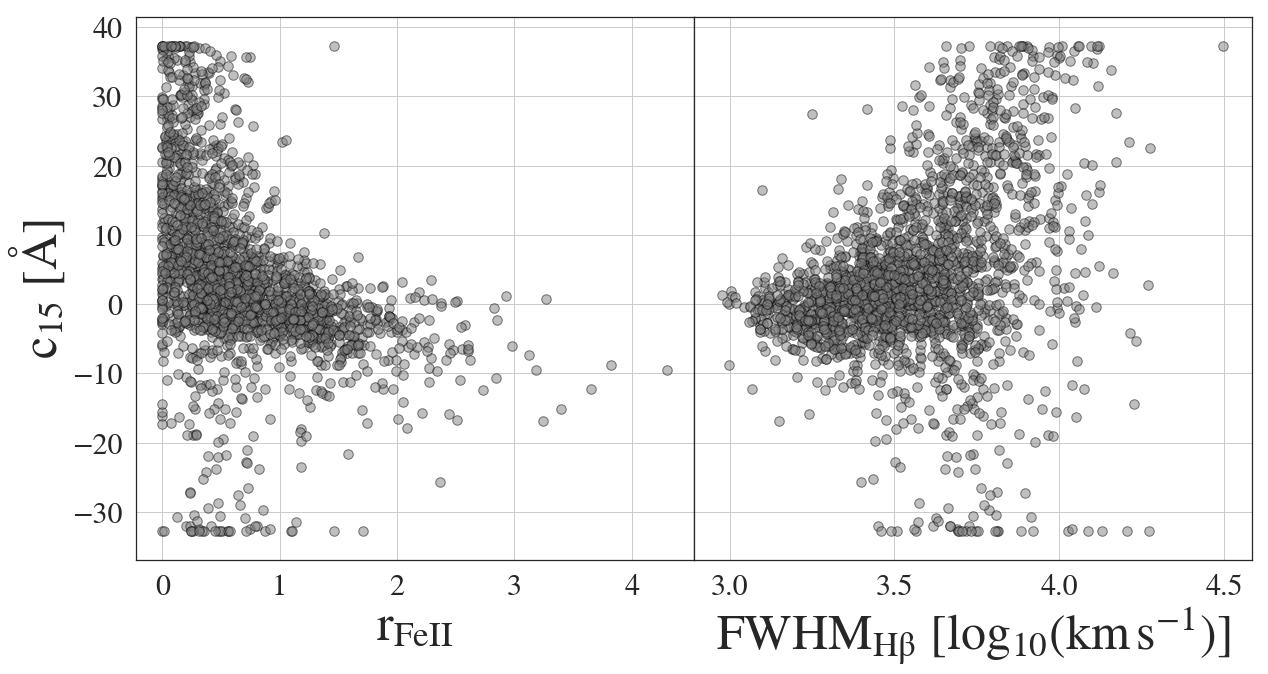}

    \caption{The individual correlations of $\rm \mathbf{c_{15}}$ with the two optical 4DE1 dimensions.  The absolute centroid shifts decrease along the EV1 sequence.}

    \label{fig:c80ev1}
\centering
\end{figure*}

A clear picture arises from the $\rm \mathbf{c_{15}}$ distribution in the EV1 plane (Fig. \ref{fig:c80ev1}). The absolute value of the shifts at the profile base appears to decrease along the EV1-sequence, while $\rm \mathbf{c_{15}}$ shows clear trends with both $\rm r_{FeII}$ and $\rm FWHM_{H\beta}$.  This picture is consistent with the presence of a redshifted VBLR.

\begin{figure}
\includegraphics[width=8.0cm]{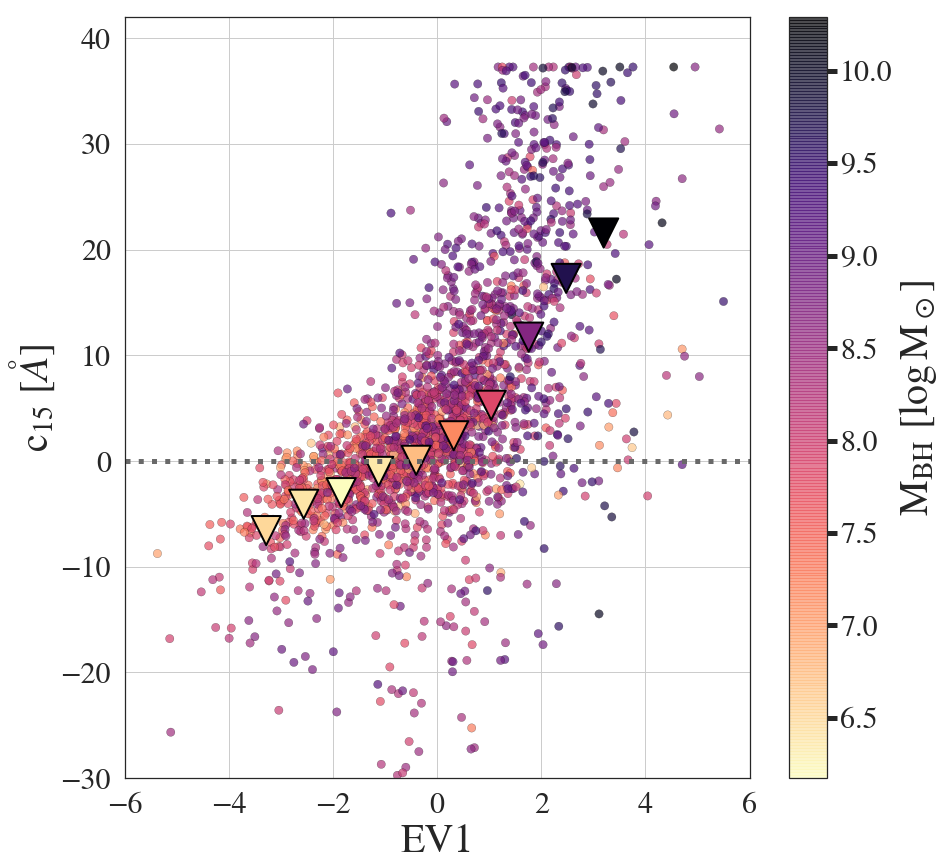}
\caption{$\rm H\beta$ base centroid shifts ($\rm \mathbf{c_{15}}$) projected onto EV1. The points are colour-coded according to the black hole masses estimated with the calibration from \citet{Assef11}. Positive centroid shifts (redshifts) correlate with EV1. Coloured triangles represent the median of  $\rm \mathbf{c_{15}}$ in EV1 bins and have been added to improve the visualisation of this trend. Their colour represents the median black hole masses in the bins, with darker colours corresponding to higher black hole masses. The highest black hole masses are found for redshifted base components.  }
\label{fig:c25ev1}
\centering
\end{figure}

The base shifts of the $\rm H\beta$ profiles in our sample are tightly correlated to the asymmetry index $\rm \Delta \lambda_{H\beta}$. The largest shifts are observed for the most red-asymmetric profiles. The highest base component redshifts are found at the top of the Type 1 AGN main sequence, i.e., at the highest widths of $\rm H\beta$: the broadest components in our sources' $\rm H\beta$ profiles are preferentially redshifted. The correlation of $\rm \mathbf{c_{15}}$ with EV1 provides additional information: as can be seen in Fig. \ref{fig:c25ev1}, the positive base centroid shifts appear to correlate with EV1. Interestingly, the broad redshifted $\rm H\beta$ components are found for the sources with the highest black hole masses.

The broad component redshifts are thus a source of variance in our sample. If the redshifted, very broad components are interpreted as the signature of a VBLR, we can argue that the presence or absence of a distinct, inner  emitting shell in the BLR (or more precisely its redshift) is a strong source of diversity in our sample of Type 1 AGN. Furthermore we report that our results are consistent with \citet{marziani09}, who find a systematic increase of $\rm FWHM_{H\beta}$ with source luminosity. More precisely, \citet{marziani09} find a scaling relation between the luminosity and the contribution of the VBLR to the full $\rm H\beta$ flux in population B ($\rm >4000 \, km \, s^{-1}$).
The absence of red asymmetric $\rm H \beta$ lines at higher $\rm r_{FeII}$ shown in Fig. \ref{fig:rfeii_asym} might be due to strong accretion disk winds preventing the formation of a VBLR.

In Appendix \ref{sec:obscuration}, we explore a toy model of BLR obscuration, which accounts for the presence of a VBLR.

%% file: 07-BRSep.tex
\section{Blue-asymmetric \texorpdfstring{$\rm H\beta$}{Hbeta}: Outflows in a flattened and stratified BLR model }

 \label{BRSEPAB}

\subsection{Blue asymmetries, outflows and self-shielding}

\begin{figure*}
    \includegraphics[width=8 cm]{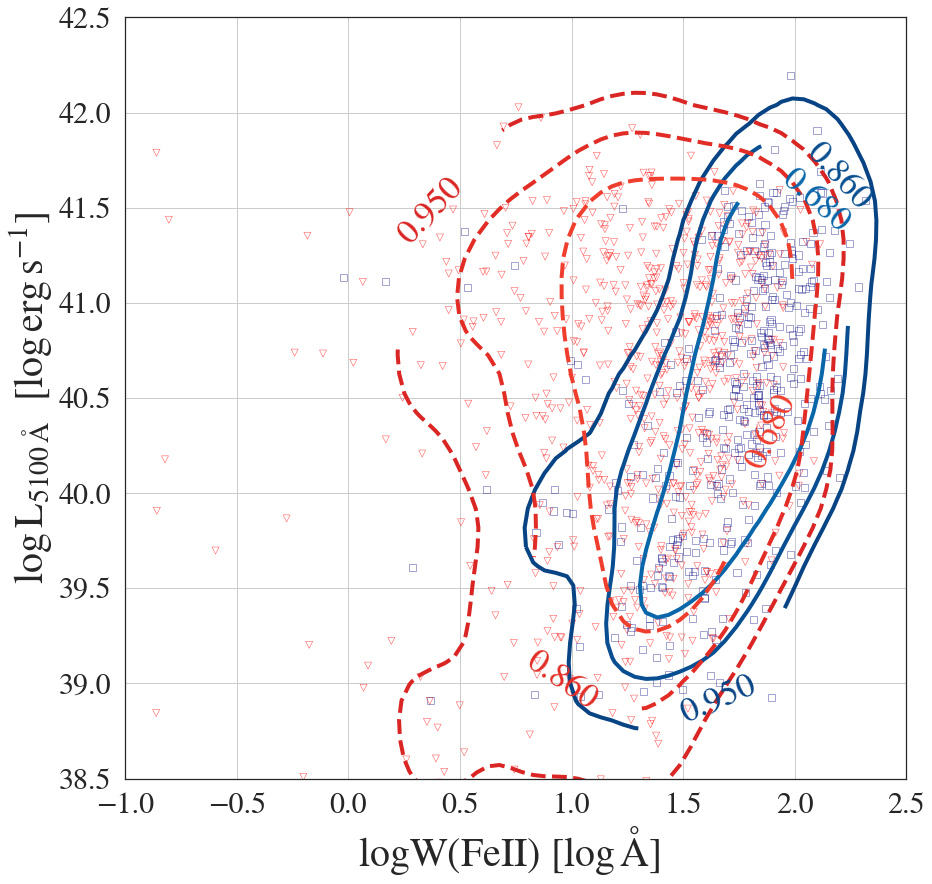}
    \includegraphics[width=8 cm]{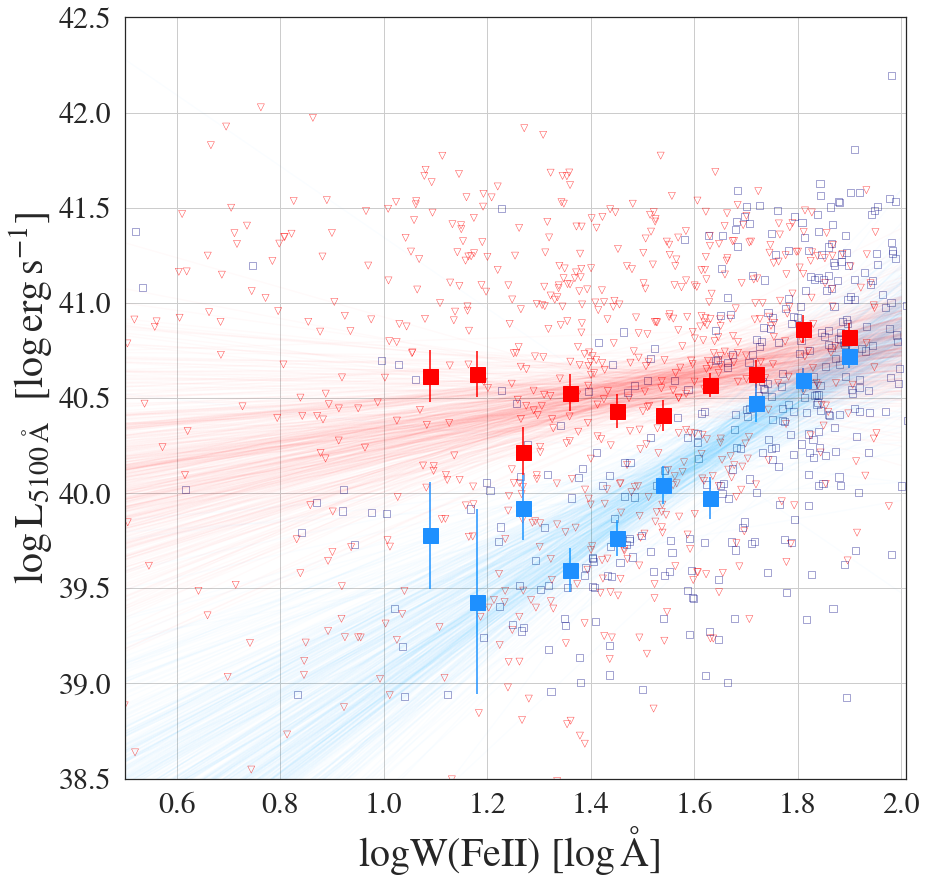}

    \caption{ \textit{Left panel} : The monochromatic luminosity at $5100 \, \angstrom$ as a function of the equivalent width of the $\rm FeII$-blend at $4570$ {\AA}. The red triangles and blue squares distinguish red- ($\rm \Delta \lambda > 0.07$)  from blue-asymmetric ($\rm \Delta \lambda < - 0.07$) $\rm H\beta$ emitters. Kernel density contours have been included. A tighter relation is observed for sources with blue-asymmetric Balmer emission. A similar picture arises in the case of $\rm L_{0.1-2 \, keV}$-$\rm W(FeII)$. 
    \textit{Right panel} : The $\rm L_{5100}$-$\rm W(FeII)$ plane binned in 10 $\rm \log W(FeII)$-bins between 1.0 and 1.9. For each bin, 10.000 bootstrap samples were obtained, for which the mean $\rm L_{5100\,\angstrom}$ was derived. The points and error bars represent the mean of the sampling distribution of the mean. The error bars correspond to the $1\sigma$ error derived from the sampling distribution of the mean, using the percentile method. The linear regression was performed with \textit{Linmix} \citep{kelly07}.}

    \label{fig:L5100Wii_bootstrap}
\centering
\end{figure*}

If we consider that blue- and red-ward asymmetric $\rm H\beta$ profiles are signatures of two distinct kinematic states of the BLR, we should divide our sample according to the sources' asymmetry indices to investigate the properties of each sub-sample. As in the previous section, we formed two sub-samples with $\rm \Delta\lambda_{H\beta}<-0.07$ (blue asymmetric) and $\rm \Delta\lambda_{H\beta} > 0.07$ (red asymmetric). We constructed a Spearman rank correlation matrix for the parameter subset, as defined in Sec. \ref{sec:sample_construction}, of each sub-sample and directly compared their correlation spaces. A striking difference is observed for the correlation of the optical and X-ray luminosity and the iron emission. 
For blue asymmetric $\rm H\beta$ emitters, $\rm L_X$ and $\rm L_{5100\angstrom}$ correlate positively with the equivalent width of the iron emission $\rm W(FeII)$ (Spearman  correlation coefficients: $\rho_S = 0.47$ and $\rho_S = 0.55$,  respectively).
The sources with red asymmetric $\rm H\beta$ have a much weaker correlation between the source luminosities and the equivalent width $\rm W(FeII)$ ($\rho_S = 0.023$ for $\rm L_X$ and $\rho_S=0.097$ for $\rm L_{5100\angstrom}$). The correlation of $\rm L_{5100\angstrom}$ and $\rm W(FeII)$ for red-asymmetric sources is $p = 0.53$ and can safely be considered as insignificant.  
This contrasting behaviour is presented in the left panel of Fig. \ref{fig:L5100Wii_bootstrap}. In order to confirm the different clustering of red- and blue-asymmetric $\rm H\beta$ emitters in the plane spanned by $\rm L_{5100\,\angstrom}$ and $\rm W(FeII)$, we binned $\rm \log \, W(FeII)$ in the range $1.0-1.9$ and bootstrapped the sub-samples in each bin (10.000 resamples). We obtained the mean $\rm L_{5100\,\angstrom}$  for each bin from the sampling distribution of the means. The $68 \%$ confidence intervals were derived using the percentile method (from the $16.0 \%$ and $84.0 \%$ percentiles). This method should yield relatively good estimates of the error, given the symmetric shape of the sampling distribution of the mean in each bin (shown in the right panel of Fig. \ref{fig:L5100Wii_bootstrap}). The positive correlation between $\rm L_{5100\angstrom}$ and $\rm W(FeII)$ for sources with blue asymmetric $\rm H\beta$ is confirmed. No trend is found for sources with red-asymmetric $\rm H\beta$. The error bars, showing the standard deviation of the the bootstrap samples' means, do not overlap.  The correlation of $\rm L_{5100\angstrom}$ and $\rm W(FeII)$ for the blue asymmetric $\rm H\beta$ population withstood a partial correlation test, where we controlled for the confounding  $\rm W(H\beta)$. Similarly, when we control for the $\rm FWHM_{FeII}$ of the Gaussian kernel used to fit FeII, $\rm L_{5100\angstrom}$ and $\rm W(FeII)$ remain positively correlated. The dependency of the FeII flux on the continuum luminosity in blue-asymmetric $\rm H\beta$ emitters is only marginally related to the broadening of the iron lines.

This contrasting behaviour also manifests itself in the correlation of the luminosity parameters with the flux ratio $\rm r_{FeII}$. Blue-asymmetric asymmetric $\rm H\beta$ sources are significantly correlated $\rm r_{FeII}$ and source luminosities ($\rho_S = 0.20$ for $\rm L_{5100\angstrom}$ and  $\rho_S = 0.19$ for $\rm L_X$). For red asymmetric $\rm H\beta$ sources, these quantities are anti-correlated ($\rho_S = -0.16$ for $\rm L_{5100\angstrom}$ and  $\rho_S = -0.25$ for $\rm L_X$).  

Another noticeable difference is observed for the $\rm \Delta \lambda_{[OIII]}$ parameter. In the sub-sample of blue-asymmetric Balmer emitters, $\rm \Delta \lambda_{[OIII]}$ and $\rm F([OIII])/F(H\beta)$ are anti-correlated ($\rho_S = -0.26$). If blue shifted wings are interpreted as the signature of NLR outflows, this result would imply that outflow velocities increase with increasing $\rm [OIII]/H\beta$ flux ratios. In the red-asymmetric sub-sample, these two parameters are positively correlated ($\rho_S = 0.17$).
 
Blue-ward asymmetries of low-ionisation lines have been related to radiation driven outflows \citep{marziani13a}. They occur in the high $\rm r_{FeII}$ bins of population A. (e.g. \citealt{ganci19}), which contain the highest accretion rate sources along the main sequence (e.g. \citealt{sun15, sulentic17, panda19}). The origin  of $\rm FeII$ emission in the BLR has been a long-standing matter of research. Several lines of evidence support an emission of $\rm FeII$ in the outer parts of the BLR, while $\rm H\beta$ might be emitted closer the the black hole (e.g. \citealt{rodriguez02,barth13,marinello16}). The physical conditions for iron ionisation have been investigated in detail. \citet{wills85} showed that photoionisation models such as the locally optimally emitting clouds (\citealt{baldwin95}) might not suffice to account for the total observed FeII flux in AGN. For a review of necessary conditions for the formation of the observed FeII in photoionisation models see \citet{collin00}. Models of collisional excitation (e.g. \citealt{baldwin04, joly07}), in addition to continuum and line fluorescence (e.g. \citealt{sigut98, marinello16}), have been considered. These models make predictions on the physical conditions of the FeII emitting region, such as shielding from the continuum source and high densities. 

In a flattened, horizontally stratified cloud distribution in Keplerian motion around the central black hole, the difference in sensitivity to continuum of the $\rm FeII$ emitting region might arise from different degrees of exposure to the central continuum source of the dense clouds from which the iron is emitted. Radiation driven winds produce outflows of $\rm H\beta$ emitting gas, thereby exposing the previously shielded $\rm FeII$ regions more directly to continuum emission. The scaling between $\rm W(FeII)$ and the $\rm L_{5100\angstrom}$ might thus be due to the contribution of UV fluorescence to the excitation of Fe+ levels. Investigating FeII fluorescence in HII regions, \citet{rodriguez99} uses the sensitivity to fluorescence of $\rm [FeII]\lambda 4287$, as well as the continuum insensitive $\rm [FeII]\lambda 8617$ to investigate the role played by the UV radiation field in the formation of this line. If the presented model of unshielding holds, the intensity ratio $\rm I([FeII]\lambda 4287)/I([FeII]\lambda 8617)$ should scale differently with the UV/optical continuum luminosity for the red- and blue-asymmetric $\rm H\beta$ populations. This may be investigated in further work. 
 
In this scenario, the distance of the FeII region to the SMBH should not depend on the presence of outflows. We performed a two-sample Anderson-Darling test on the distributions of the $\rm FWHM_{FeII}$ of the Gaussian kernel which was convolved with the FeII template in the red- and blue-asymmetric $\rm H\beta$ subsets. 55 sources in our sample have no FeII flux detection and were not considered for this test. The obtained p-value is $p>0.25$. We thus cannot reject that the $\rm FWHM_{FeII}$ for red- and blue-asymmetric $\rm H\beta$ sources were sampled from the same distribution. Under the assumption of a Keplerian velocity field, the radial distribution of the $\rm FeII$ emitting clouds is not affected by the presence of outflows.

We conclude that our data is consistent with a flat, stratified and self-shielding model of the BLR, in which the FeII and $\rm H\beta$ emissions originate at different radii. When the inner clouds are driven out of the plane by outflows during strong accretion events, the previously shielded, neutral and dense FeII clouds are more exposed to the ionizing continuum radiation, which is clearly seen in the FeII flux-luminosity scaling.

\subsection{Evidence for model degeneracy: FeII vs. \texorpdfstring{$\rm H\beta$}{Hbeta}}

We note that C19 fitted the emission of the iron complex over the full $\rm H\beta$ centered region, using the I Zw 1 template \citep{Boroson92}. Since all the emission features of $\rm FeII$ in the spectral window were taken in account during the fits, the effect of contamination by $\rm H\beta$ should be marginal on the total $\rm FeII$ flux, i.e. $\rm W(FeII)$ should not be strongly affected by increasing $\rm H\beta$ line flux. However we can investigate if the $\rm FeII$ contaminates the $\rm H\beta$ fit on its blue side. 

For blue asymmetric $\rm H\beta$ emitters, we seek to test if the the equivalent width of $\rm FeII$ correlates with the $\rm H\beta$ asymmetry index, i.e. we want to determine if the excess flux on the blue side of $\rm H\beta$ is due to contamination by iron.

On the sub-sample with, $\rm \Delta\lambda_{H\beta}<-0.07$, we measure the Spearman rank correlation coefficient of $\rm \mid \Delta\lambda H\beta \mid$ and $\rm W(FeII)$. We obtain $\rho = 0.087$ with a p-value of $p=0.08$. Even if we do not correct the significance threshold for multi-hypothesis testing, one can state that there is no significant correlation between $\rm \Delta\lambda_{H\beta}$ and $\rm W(FeII)$. We note however that the equivalent width of $\rm H\beta$ is a confounding variable for both: $\rm \Delta \lambda H \beta$ ($p=-0.26$) and $\rm W(FeII)$ ($p=0.29$). We thus have to once again perform a partial correlation analysis, marginalising over $\rm W(H\beta)$ in order to obtain the real correlation behaviour of the blue-asymmetries and the iron strength. The partial correlation coefficient is $p_P= 0.17$ with a p-value of $p \sim 10^{-4}$. There is a significant correlation between $\rm\Delta\lambda_{H\beta}$ and  $\rm W(FeII)$ once we control for $\rm W(H\beta)$. We thus cannot exclude that the iron complex contaminates the blue wing of $\rm \Delta H\beta$.

%% file: 08-Conclusions.tex
\section{Conclusions}
\label{sec:conclusions}
In this work we have performed a statistical analysis of a high signal-to-noise ratio ($\rm S/N > 20$) sub-sample of the SDSS-IV/SPIDERS DR14 VAC of X-ray selected Type 1 AGN. The sample included 2043 sources spanning an X-ray luminosity range of $\rm L_X = 1.9 \times 10^{41} - 9.9 \times 10^{45} \, erg \, s^{-1}$, up to redshift $z = 0.80$. It is the largest sample of high signal-to-noise ratio sources used for the PCA-based, statistical study of Type 1 AGN spectral properties to date.

\begin{itemize}
    \item We used PCA as a central tool to determine the source of variance in our data and have mapped a correlation space which is remarkably consistent with previous studies of the optical Eigenvector 1 (e.g., \citealt{Boroson92, sulentic00a, Grupe04,Shen14}), while probing a larger cosmological volume. 
    \item  We confirm that the Eddington ratio and the black hole mass are significantly related to the observed diversity of Type 1 AGN, through their correlation to EV1 and EV2. 
    \item Studying $\rm H\beta$ line shapes in this context, we find blue asymmetric emission profiles for the full Type 1 AGN main sequence, while red asymmetries only appear for low accretors. Z10, investigating $\rm H\beta$ profile asymmetries in the 4DE1 context,  have suggested that profiles with lower $\rm FWHM_{H\beta}$ tend to be more symmetric while profiles with larger $\rm FWHM_{H\beta}$  are preferentially red-asymmetric. The larger number of sources in our sample enabled us to complete this picture: While we do observe a larger portion of sources with red asymmetry index at high $\rm FWHM_{H\beta}$, lower width profiles appear to cover the full range of asymmetries observed in our sample, i.e., lower $\rm FWHM_{H\beta}$ do not lead to more symmetric profiles. We can, however, confirm a strong trend between $\rm r_{FeII}$ and $\rm \Delta \lambda_{H\beta}$. In their discussion of the physical origins of $\rm H\beta$ profile shapes, Z10 identify VBLR-emission and disk winds as main candidates.
    \item A sub-class of our sources does indeed show very shifted, broad  components in their  $\rm H\beta$ emission, possibly due to the presence of a distinct emitting gas distribution in the inner regions of the BLR. The redshift of this VBLR correlates with Eigenvector 1 and might thus be related to Type 1 AGN diversity.
    \item Exploring parameter correlations for blue- and redward asymmetric $\rm H\beta$ emitters separately, we observed that $\rm FeII$ line flux  correlates differently with source luminosity for red- and blue-ward asymmetric $\rm H\beta$ emitters. We discussed this effect in the light of a flattened, self-shielding BLR, in which the FeII emitting clouds are located are larger radii than the ones emitting the Balmer lines. The outflows, we associate $\rm H\beta$ blue-ward asymmetries with, might in such a configuration strip the FeII region of its shield, i.e. the innermost parts of the BLR. However, we find evidence for inter-line contamination between FeII and $\rm H\beta$, which might play a confounding role in this effect.

\end{itemize}


%% file: 09-Acknowledgements.tex
\section*{Acknowledgements}
 
We thank the anonymous referee for comments that significantly improved this paper.

We would like to thank Hagai Netzer for insightful and stimulating conversations on the physics of the BLR. We also thank Claudia Scarlata for her precious input on Machine Learning techniques and Tom Dwelly for helpful advice.

FJC acknowledges financial support through grant AYA2015-64346-C2-1P (MINECO/FEDER).

FJC also acknowledges funding from the European Union's Horizon 2020 Programme under the AHEAD project (grant agreement n. 654215).


Funding for the Sloan Digital Sky Survey IV has been provided by the Alfred P. Sloan Foundation, the U.S. Department of Energy Office of Science, and the Participating Institutions. SDSS acknowledges support and resources from the Center for High-Performance Computing at the University of Utah. The SDSS web site is www.sdss.org.

SDSS is managed by the Astrophysical Research Consortium for the Participating Institutions of the SDSS Collaboration including the Brazilian Participation Group, the Carnegie Institution for Science, Carnegie Mellon University, the Chilean Participation Group, the French Participation Group, Harvard-Smithsonian Center for Astrophysics, Instituto de Astrof\'{i}sica de Canarias, The Johns Hopkins University, Kavli Institute for the Physics and Mathematics of the Universe (IPMU) / University of Tokyo, the Korean Participation Group, Lawrence Berkeley National Laboratory, Leibniz Institut f\"{u}r Astrophysik Potsdam (AIP), Max-Planck-Institut f\"{u}r Astronomie (MPIA Heidelberg), Max-Planck-Institut f\"{u}r Astrophysik (MPA Garching), Max-Planck-Institut f\"{u}r Extraterrestrische Physik (MPE), National Astronomical Observatories of China, New Mexico State University, New York University, University of Notre Dame, Observat\'{o}rio Nacional / MCTI, The Ohio State University, Pennsylvania State University, Shanghai Astronomical Observatory, United Kingdom Participation Group, Universidad Nacional Aut\'{o}noma de M\'{e}xico, University of Arizona, University of Colorado Boulder, University of Oxford, University of Portsmouth, University of Utah, University of Virginia, University of Washington, University of Wisconsin, Vanderbilt University, and Yale University.

%% file: Appendix.tex
\section{Equivalent widths estimate}
\label{sec:eqw}

 The equivalent width offers a measure of an emission line's strength.  Equivalent widths were not included in C19 and had to be derived. 
 
 The equivalent width of a spectral line is defined as:
 
 \begin{equation}
 W_\lambda = \int  \frac{S_1(\lambda)-S_c(\lambda)}{S_c(\lambda)} d\lambda
 \label{EqWidth}
 \end{equation}
$S_1$ denotes the total flux density at a given wavelength and $S_c$ the continuum flux density. 

To obtain the interpolated $S_1$ and $S_c$ fluxes, the lines were reconstructed from the catalogue's fitting parameters, the associated Gaussian fitting functions, the continuum power law, and the templates for the iron and galaxy contributions. Each Gaussian component $G(\lambda)$ was reconstructed according to:

\begin{equation}
G_{\rm line}(\lambda) = N_{\rm line}\cdot \exp \left[ - \frac{1}{2} \left ( \frac{\lambda - P_{\rm line}}{W_{\rm line}} \right)^2 \right]
\end{equation}

Here $N_{line}$, $P_{line}$ and $W_{line}$ denote the norm, peak wavelength and width for the fit as listed in C19. For multiple Gaussian components $G_j(\lambda)$  the total flux in the line is then obtained from $G_{total}(\lambda) = \sum_j G_j(\lambda)$. In C19, the continuum model consists of a power law, a host galaxy component and the iron emission.

The power law model is reconstructed from :

\begin{equation}
 PL(\lambda) = N_{\rm PL}\cdot \lambda^{S_{\rm PL}}
\end{equation}
$N_{\rm PL}$ and $S_{\rm PL}$ are the norm and the slope of the power law for each spectrum as listed in the catalog.
 Following the catalogue's author's prescription a Gaussian filter was applied to the $\rm FeII$ emission to reproduce the blending of the multiplets. The continuum emission is calculated by:

\begin{equation}
S_c(\lambda) = PL(\lambda) + FEII(\lambda) + GAL(\lambda)
\end{equation}

The total flux in a given line is: $S_1(\lambda)= S_c(\lambda) + G_{\rm total}(\lambda)$. 

\begin{figure*}
    \includegraphics[width=6 cm]{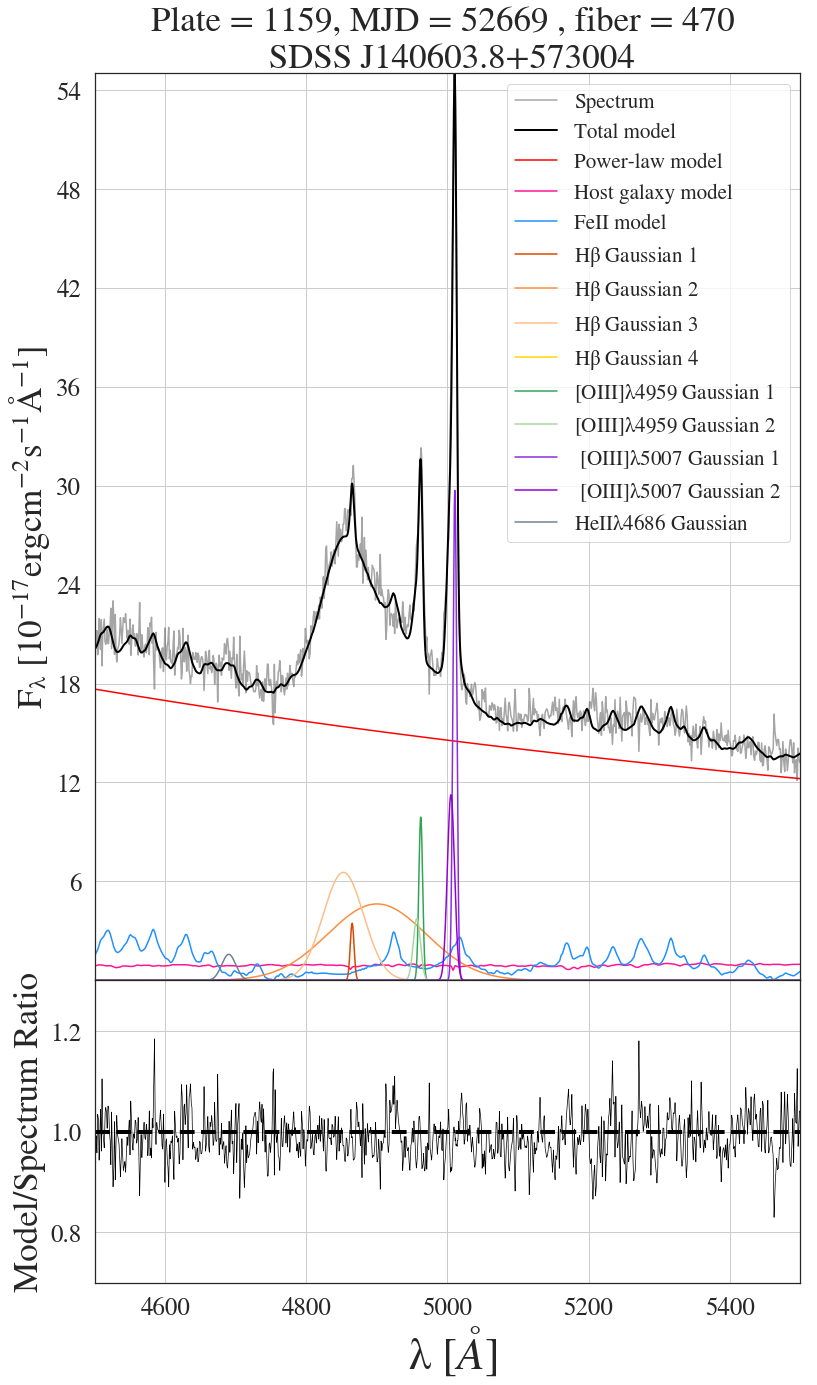}
    \includegraphics[width=6.1 cm]{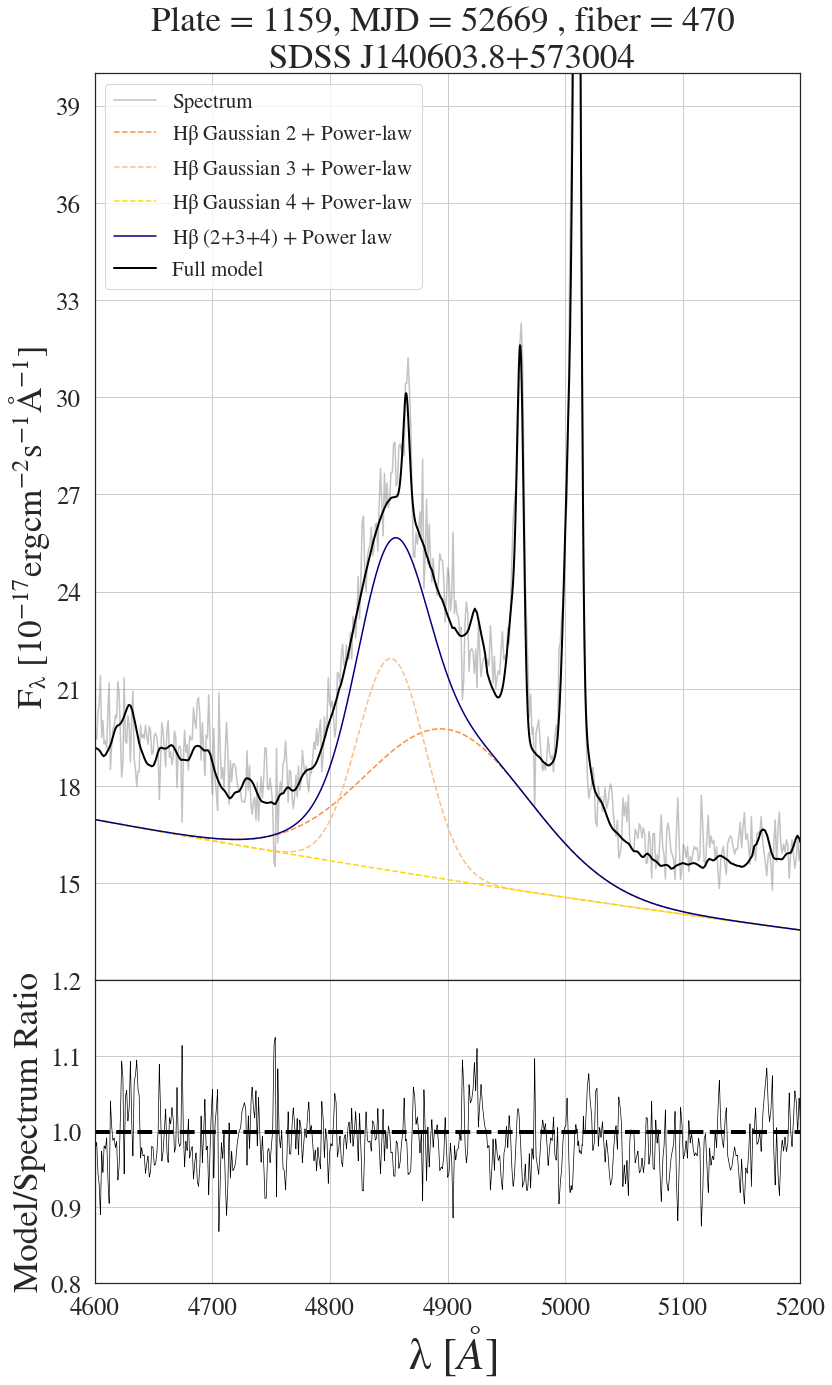}
\caption{\textit{Left panel}: Reconstructed $\rm H\beta$ fitting region using the model used by C19. The $\rm [OIII]$, $\rm HeII$ and $\rm H\beta$ model components are labeled. The components of the continuum model (a power-law, a host galaxy model and the iron pseudo-continuum) are also displayed. The original observed spectrum is shown in light grey. \textit{Right panel:} A zoom onto the $\rm H\beta$ Gaussian line decomposition. In order to improve the visualisation, the power-law contribution was added to the Gaussian components.  Only two broad components were required to fit $\rm H\beta$. }
    \label{fig:line}
    \centering
\end{figure*}

\begin{figure*}
    \includegraphics[width=6 cm]{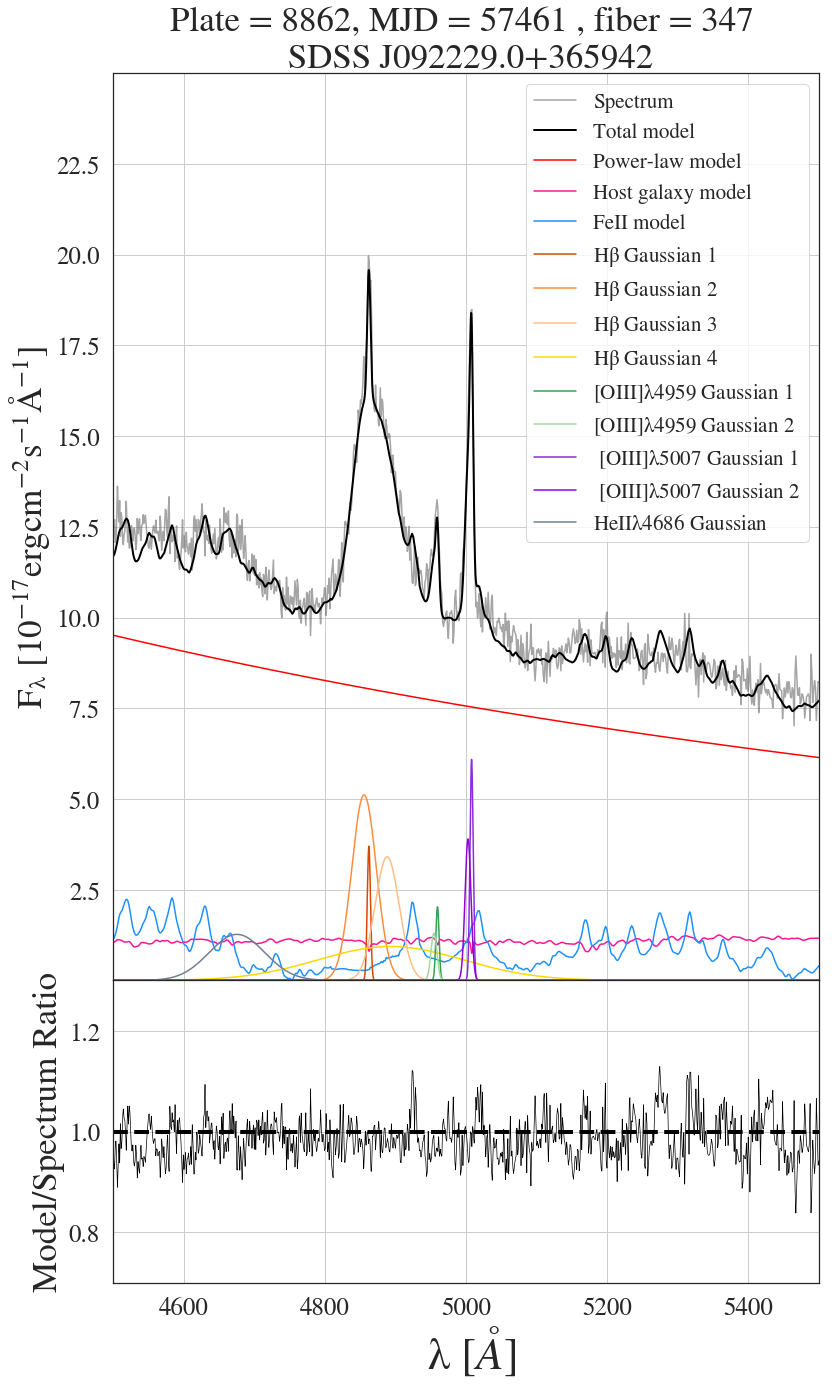}
    \includegraphics[width=6.1cm]{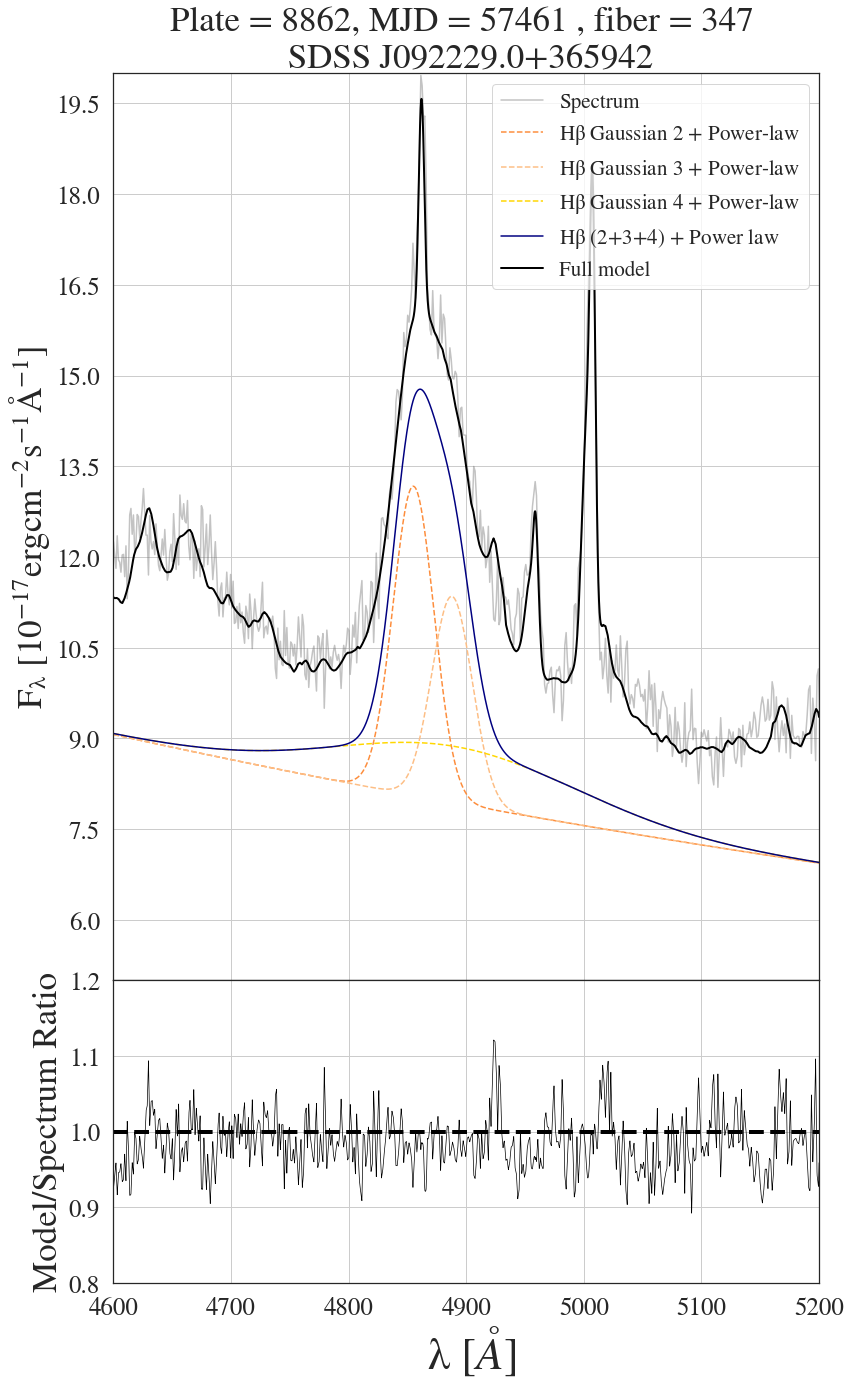}
\caption{Same as in Fig. \ref{fig:line} for another source in our sample. Here three broad Gaussian components were needed to fit $\rm H\beta$}.
    \label{fig:line2}
    \centering
\end{figure*}

Fig. \ref{fig:line}-\ref{fig:line2} displays two fits centered around the $\rm H\beta$/$\rm [OIII]$ complex. reconstructed  using the above model. Interpolating $D_1$ and $S_c$ over the fitting range of $\rm H\beta$ or $\rm MgII$ and performing the integration in (\ref{EqWidth}) yields the equivalent width for the chosen lines. 

\section{Examples of \texorpdfstring{$\rm H\beta$}{Hbeta} profiles: red- and blue asymmetric \texorpdfstring{$\&$}{and} very broad components}
\label{sec:red_blue_vbc}

In the top (middle) three panels of the figure in Appendix \ref{sec:red_blue_vbc} we present examples of red- (blue-) asymmetric broad $\rm H\beta$ profiles and the Gaussian fits performed in C19. The corresponding asymmetry indices are indicated in the caption. Similarly in the bottom three panels of this same figure, $\rm H\beta$ lines with very broad Gaussian components are presented. The VBC is highlighted in red.

\begin{figure*}

    \includegraphics[width=4.4 cm]{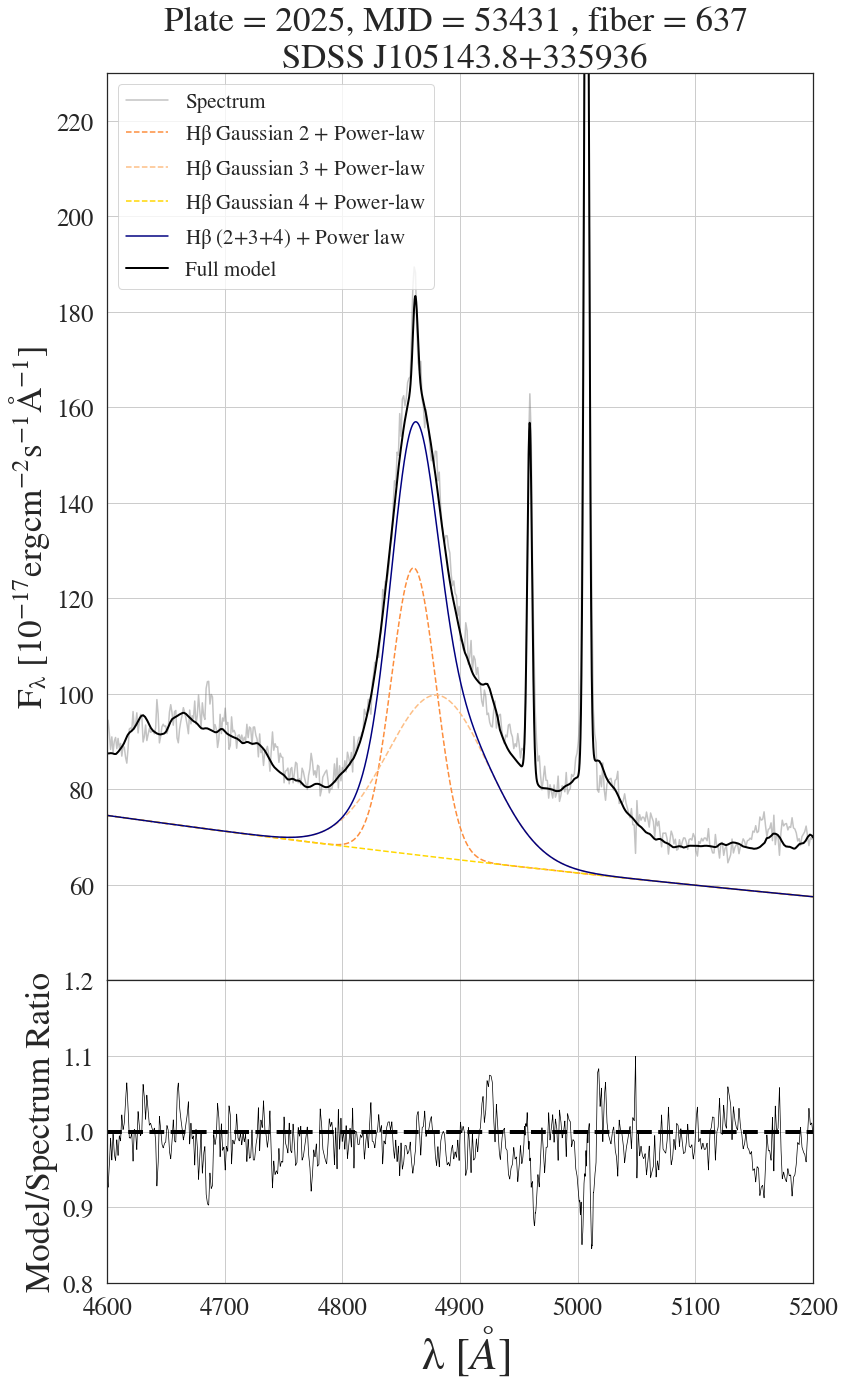}
    \includegraphics[width=4.4 cm]{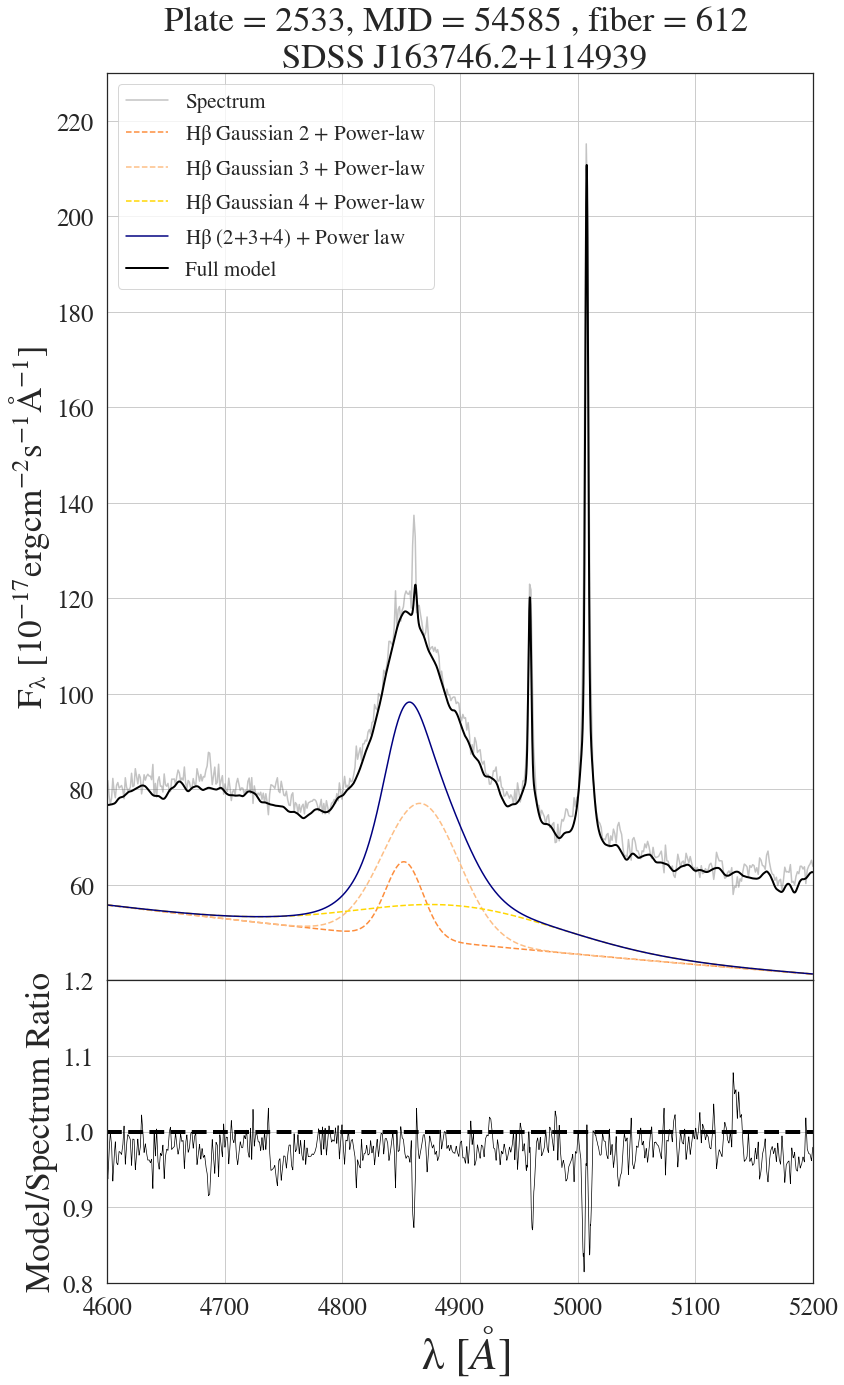}
    \includegraphics[width=4.4 cm]{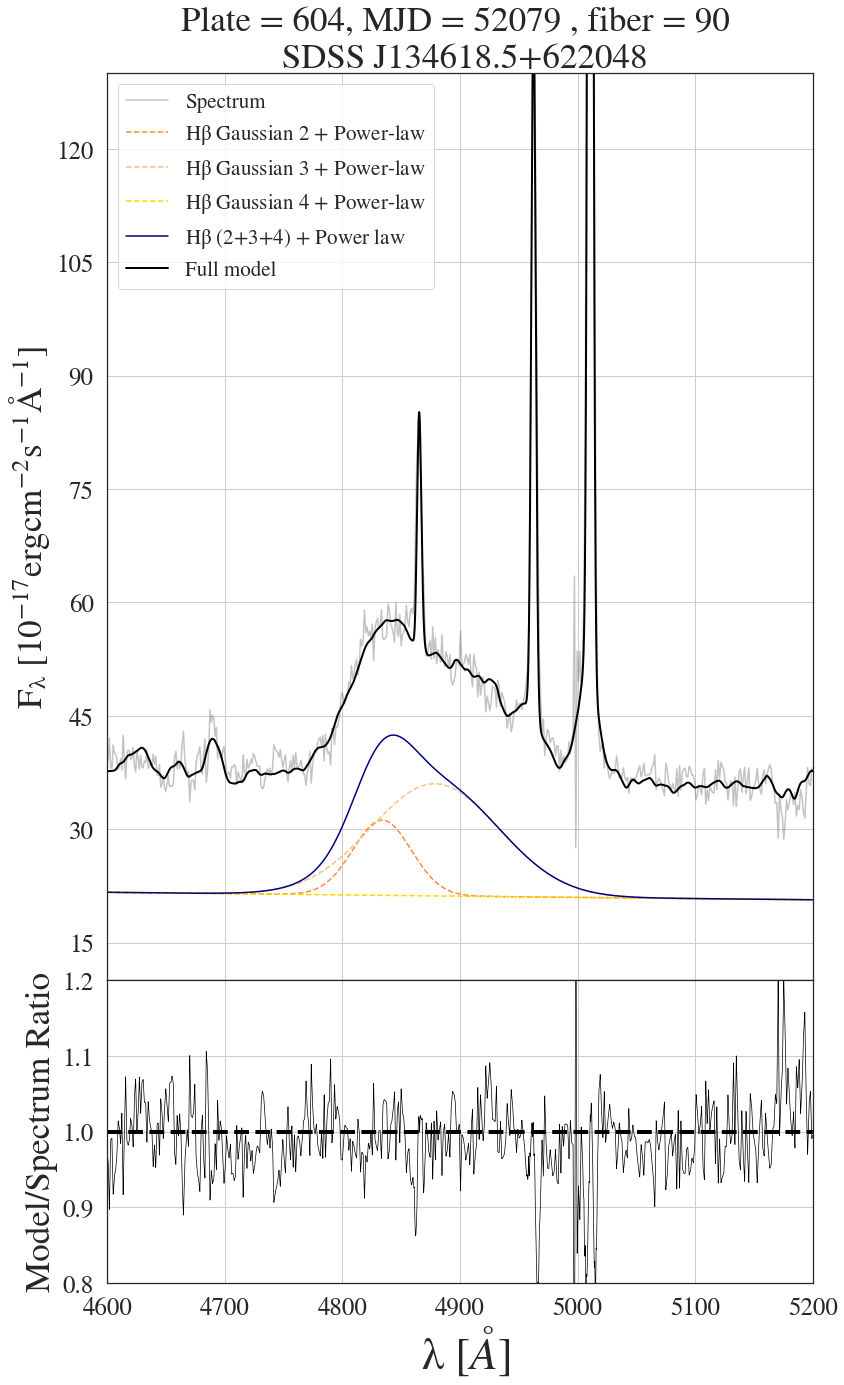}

    \includegraphics[width=4.4 cm]{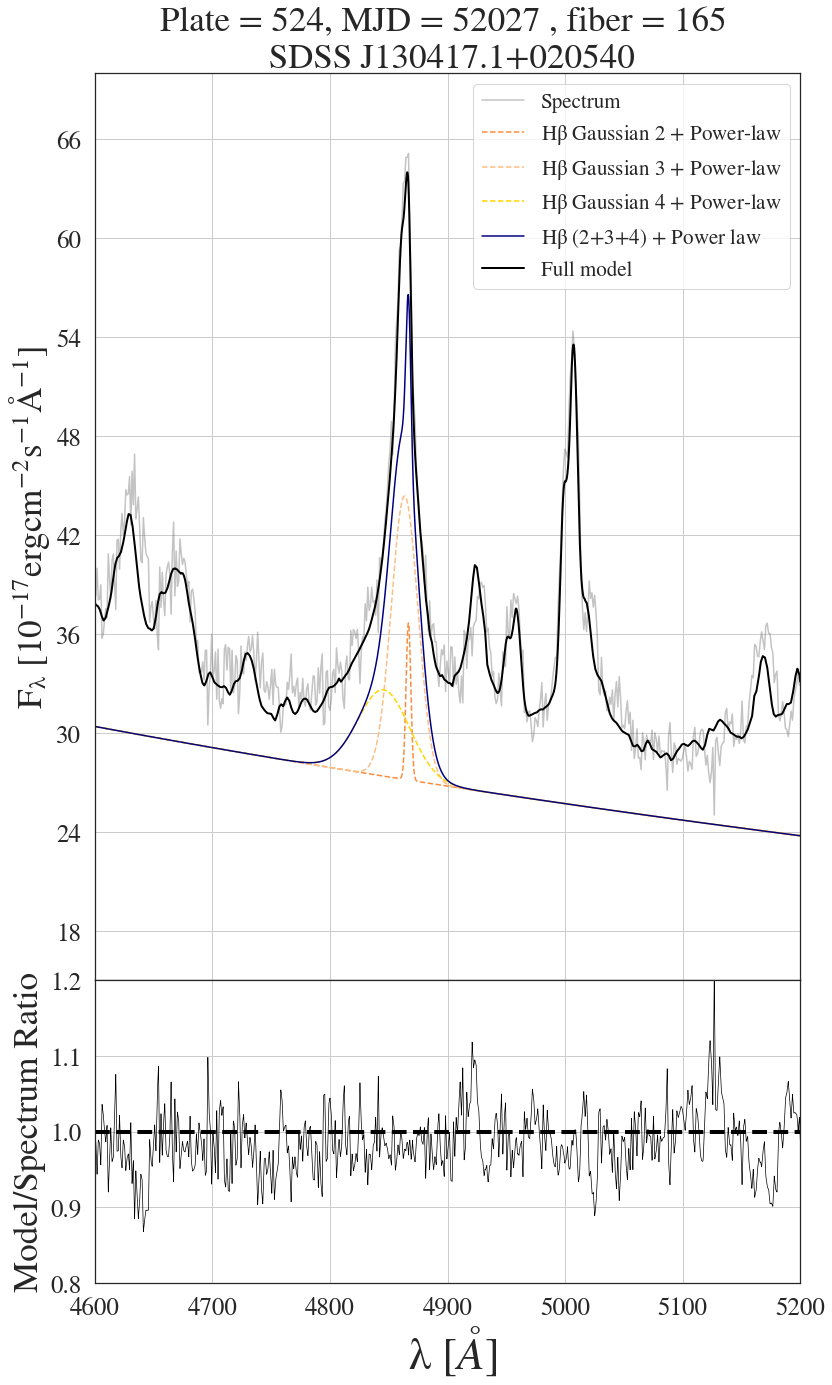}
    \includegraphics[width=4.4 cm]{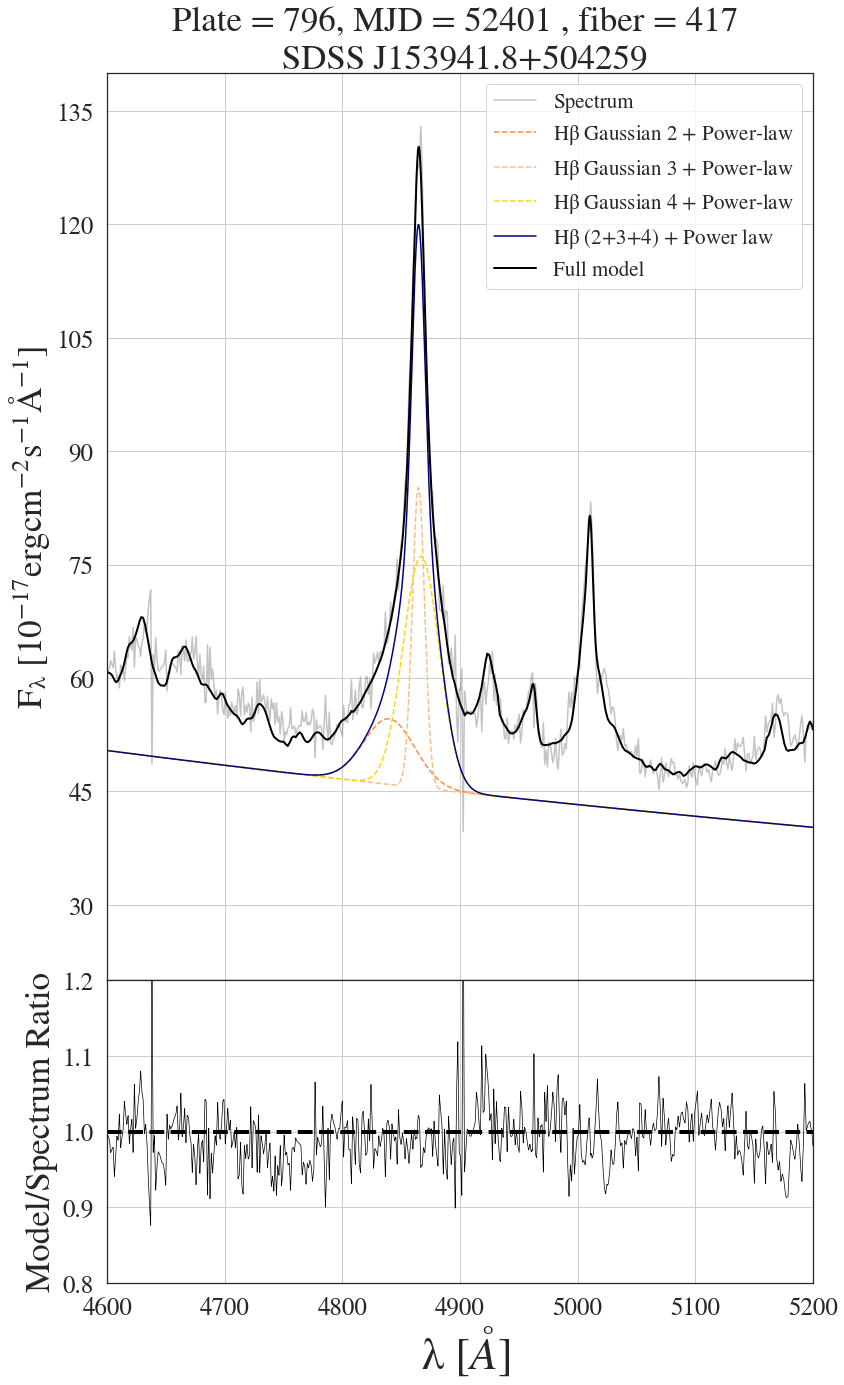}
    \includegraphics[width=4.4 cm]{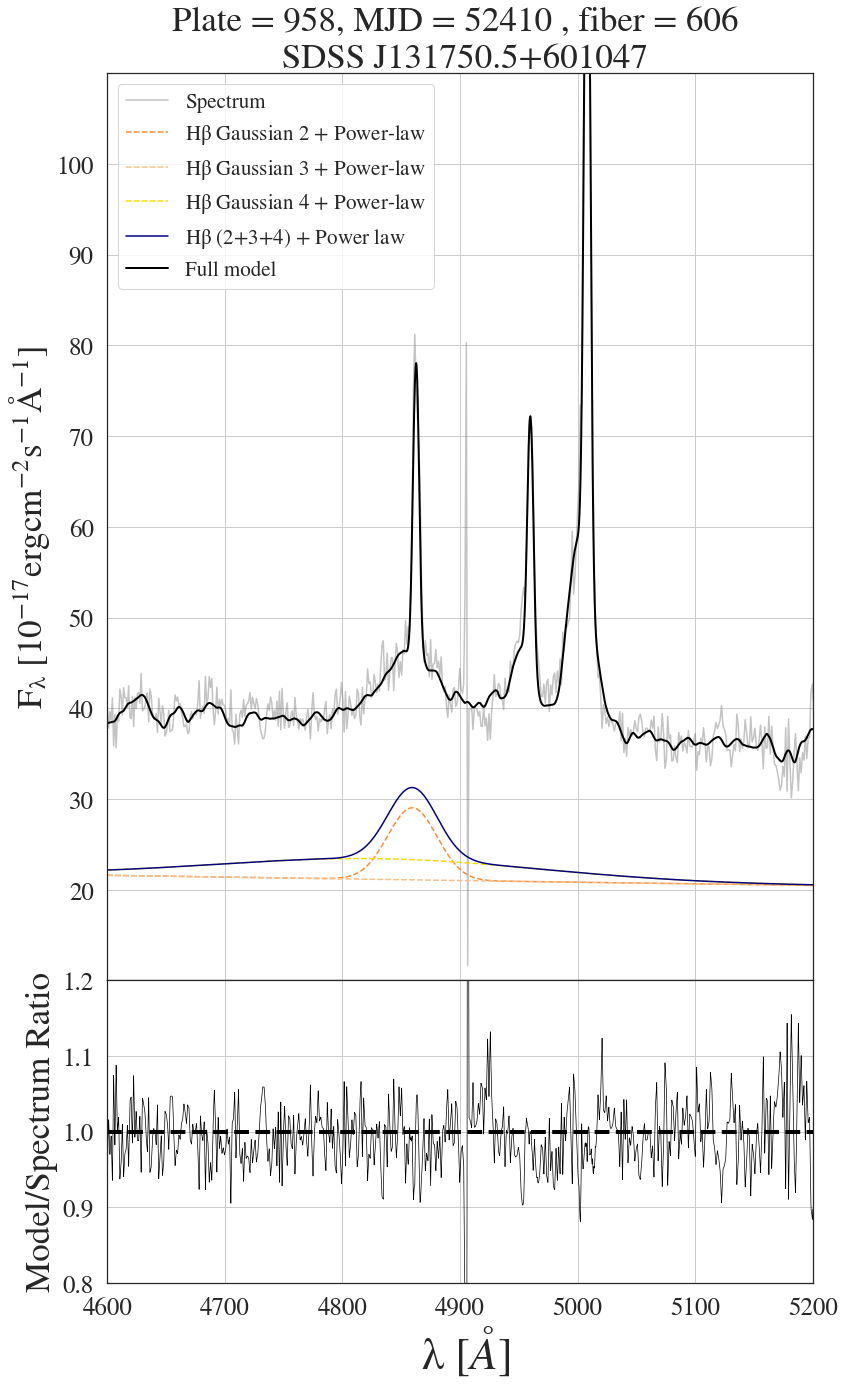}


    \includegraphics[width=4.4 cm]{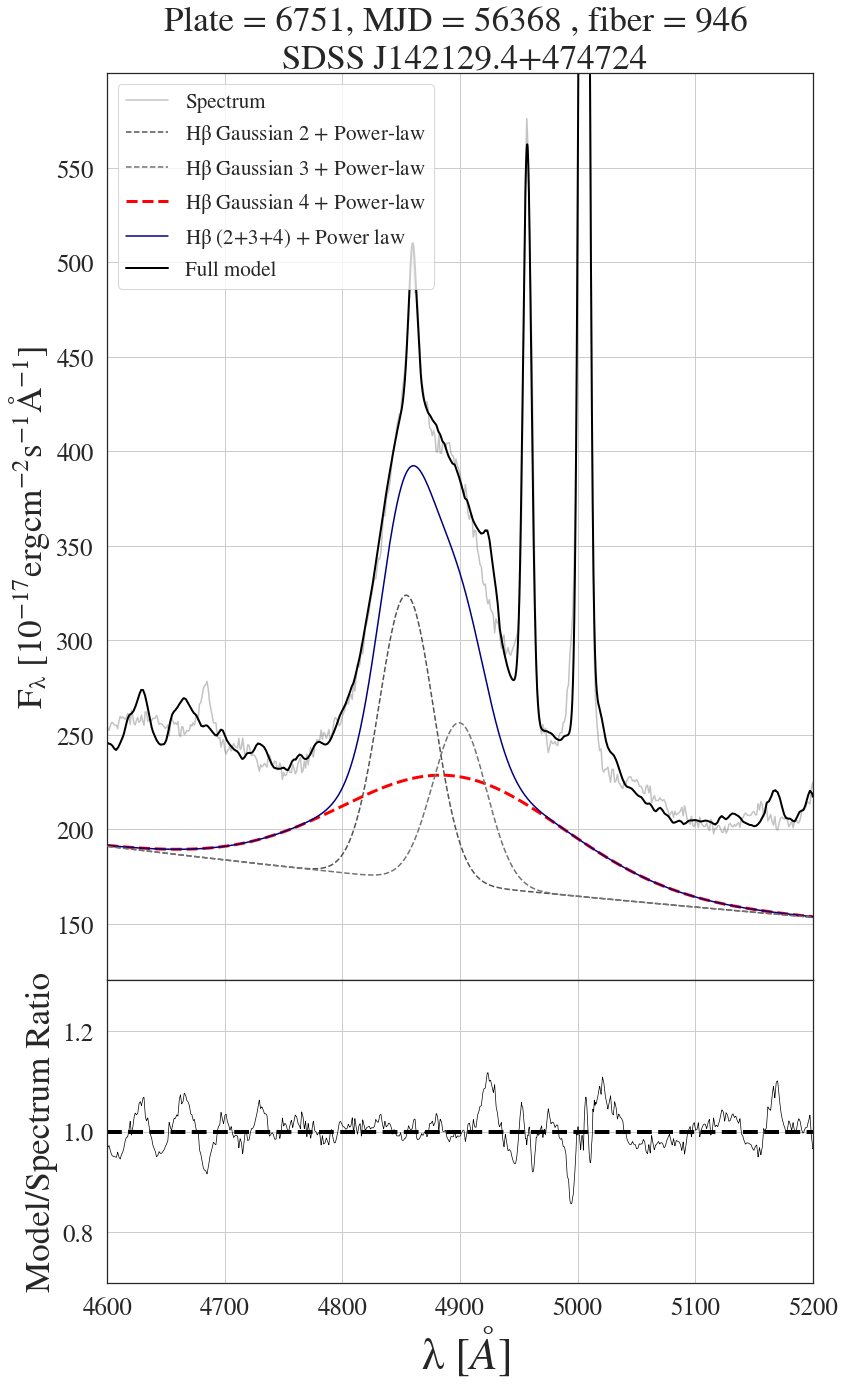}
    \includegraphics[width=4.4 cm]{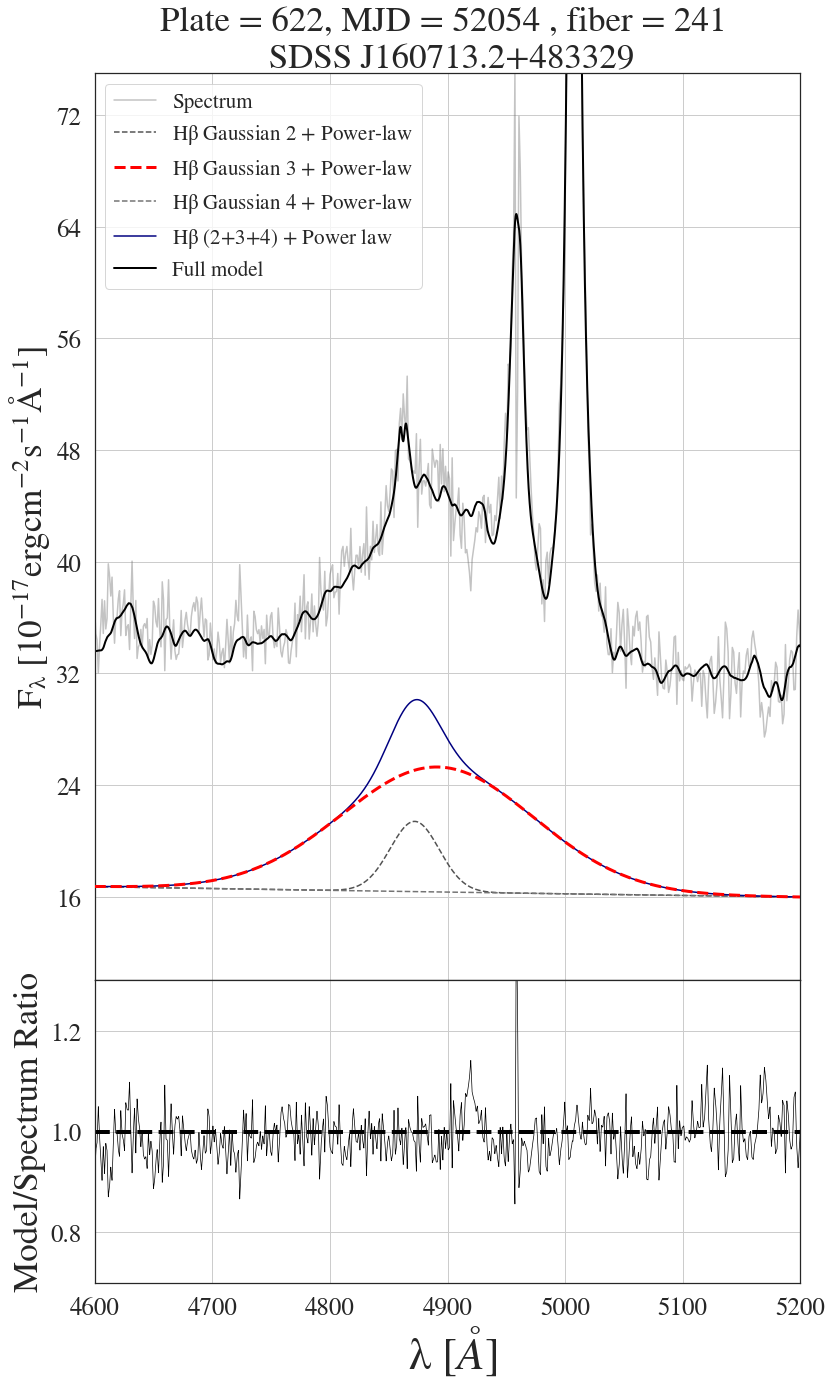}
    \includegraphics[width=4.4 cm]{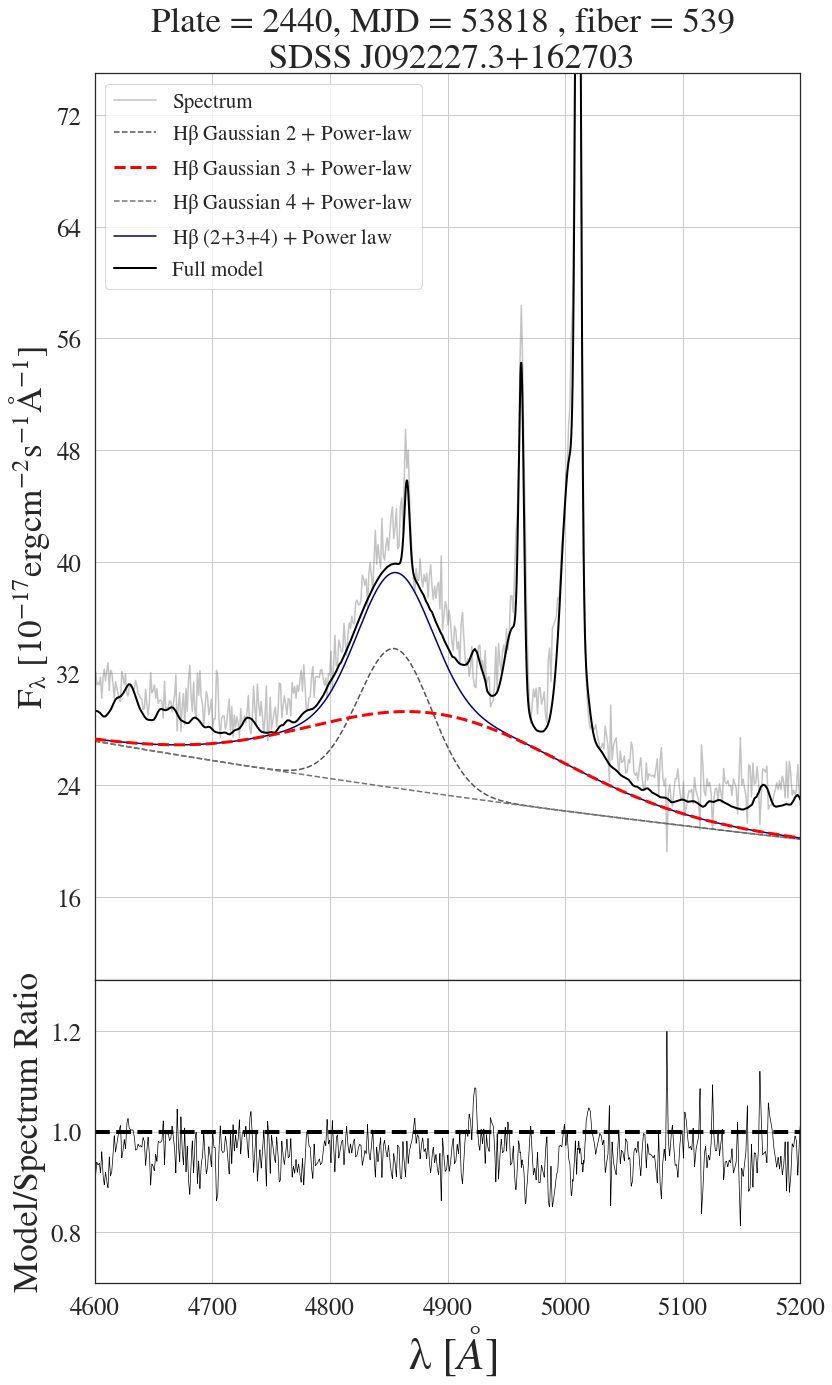}
    \label{fig:vbc_example}

\caption{{\bf Top:} The Gaussian line decomposition and spectra of the broad $\rm H\beta$ from three sources with red-asymmetric $\rm H\beta$ profiles. From left to right, the asymmetry indices $\rm \Delta \lambda_{H\beta}$ are: +0.22, +0.22 and +0.27.
{\bf Middle:} The Gaussian line decomposition and spectra of the broad $\rm H\beta$ from three sources with blue-asymmetric $\rm H\beta$ profiles. From left to right, the asymmetry indices $\rm \Delta \lambda_{H\beta}$ are: -0.23, -0.20 and -0.24. 
{\bf Bottom:} The Gaussian line decomposition and spectra of the broad $\rm H\beta$ from three sources which have one very broad Gaussian component ($\rm FWHM > 10.000 \, km.s^{-1}$. The very broad component is marked in red. }
    \centering
\end{figure*}

\section{Testing a simple obscuration scenario}
\label{sec:obscuration}

\citet{richards02} proposed that the spatial configuration resulting from the BLR's inclination and obscuration might explain the asymmetries in the observed profile of $\rm CIV\lambda1549\angstrom$ (see also 4.3. in Z10). In this picture, the asymmetry arises from the attenuation of redshifted photons by obscuring material. Similarly \citet{gaskell18} developed a model of outflowing clouds of dust, blocking the line of sight to the inner disk-shaped BLR, which predicts broad emission profiles similar to those observed.  An alternative view of the effect of the BLR and outflows is developed by \citet{czerny17} (see also \citealt{czerny11}). In this model the dusty clouds are driven out of an optically thick disk, and are eventually exposed to radiation by the central source. The dust evaporates, removing the matter/radiation interaction and results in the fallback of gas clouds. This scenario is referred to as Failed Radiatively Accelerated Dusty Outflow (FRADO). The physics of dust sublimation and the consequences for the BLR disk structure were further investigated by \citet{baskin18}.

Fig. \ref{fig:centroid} allows us to investigate a simple obscuration scenario in which the observed asymmetry of the line shapes of $\rm H\beta$ would result from partial obscuration of a flattened BLR in Keplerian motion around the black hole. If the receding or incoming region (with respect to our line of sight) of a disk like structure is obscured by optically thick material, the emitted broad lines would be respectively red- or blue asymmetric (see for example Fig. 2 in \citealt{gaskell18}), since one peak of the double-peaked emission profile would be attenuated. In this scenario, the low-intensity, `excess' emission measured on blue (or red) side of the line would correspond to the attenuated blue (or red) horn of the disk-like emission. 

Furthermore, if we assume no further kinematics in our BLR model, the attenuation of the red of blue peak of the disk like emission would lead to a centroid shift configuration :
\begin{itemize}
   \item Blue-ward asymmetric $\rm H\beta$ ($\rm \Delta \lambda_{H\beta}<0$): Redshift of the peak component ($\rm c_{80}>0$) and blueshift of the base component ($\rm c_{15}< 0$)
   \item Red-ward asymmetric $\rm H\beta$ ($\rm \Delta \lambda_{H\beta}>0$): Blueshift of the peak component ($\rm c_{80}<0$) and redshift of the base component ($\rm c_{15}> 0$)
\end{itemize}

Fig.\ref{fig:centroid} supports this scenario. The results in the previous section suggest that the broad Balmer shapes might be affected by the presence of a distinct emitting region: the VBLR. The next step is to test if the obscuration scenario is compatible with the presence of a very broad component related to the stratification of the BLR. To this effect we identify sources for which the broad $\rm H\beta$ has been fit broad Gaussians. 
 
\begin{figure}
   \includegraphics[width=8 cm]{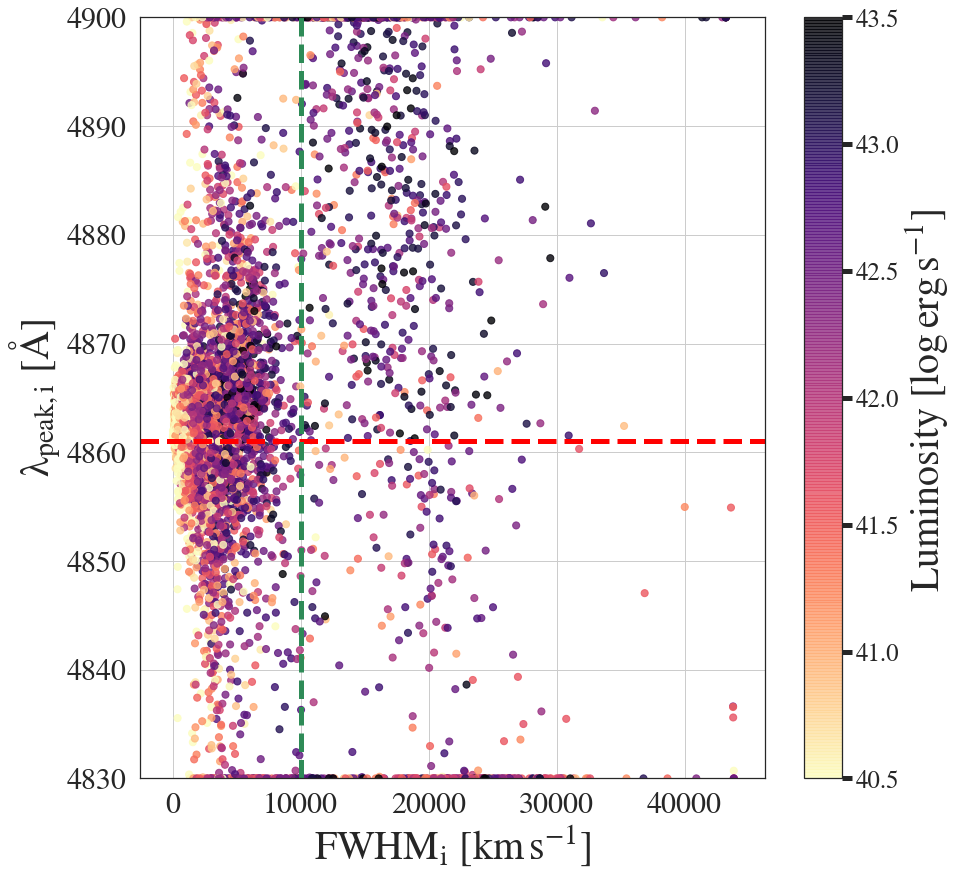}
   \caption{The peak wavelength of the Gaussians used to fit the $\rm H\beta$ components as a function of their FWHM. The red-dotted line indicates the systemic wavelength of $\rm H\beta$. The green dotted lines shows an empirical separation of broad and very broad Gaussian components at $\rm 10.000 \, km \, s^{-1}$. The right side of this separation clearly reveals a sequence of very broad, mainly redshifted Gaussians. The points are colour-coded according to the luminosity from each Gaussian. Each source in our sample has four points in this figure (one for each of its Gaussian).}
    \label{fig:width1}
\centering
\end{figure}

Fig. \ref{fig:width1} presents the peak wavelength of all the Gaussians which have been used to fit the components of $\rm H\beta$ as a function of their FWHM. We clearly identify a sequence of very broad Gaussians at velocities $\rm > 10.000 \, km \, s^{-1}$ which are mostly redshifted with respect to systemic redshift.  We investigated the reduced $\chi^2$ distribution of the fits of the sources which have at least one Gaussian component of FWHM  $\rm >10.000 \ km.s^{-1}$. The sources which have a very broad component are not biased towards higher reduced $\chi^2$ (and thus higher BIC) values, indicating that they are not preferentially found in less secure fits. The statistical significance of a second mode in the distribution of the Gaussian FWHM in our sample at $\rm FWHM > 10.000 \, km \, s^{-1}$ was assessed with a Silverman test of multimodality (\citealt{silverman81}, for its calibration see e.g. \citealt{hall01,ameijeiras16}). We could confirm that this second mode of very broad Gaussian components is indeed significant. The test is presented in details in Appendix \ref{sec:silv}. We empirically identify this sequence as the sub-class of sources which contain a VBLR. As expected the broadest Gaussians have the largest contributions to the full line luminosity.
Fig.\ref{fig:centroid} includes the density contours containing the sources for which $\rm H\beta$ has not been fitted with a very broad Gaussian ($\rm FWHM < 10.000 \, km \, s^{-1}$, i.e., the sources to the left of the vertical dashed line in Fig.\ref{fig:width1}). The trend observed for all objects in our sample is preserved for this subset, supporting the possibility of a partially obscurated flattened BLR + VBLR model. The red- and blueshifts of the $\rm H\beta$ base and peak components are symmetrically distributed, which is consistent with the obscuration scenario.

\begin{figure*}
    \includegraphics[width=8 cm]{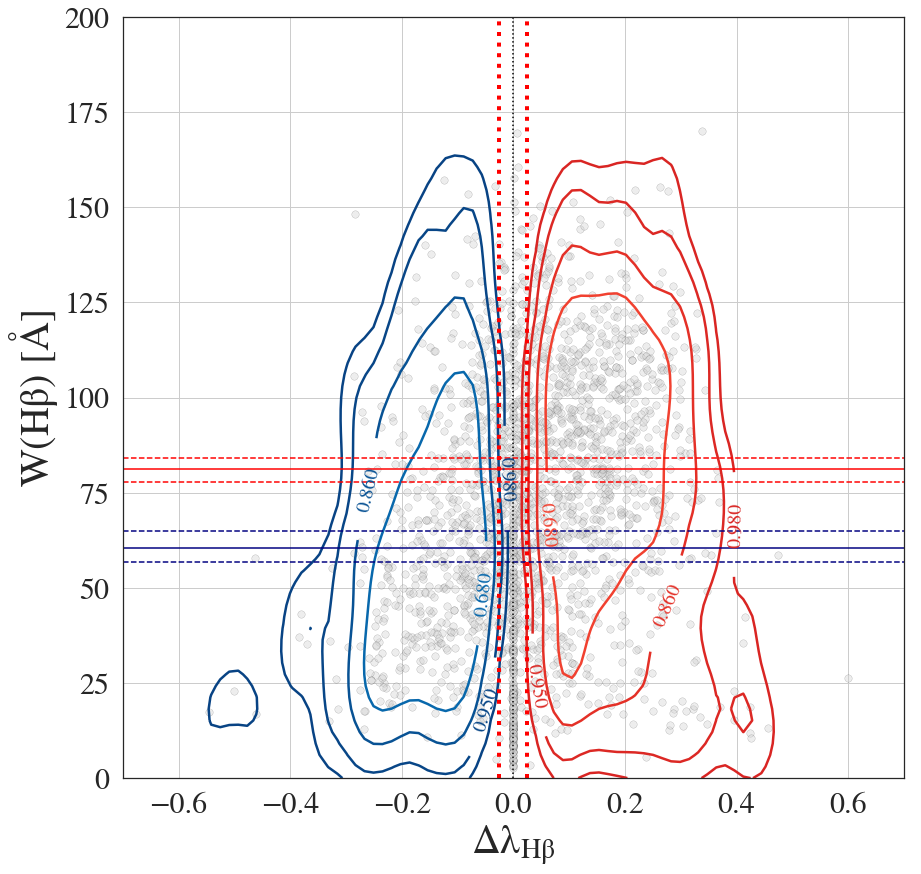}
    \includegraphics[width=8cm]{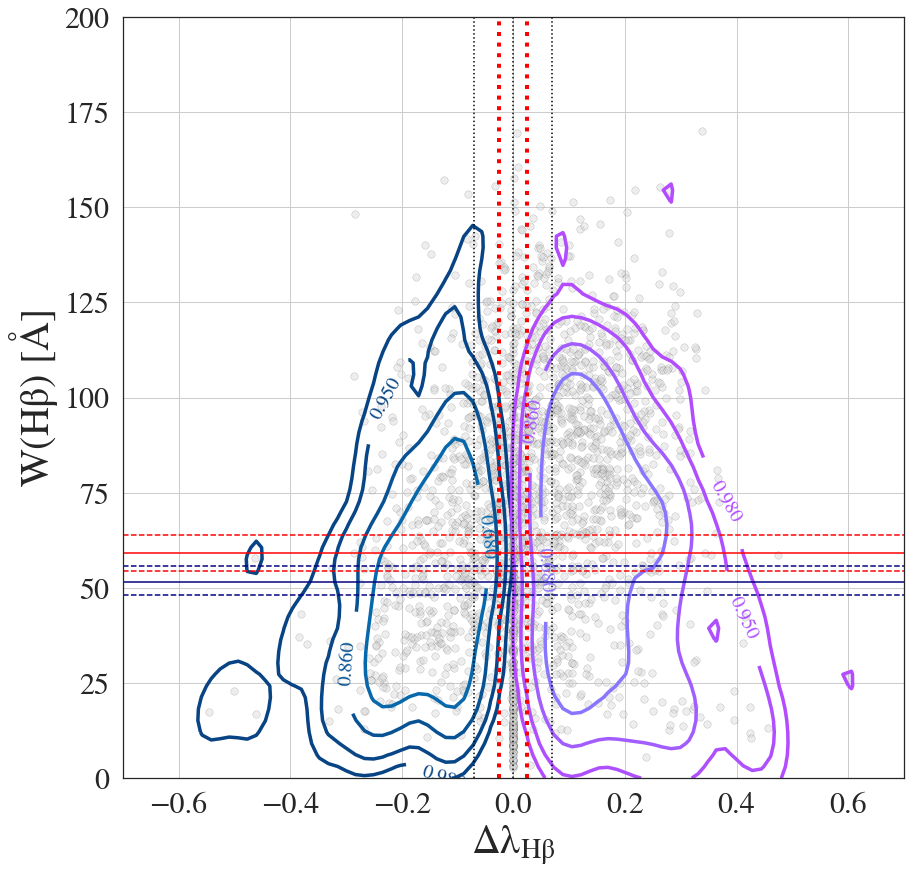}
    \caption{\textit{Left panel}: The equivalent width of the broad $\rm H\beta$ emission as a function of its asymmetry Index. The blue and red density contours are included to improve the visualisation of the red- and blue asymmetric sub-samples. An offset between the two populations is observed. The horizontal red (blue) line mark the mean of the bootstrapped mean $\rm W(H\beta)$ of the the red- (blue-)ward asymmetric sources. The red and blue, horizontal dashed lines correspond to the $3\sigma$ confidence contours, as derived from the percentile method at $0.13 \%$ and $99.87 \%$  of the sampling distribution of the mean. \textit{Right panel}: Same format as on the left. The blue and purple contours respectively represent the sources with blue- and red-asymmetric $\rm H\beta$, for which the Gaussian models had $\rm FWHM < 10.000 \, km \, s^{-1}$.} 
    \label{fig:whb_asym}
    \centering
\end{figure*}

In a simple model in which wing attenuation via partial obscuration is the sole source of asymmetry of the broad emission lines, we expect the $\rm H\beta$ line fluxes to be symmetrically distributed for $\rm \Delta\lambda_{H\beta}<0$ and $\rm \Delta\lambda_{H\beta}>0$, as no wing of the emission line profile should be preferentially attenuated. 
 The left panel in Fig.\ref{fig:whb_asym} shows the equivalent width of $\rm H\beta$ plotted against its asymmetry index for our full sample. We observe an offset in $\rm W(H\beta)$ between the population of red- and blue-asymmetric emitters. We argue that this offset might once again be the signature of the presence of a VBLR in some of our sources. If we display the kernel density contours for the subset of sources which were not fitted with a very broad Gaussian, a more symmetric distribution of equivalent widths is produced (i.e., line fluxes), as shown in the right panel of Fig. \ref{fig:whb_asym}.

 These simple tests have demonstrated that our sample is consistent with a model of partial obscuration coupled to a stratified BLR (one that might or might not contain a VBLR).

\section{VBC mode significance with a Silverman test of multi-modality}
\label{sec:silv}
The FWHM of broad Gaussian components ($\rm > 800 \, km\,s^{-1}$) that were fit to the $\rm H\beta$ emission lines show a bimodal distribution.  The two modes appear to distinguish broad components ($ \rm 800 \, km/s \, < \,  FWHM \,  < \, 10.000 \, km/s$) and very broad components ($ \rm FWHM > 10.000 km\,s^{-1}$). Out of the 8400 Gaussian components used to fit the 2100 sources of the sample, 4936 ($58 \%$) can be considered broad, with $\rm FWHM > 800 \,  km \, s$ and 1080 ($13 \%$) are very broad components with $\rm FWHM > 10.000 \, km \, s$. The significance of the second mode was assessed using the Silverman test of multimodality \citep{silverman81}.  We present the outline of this test here.

\begin{itemize}
        \item We extract all the Gaussians used in the fits of our sample with the condition: $\rm FWHM > 800 \, km\,s^{-1} \, \,  \& \, \,  \lambda_{peak} \neq 4830 \angstrom \, \, \& \, \, \lambda_{peak} \neq 4900 \angstrom$. This cut removes narrow Gaussians, as well as Gaussian which central wavelength was shifted to the edge of the fitting window. 4275 Gaussian components remain. We show the resulting density histogram of FWHM in Fig. \ref{fig:dens}.
        
        \begin{figure}[h]
           \centering
           \includegraphics[width=3.0in]{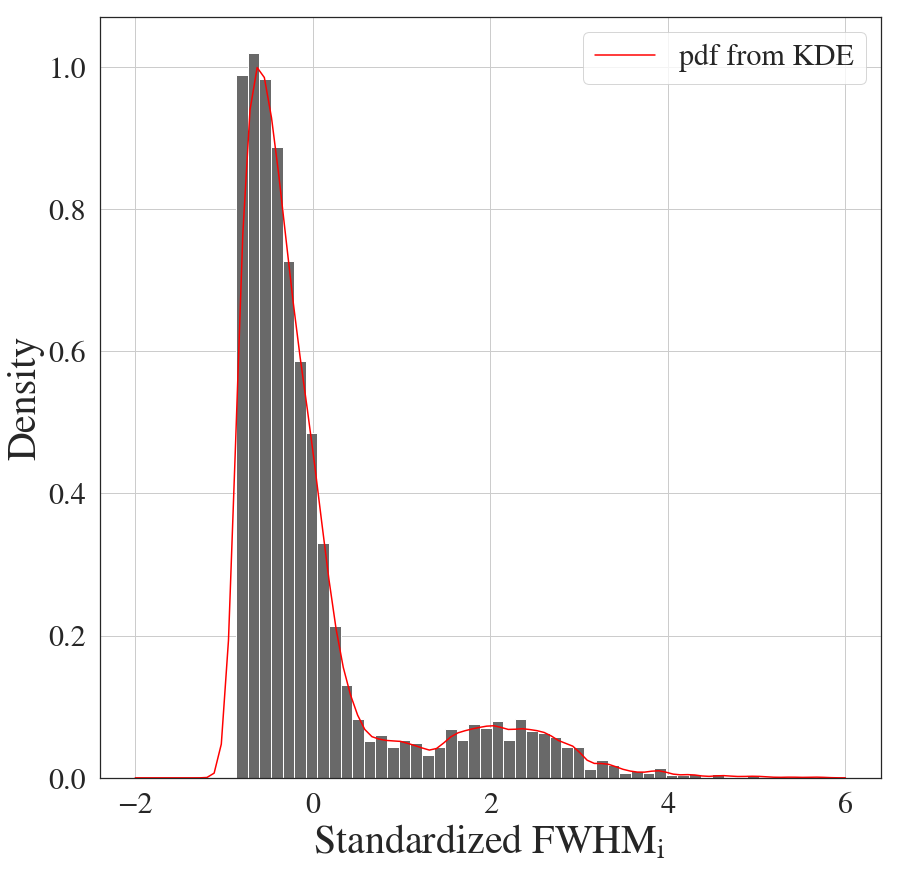}
           \caption{ Density of standardadised Gaussian component FWHM. }
            \label{fig:dens}
          \end{figure}
          
        \item For a range of bandwidths $h = [0.01,2.00]$ (step size s= 100), we performed a Gaussian kernel density estimate (KDE) of the distribution of $\rm FWHM$ (centered at zero and scaled to unit variance).  An example probability density function (pdf) from KDE is displayed in Fig. \ref{fig:dens} (red line). For $n$ observations in our sample, the bandwidth $h$ smooths the Gaussian kernel estimate as follows:
        
        \begin{equation}
            \hat{f}(x,h) = \frac{1}{nh}\sum_{i = 1}^n \frac{1}{\sqrt{2\pi}}e^{-\frac{(x-X_i)^2}{2h^2}}
        \end{equation}
        
        \item We then detect the critical bandwidths $h_{c,j}^{\text{DATA}}$ , at which new local maxima appear in the pdf, when $h$ is decreased. For $1<j<n$ (where n corresponds to the number of observations),  the critical bandwidths $h_{c,j}^{\text{DATA}}$  are formally defined as the minimum $h$, at which the Gaussian density estimate has no more then $j$ maxima.
        In Fig. \ref{fig:kh}, the bandwidth $h$ and the corresponding number of detected local maxima $j$ are displayed. Critical widths are symbolized by vertical lines.
        
        \begin{figure}[h]
           \centering
           \includegraphics[width=3.0in]{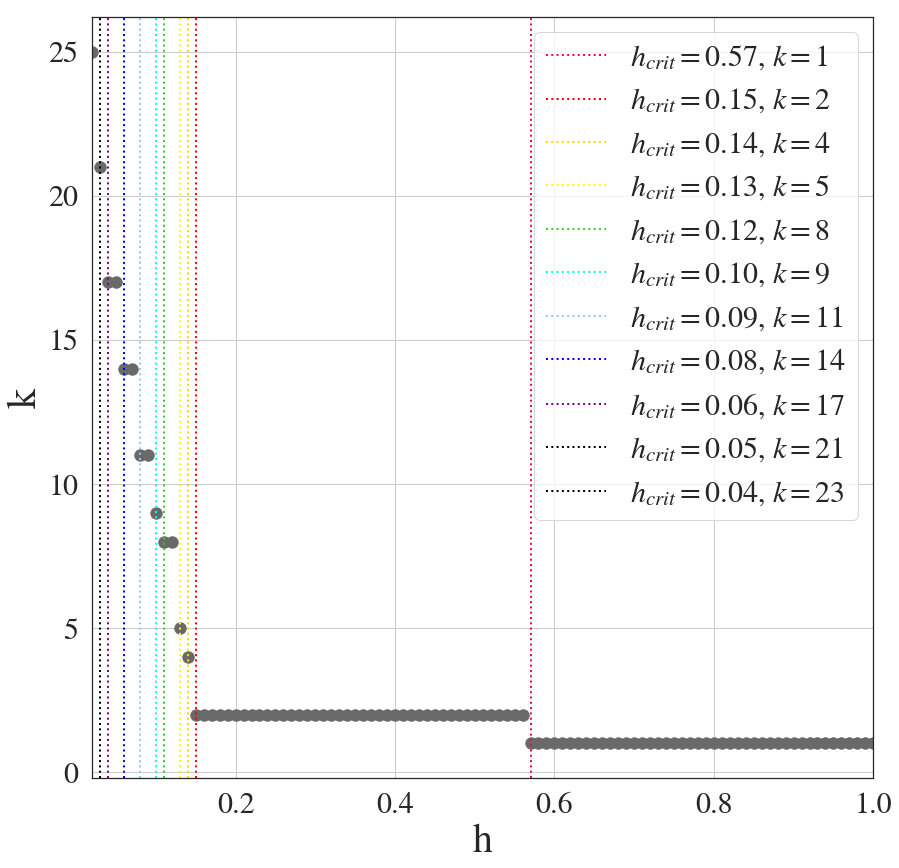}
           \caption{ Number of local maxima $k$  of the estimated pdf as a function of the smoothing $h$, the bandwidth of the Gaussian KDE. The critical values $h_{c,j}^{\text{DATA}}$  are marked as vertical lines.  }
            \label{fig:kh}
          \end{figure}
          
        \item In order to assess the significance of a mode detection,  we follow \citet{fisher94} and set up a null hypothesis test, which is based on the observation that the number of modes of $\hat{f}(x)$ decreases with increasing bandwidth $h$. If $k$ is the true number of modes: \newline 
        
        \begin{center}
        Null hypothesis $H_0: k = j$\newline
        Alternative $H_1: k \geq j +1$ for $j=1,2...$            
        \end{center}
        
        For each $h_{c,j}^{\text{DATA}}$ detected in our data, we perform a Gaussian kernel density estimate on 100 smoothed bootstrapped samples, using $h_{c,j}^{\text{DATA}}$  as bandwidth. These resamples are obtained, by sampling smoothed $\rm FWHM_{s}$, using bootstrapping (with replacement). The exact computation of $\rm FWHM_{s}$  is as follows:
        
        \begin{equation}
           \rm{FWHM_{s}} = \frac{1}{\sqrt{1+{h_{c,j}}^2/\sigma^2}}\left(\rm{FWHM_{r}} + h_{c,j} \epsilon \right)
        \end{equation}
        
        were $\rm FWHM_{r}$ are resampled FWHM, $\rm \sigma$, the standard deviation of the bootstrapped sample and $\epsilon$ is a random standard variable.
        For the null-hypothesis test \citet{silverman81} motivates a p-value defined for a given number of maxima j as:
        
        \begin{equation}
            p = P\{h_{c,j}^\text{NEW} > h_{c,j}^{\text{DATA}}\}  = P\{\hat{f}(x,h_{c,j}^{\text{DATA}}) \text{ has } > j \text{ modes}\},
        \end{equation}{}
        
        where $h_{c,j}^\text{NEW}$ is a the critical value for j maxima derived from the resamples, $h_{c,j}^{\text{DATA}}$ the initial critical width from the data, $\hat{f}(x,h_{c,j}^{\text{DATA}})$ the kernel density estimate of the resamples using $h_{c,j}^{\text{DATA}}$ as critical density.
        The null hypothesis is rejected if $\rm P\{h_{c,j}^\text{NEW} > h_{c,j}^{\text{DATA}}\} < \alpha$.
        For each $h_{c,j}^{\text{DATA}}$, and thus each mode detection, we compute a p-value. 
        We assess how many times the density estimates with fixed bandwidths  of the 100 resamples have at most $j$ detected modes. The ratio of these realizations and the total number of simulations are taken as p-value. The resulting p-values are listed in Table \ref{Tab:tcr}.  

        \begin{table}
        \centering

        \begin{tabular}{|l|l|l|lll}
        \cline{1-3}
        $\rm h_j^{DATA}$ & Number of local maxima & p-value &  &  &  \\ \cline{1-3}
        $0.57$           & 1                      & 0.100   &  &  &  \\ \cline{1-3}
        $0.15$           & 2                      & 0.001   &  &  &  \\ \cline{1-3}
        $0.14$           & 3                      & 0.008   &  &  &  \\ \cline{1-3}
        $0.13$           & 5                      & 0.041   &  &  &  \\ \cline{1-3}
        $0.12$           & 8                      & 0.095   &  &  &  \\ \cline{1-3}
        ...              & ...                    & ...     &  &  &  \\ \cline{1-3}
        \end{tabular}
         \caption{p-values for mode detections for decreasing critical bandwidths.}
          \label{Tab:tcr}

        \end{table}

\item Following \citet{silverman81}, we read this table as hierarchical successive significance test. As soon as for a new $j$, the p-values drop below an $\alpha$ significance threshold, we interpret the detection of $j$ modes as significant. From Table \ref{Tab:tcr}, if we adopt a significance threshold $\alpha = 0.05$, it appears that the bimodality of the distribution of our Gaussian components FWHM is significant, i.e. the very broad mode of Gaussian components is significant.

    \end{itemize}
    